\documentclass{aa}  

\usepackage{graphicx}

\usepackage[colorlinks,
            linkcolor=blue,
            citecolor=blue,  
            ]{hyperref}
\usepackage{txfonts}
\begin{document}

   \title{Influence of a mass transfer stability criterion on double white dwarf populations}
   \titlerunning{Influence of $q_{\rm c}$ on DWD populations}

   \author{Li Zhenwei
          \inst{1,2,3}
          \and
          Chen Xuefei\inst{1,2,3} 
          \and 
          Ge Hongwei\inst{1,2,3}
          \and
          Chen Hai-Liang\inst{1,2,3}
          \and
          Han Zhanwen\inst{1,2,3,4} 
          }
   \authorrunning{Li Z. et al.}

   \institute{Yunnan Observatories, Chinese Academy of Sciences, Kunming, 650011, People’s Republic of China\\ 
              \email{lizw@ynao.ac.cn;cxf@ynao.ac.cn}
         \and
             Key Laboratory for the Structure and Evolution of Celestial Objects, Chinese Academy of Science, People’s Republic of China
         \and 
             Center for Astronomical Mega-Science, Chinese Academy of Science, 20A Datun Road, Chaoyang District, Beijing 100012, People’s Republic of China
         \and
             University of the Chinese Academy of Sciences, 19A Yuquan Road, Shijingshan District, Beijing 100049, People's Republic of China
             }

   \date{Received XXX; accepted XXX}

% \abstract{}{}{}{}{} 
% 5 {} token are mandatory
 
  \abstract
  % context heading (optional)
  % {} leave it empty if necessary  
   {Mass transfer stability is a key issue in studies of binary evolution.
    Critical mass ratios for dynamically stable mass transfer have been analyzed on the basis of an adiabatic mass loss model, finding that the donor stars on the giant branches tend to be more stable than that based on the composite polytropic stellar model. Double white dwarfs (DWDs) are of great importance in many fields and their properties would be significantly affected under the new mass transfer stability criterion.}
  % aims heading (mandatory)
   {We seek to investigate the influence of mass transfer stability on the formation and properties of DWD populations and discuss the implications in supernova Type Ia (SN Ia) and gravitational wave (GW) sources.
}
  % methods heading (mandatory)
   {We performed a series of binary population synthesis, adopting the critical mass ratios from the adiabatic mass loss model (i.e., Ge's model) and that of the composite polytropic model, respectively.
   In each simulation, $5\times 10^6$ binaries were included and evolved from zero-age main sequence to the end of their evolution and the DWDs were gradually obtained. 
   %Two mass transfer stability criteria, which are obtained from realistic stellar models (Ge's model) and composite polytropic model (polytropic model) respectively, are considered for comparison.
   }
  % conclusions heading (optional), leave it empty if necessary 
   {
   %By using the new dynamical stability criterion of Ge's model, 
   For Ge's model, {most of the DWDs are produced from the stable non-conservative Roche lobe (RL) overflow, along with a common-envelope (CE) ejection channel (RL+CE channel), regardless of the CE ejection efficiency, $\alpha_{\rm CE}$. Conversely, the results of the polytropic model strongly depend on the adopted value of $\alpha_{\rm CE}$. We find DWDs produced from the RL+CE channel have comparable WD masses and the mass ratio distribution peaks at around 1.} Based on the magnitude-limited sample of DWDs, the space densities for the detectable DWDs and those with extremely low-mass WD (ELM WD) companions in Ge's model is: $1347\;\rm kpc^{-3}$ and $473\;\rm kpc^{-3}$, respectively, which is close to what has been shown in observations. {On the other hand, the polytropic model overpredicts space density of DWDs by a factor of about 2-3.}
   We also find that the results of DWD merger rate distribution {per Galaxy} in Ge’s model reproduce the observations better than that of the polytropic model, and the merger rate of DWDs with ELM WD companions {in the Galaxy} is about $1.8\times 10^{-3}\;\rm yr^{-1}$ in Ge's model. This result is comparable to the observation estimation of $2\times 10^{-3}\;\rm yr^{-1}$. The findings from Ge's model predict a Galactic {SN} Ia rate of $\sim 6\times 10^{-3}\;\rm yr^{-1}$ from DWDs, supporting observations of $(5.4\pm1.2)\times 10^{-3}\;\rm yr^{-1}$. For the fiducial model {of $\alpha_{\rm CE}=1$}, the number of detectable GW sources in the polytropic model is larger than that in Ge's model by about $35\%$.}
   {We confirm that mass transfer stability plays an important role in the formation and properties of DWD populations as well as in the progenitors of SNe Ia and detectable GW sources. {The results of Ge's model support the observational DWD merger rate distribution per Galaxy and the space density of DWDs in the Galaxy.}}


   \maketitle
%
%________________________________________________________________

\section{Introduction}
\label{sec:1}

White dwarfs are the most common stellar remnants in the Universe and they provide effective ways to study stellar evolution and the star formation history of the Galaxy \citep{althaus2010,tremblay2014}. Double white dwarfs (DWDs) are of great importance in many fields. The merger of DWDs can produce type Ia supernovae (SNe Ia),
which are crucial for studying galaxy evolution and cosmology \citep{riess1998,perlmutter1999}. Some close DWDs with orbital periods of several minutes, for instance, ZTF J1539+5027 with an orbital period of 6.91 minutes \citep{burdge2019} or SDSS J0651+2844 with an orbital period of 12.75 minutes \citep{brown2011b}, 
are the main detectable sources for future space-based gravitational-waves detectors, such as Laser Interferometer Space Antenna \citep{LISA2017}, TianQin \citep{luo2015}, and TaiJi \citep{taiji2018}. The formation of DWDs is of particular interest in improving our understanding of the details of binary evolution, including stellar evolution, tidal effect, mass transfer, and common envelope (CE) evolution, and so on (\citealt{han1998,nelemans2000,fuller2013}; see also \citealt{han2020} for a recent review). 

The exploration of DWDs in observations has been ongoing over several decades (e.g., \citealt{saffer1988,marsh1995,moran1997,marsh2011,chandra2021,kosakowski2020}). With the operation of large-scale survey projects, such as Sloan Digital Sky Survey (SDSS, \citealt{york2000,badenes2009,mullally2009,badenes2012,breedt2017,chandra2021,kosakowski2021}), ZTF Survey \citep{ZTF2019,burdge2019,burdge2020a,burdge2020b,coughlin2020,keller2022}, Gaia \citep{GAIA2016,elbadry2018,elbadry2021,torres2022}, 
and the extremely low-mass (ELM) survey \citep{brown2010,brown2012,brown2013,brown2016a,kilic2011a,kilic2012,gianninas2015,brown2020,brown2022}, the number of known DWDs has increased exponentially. 
However, since WDs are intrinsically faint, only about $\sim 150$ DWDs have had their precise orbital parameters measured 
and most of them include one ELM component, namely, with a mass $\lesssim 0.3M_\odot$, and have been found by the ELM survey. 
Nevertheless, the samples are sufficiently large to put some limits on theoretical studies.

In theory, binaries are expected to experience at least one mass transfer phase in order to produce close DWDs \citep{webbink1979}. 
{Many factors may affect mass transfer stability, for instance:\ the mass ratio of the binary, structures of both components at the onset of mass transfer, mass accretion efficiency, angular momentum loss, and others. For the case of unstable mass transfer, 
the binary will enter into the common envelope evolution phase. 
A close binary will be remained if enough energy is produced to eject the CE; otherwise the binary merges.} 
The classical treatment for the CE evolution is known as standard energy mechanism (or $\alpha-$mechanism), namely, it is a part (namely $\alpha$) of the orbital energy {released due to orbital shrinkage} is used to eject the CE \citep{webbink1984,livio1988,dekool1990}. 
{Based on the $\alpha-$mechanism, 
Han (1998) systematically investigated the formation of DWDs via binary interactions and explained, to a satisfactory extent, the distributions of masses, mass ratios, orbital periods, and birth rate of the observed DWDs at that time. When reconstructing the evolution of double helium WDs, however, \citet{nelemans2000} found that the orbital periods of the progenitors after the first phase of mass transfer cannot be explained by the standard CE ejection, nor the stable conservative mass transfer. They then proposed a revised CE ejection mechanism, namely, the $\gamma-$mechanism, based on angular momentum balance for the first mass transfer phase to reproduce the observed properties of these DWDs.} 
Here, $\gamma$ is a dimensionless parameter and indicates the specific angular momentum of the envelope related to the average specific angular momentum of the total mass.
Assuming that the first CE process is based on the $\gamma-$mechanism while the second CE process is the $\alpha-$mechanism, 
\citet{nelemans2001b} studied the formation of DWDs by a binary population synthesis (BPS) method and found that the Galactic DWDs are well reproduced with $\gamma$ in the range of 1.5-1.75 (see also \citealt{nelemans2005}). 
Following these works, \citet{toonen2012,toonen2017} showed that 
the results based on the $\gamma-$mechanism support the local populations of DWDs better than that of $\alpha-$mechanism. 
However, there is no clear physical explanation for the $\gamma-$mechanism \citep{webbink2008,davis2010,zorotovic2010,woods2011}. 
Furthermore, {the study of \citet{woods2012} showed that substituting the first CE with $\gamma-$mechanism for stable non-conservative can also create most of the observed DWDs.} This indicates that the criterion for mass transfer stability could have a significant influence on the formation of DWDs.

The critical mass ratio for dynamically unstable mass transfer is a {long-standing} problem in binary evolution.
Whether or not the mass transfer is dynamically stable can be understood in terms of the response of donor star radius and Roche lobe (RL) radius to the mass transfer rate. It is convenient to define the following:\ 
\begin{eqnarray}
  \zeta_{\rm ad} = \left(\frac{{\rm d}\ln R_1}{{\rm d}\ln M_1}\right)_{\rm ad},
\label{eq:1}
\end{eqnarray}

\begin{eqnarray}
\zeta_{\rm RL} = \frac{{\rm d}\ln R_{\rm RL,1}}{{\rm d}\ln M_1},
\label{eq:2}
\end{eqnarray}
where $M_1$ is the donor mass, $R_1$ and $R_{\rm RL,1}$ is the donor radius and its Roche lobe radius, respectively. In addition, $\zeta_{\rm ad}$ and $\zeta_{\rm RL}$ represent, respectively, the adiabatic response and Roche lobe response of the donor to mass loss \citep{hjellming1987,soberman1997,tout1997}. 
The case of $\zeta_{\rm ad}\leq \zeta_{\rm RL}$ means that the donor departs from hydrostatic equilibrium and the mass transfer proceeds on a dynamical timescale. Then the donor envelope engulfs the companion star and the binary enters into CE evolution \citep{paczynski1976}. The tangency condition of $\zeta_{\rm ad}=\zeta_{\rm RL}$ defines a critical mass ratio, $q_{\rm c}$, above which the mass transfer is unstable to dynamical timescale mass transfer \citep{hjellming1987,tout1997,hurley2002}. Based on the polytropic stellar models, \citet{webbink1988} estimates the critical mass ratio for a giant with $M_{\rm c}/M_{\rm 1}\gtrsim 0.2$, as:
\begin{eqnarray}
q_{\rm c} = 0.362+\frac{1}{3(1-M_{\rm c}/M_{\rm 1})},
\label{eq:3}
\end{eqnarray}
where $M_{\rm c}$ is the core mass of the donor. The composite polytropic model is a good approximation for low-mass zero-age main sequence (MS) stars with a mass lower than $1M_\odot$. However, it may underestimate the critical mass ratio for stars on the giant phase \citep{chenx2008,woods2011}. In addition, Eq. \ref{eq:3} is obtained based on the assumptions of conservation evolution and the mass transfer stability should also depend on the mass and angular momentum loss during the Roche lobe overflow (RLOF) phase \citep{pastetter1989,soberman1997,chenx2008,vos2019}. Therefore, the calculation of critical mass ratio with {realistic stellar models in detailed adiabatic mass-loss model} is desirable.

By adopting the realistic stellar structure, the stability thresholds of stars on the Hertzsprung gap (HG) and the giant branches have been derived in many works (e.g., \citealt{chen2003,chenx2008,pavlovskii2015}). Moreover, \citet{geh2010,geh2015,geh2020} carried out a systematic survey of the thresholds for dynamical timescale mass transfer with the adiabatic mass loss model (see also \citealt{han2020}). They obtained a grid of critical mass ratio for donors with masses of $0.1-100M_\odot$ and from zero-age MS to the tip of the asymptotic giant branch (AGB), which can be put into the BPS studies. 

In this work, we attempt to investigate the influence of mass transfer stability on DWD populations. To visualize the results, we perform two sets of simulations. In the two models, the critical mass ratios are obtained from the composite polytropic model (hereafter, the polytropic model) and {detailed adiabatic mass-loss model with realistic stellar structures} (hereafter, Ge’s model), respectively. This paper is structured as follows. The formation channels of DWDs are introduced in Section \ref{sec:2} and the model inputs are given in Section \ref{sec:3}. The population synthesis results are presented in Section \ref{sec:4} and the implications for the observations are addressed in Section \ref{sec:5}. {In Section \ref{sec:6}, we discuss our results in comparison with the $\gamma-$mechanism, followed by a summary and conclusions in Section \ref{sec:7}}.

%__________________________________________________________________

\section{Formation scenarios of DWDs}
\label{sec:2}
In general, a binary will experience one or two mass transfer phases to produce a DWD \citep{han1998}. The formation of DWDs is summarized as follows for convenience. 
For brevity, we denote the mass transfer phase with RL if it is stable, that is, the mass ratio, $q,$ (the donor/the loser) is less than $q_{\rm c}$ at the onset of mass transfer,
and with CE if it is dynamically unstable ($q>q_{\rm c}$)
\footnote{{Accoding to \citet{han1998}, there is one more formation channel which has not been shown here 
i.e., exposed core + CE, which only works when the tidally enhanced stellar wind has been included. 
In this channel, the envelope of the initially more massive component is lost by the enhanced stellar wind. The core is exposed after all the envelope is lost and becomes a WD. The secondary evolves and fills its Roche lobe during the HG or red giant branch (RGB) stage, and the binary enters into the CE phase, leaving a DWD system after the ejection of the CE. 
We have not considered the tidally enhanced stellar winds in our studies.}}.

Starting with the RL + CE channel: the more massive star evolves first and fills its Roche lobe, transfers mass to the secondary stably, and leaves a WD with an unevolved secondary after this process. The secondary then evolves and fills its Roche lobe in the subsequent evolution, leading to the second mass transfer phase. On this occasion, the mass transfer is dynamically unstable, and the binary enters the CE stage. The DWD is formed after the successful ejection of CE.

As for the CE + CE channel: the process is similar to {the first channel}, but the first mass transfer phase is unstable. For the CE + RL channel: the first mass transfer phase is dynamically unstable, and the binary enters into CE evolution and consists of a WD and MS star after the ejection of the CE. The MS star evolves and fills its Roche lobe and transfer material to the WD stably, leaving a DWD after the end of mass transfer.
For the RL + RL channel: the two mass transfer phases are stable in this channel. For the single CE: the CE phase happens when both stars are on the RGB or AGB. Here the CE is made up of the envelope of the two stars, and the DWD will be produced after the ejection of the CE. 

The first two channels are the main formation channels of DWDs, as justified in previous works (e.g., \citealt{iben1997,han1998,nelemans2001a}). The CE+RL channel is one of two main channels for the formation of DWDs with ELM WD\footnote{In this work, ELM WD is defined as He WD with mass less than $0.3M_\odot$, as suggested by \citet{lizw2019}.} companions.
%where the parameter spaces that produce ELM WDs are obtained by detailed binary evolution code Modules for Experiments in Stellar Astrophysics (\textsf{MESA}, version 9575; \citealt{paxton2011,paxton2013,paxton2015}), and the calculation methods can be found in \citet{lizw2019}. 
The RL+RL channel will not be considered in this work.
The DWDs from this channel have very wide orbits ($\gtrsim 1\;\rm au$; \citealt{korol2022b}) and 
the orbital parameters for such wide DWDs are difficult to determine. %due to the lack of corresponding spectroscopic observations. 
The last channel will also not be taken into account due to the small proportion of DWDs from this channel.
%The single CE channel is incorporated in the CE+CE channel. This does not affect the final results, since the contributions of the single CE channel for DWDs are very small. 
\section{Methods and input parameters}
\label{sec:3}

We employed the rapid binary evolution code \textsf{BSE} \citep{hurley2000,hurley2002} to test the influence of mass transfer stability on the DWD populations. The single-star models are based on the analytic formulae that approximate the star evolutions with a wide range of stellar mass and metallicity. In this study, the initial stellar masses are adopted from $0.8M_\odot$ to $10M_\odot$, and the metallicity is set to be 0.02. The most important inputs are described as follows. 

\subsection{Stellar wind}
\label{subsec:3.1}

The stellar wind during HG and beyond is adopted from \citet{reimers1975} for low- and intermediate-mass stars, which is:
\begin{eqnarray}
\dot{M}_{\rm R} = 4\times 10^{-13}\eta\frac{R_1}{R_\odot}\frac{L}{L_\odot}\frac{M_\odot}{M_1}M_\odot\rm yr^{-1},
\label{eq:4}
\end{eqnarray}
where $L$ are the radius and luminosity of the star. The dimensionless parameter, $\eta,$ is limited by the observations in Galactic globular clusters \citep{iben1983} and is taken to 0.5. The wind loss on the thermal pulse-AGB (TP-AGB) stage is based on the prescription from \citep{vassiliadis1993}:\ 
\begin{eqnarray}
\log\left(\frac{\dot{M}_{\rm V}}{M_\odot\rm yr^{-1}}\right) &=& -11.4 + \\ \nonumber
 & & 0.0125\left(\frac{P_0}{\rm d}-100\; {\rm max}\left[\frac{M_1}{M_\odot}-2.5,0\right]\right),
\label{eq:5}
\end{eqnarray}
where $P_0$ is the Mira pulsation period given by:
\begin{eqnarray}
\log\left(\frac{P_0}{\rm d}\right) = {\rm min}\left(3.3,-2.7-0.9\log\left[\frac{M_1}{M_\odot}+1.94\log\left[\frac{R_1}{R_\odot}\right]\right]\right).
\label{eq:6}
\end{eqnarray}
The wind mass loss is limited by the steady superwind phase with a maximum of $\dot{M}_{\rm V} = 1.36\times 10^{-9}(L/L_\odot)M_\odot\rm yr^{-1}$. 

For the naked helium stars, the wind prescription is given by:
\begin{eqnarray}
\dot{M}_{\rm He} = {\rm max} (\dot{M}_{\rm WR},\dot{M}_{\rm R}), 
\label{eq:7}
\end{eqnarray}
where $\dot{M}_{\rm WR}=10^{-13}(L/L_\odot)^{1.5}M_\odot\rm yr^{-1}$ is the wind of Wolf-Rayet stars \citep{hamann1995,hamann1998}. 

%\subsection{Angular momentum loss}
\subsection{{Orbital changes}}
\label{subsec:3.2}

{There are several physical processes that affect the orbital separation, for instance, gravitational wave radiation (GWR), magnetic braking, mass loss, wind interaction, and tidal friction. In this work, we consider all of the above processes, and the calculation methods are introduced below. The eccentricity is set to be zero for convenience.}

\subsubsection{{Gravitational wave radiation}}
\label{subsubsec:3.2.1}

The GWR plays a crucial role in the close binary, and the angular momentum loss due to GWR is given by \citep{landau1975}
\begin{eqnarray}
\dot{J}_{\rm GW} = -\frac{32}{5c^2}\left(\frac{2\pi G}{P_{\rm orb}}\right)^{7/3}\frac{(M_1 M_2)^2}{(M_1+M_2)^{2/3}},
\label{eq:8}
\end{eqnarray}
where $G$ is the gravitational constant, $c$ is the speed of light, and $P_{\rm orb}$ is the orbital period.

\subsubsection{{Magnetic braking}}
\label{subsubsec:3.2.2}

The magnetic braking is considered for stars with appreciable convective envelopes and the angular momentum loss is calculated as \citep{rappaport1983}
\begin{eqnarray}
\dot{J}_{\rm MB} = -5.83\times 10^{-16}\frac{M_{\rm env}}{M_{\rm 1}}\left(\frac{R_{\rm 1}\Omega_{\rm spin}}{R_\odot\rm yr^{-1}}\right)^{\gamma_{\rm MB}}M_\odot R_\odot^2\rm yr^{-2},
\label{eq:9}
\end{eqnarray}
where $\gamma_{\rm MB}=3$, $M_{\rm env}$ is the envelope mass, and $\Omega_{\rm spin}$ is the spin angular velocity.

\subsubsection{{Stable Roche lobe overflow}}
\label{subsubsec:3.2.3}

For DWDs from the RL+CE and CE+RL channels, the binaries experience one stable mass transfer phase. Following \citet{woods2011}, we considered the non-conservative mass transfer process and the accretion efficiencies, $\beta_{\rm rlof}$, were adopted with different values according to the types of accretors. For the RL+CE channel, the accretor in the stable mass transfer phase is non-degenerate, and $\beta_{\rm rlof}$ is given by \citep{hurley2002}:
\begin{eqnarray}
  \beta_{\rm rlof} \equiv \frac{\dot{M}_2}{|\dot{M}_1|}= {\rm min} \left(10\frac{\dot{M}_{\rm KH,2}}{\dot{M}_1},\beta_{\rm max}\right),
  \label{eq:10}
\end{eqnarray}
where $\dot{M}_1 (<0)$ and $\dot{M}_2 (>0)$ are the mass transfer rate and the accretion rate, respectively; $\dot{M}_{\rm KH,2}$ is the thermal timescale mass transfer rate of the accretor and $\beta_{\rm max}$ is the upper limit of the accretion efficiency, which is adopted as $0.5$ in this work, as suggested by \citet{han1998}. For the CE+RL channel, the WD is formed after the CE phase and the accretor in the stable mass transfer phase is degenerate. This channel is one of the formation channels for the DWDs with ELM WD companions. And the calculations of this part have been done with detailed binary evolution code Modules for Experiments in Stellar Astrophysics (\textsf{MESA}, version 9575; \citealt{paxton2011,paxton2013,paxton2015}) in \citet{lizw2019}. In the simulations, the transferred material can be accumulated on the surface of WDs if the mass transfer rate is above the lower limit of the stability burning region of the WDs \citep{hachisu1999,kato2004} and then $\beta_{\rm rlof}$ is given.  For both cases, the lost material is assumed to remove the specific angular momentum of the accretor\footnote{{If more efficient angular momentum loss mechanisms, e.g., the outflow from the out Lagrangian point or a circumbinary disk, are considered, the orbital separation would shrink faster, then the critical mass ratio is supposed to get smaller. As a result, it may predict more binaries entering into the CE phase. On the other hand, if the binary still experiences stable mass transfer process, the orbital separation would shrink faster, resulting in a larger mass transfer rate and a shorter timescale for the mass transfer process. The WDs produced in this way will be smaller and have shorter orbital periods in comparison to our results in Ge's model.} 
}. The angular momentum loss due to the mass loss can be expressed as: 
\begin{eqnarray}
\dot{J}_{\rm ML} = -(1-\beta_{\rm rlof})\dot{M}_1\left(\frac{M_1}{M_1+M_2}\right)^2\frac{2\pi a^2}{P_{\rm orb}},
\label{eq:11}
\end{eqnarray}
where $a$ is the binary separation. 
%In this work, we consider the non-conservation mass transfer as suggested by \citet{woods2012}, and the $\beta_{\rm rlof}$ is set to be 0.5. 

\subsubsection{{Wind accretion}}
\label{subsubsec:3.2.4}

{The secondary can accrete some of the wind material as the star orbits through it, then the mean accretion rate is estimated as \citep{bondi1944,boffin1988}: }
\begin{eqnarray}
  \dot{M}_{\rm a} = {\rm min}\left(0.5\dot{M}_{\rm W}, -\left(\frac{GM_{\rm 2}}{v^2_{\rm W}}\right)^2\frac{\alpha_{\rm W}}{2a^2}\frac{1}{(1+v^2)^{3/2}}\dot{M}_{\rm W}\right),
  \label{eq:12}
\end{eqnarray}
{where $\dot{M}_{\rm W}$ is the wind mass loss rate, $v_{\rm W}=\sqrt{0.25GM_{\rm 1}/R_{\rm 1}}$ is the wind velocity of the donor \citep{hurley2002}, $\alpha_{\rm W}$ is the Bondi-Hoyle accretion efficiency parameter and is set to $3/2$ \citep{boffin1988}, $v^2 = v^2_{\rm orb}/v^2_{\rm W}$, $v^2_{\rm orb}=GM/a$, and $M\equiv M_1+M_2$ is the total mass. To be consistent with the accretion in the stable RLOF phase, The upper limit of wind accretion efficiency is also set to be 0.5. The wind material interacts with the binary orbit and leads to the change of the orbital angular momentum, which is given by \citep{hurley2002}: 
\begin{eqnarray}
  \dot{J}_{\rm W} = \left(\dot{M}_{\rm W}a_1^2-\frac{M_2}{M_1}\dot{M}_{\rm a}a^2_2\right)\Omega_{\rm orb},
  \label{eq:13}
\end{eqnarray}
where $\dot{M}_{\rm W}$ is the stellar wind calculated in Section 3.1, $a_1^2 = (M_2/M)^2a^2$ and $a_2^2=(M_1/M)^2a^2$. The wind accretion processes may lead to the shrinkage of binary orbit, and this issue is discussed in Appendix \ref{app:A}. }

\subsubsection{{Tidal friction}}
\label{subsubsec:3.2.5}

{
The tidal torque transfers angular momentum between the stellar spin and the orbit, then leads to the orbital changes. The exchange of angular momentum can be expressed as \citep{hurley2002}: 
\begin{eqnarray}
  \triangle J_{\rm orb} = -\dot{J}_{\rm spin} \triangle t, 
  \triangle J_{\rm spin} = \dot{J}_{\rm spin} \triangle t, 
  \label{eq:14}
\end{eqnarray}
where $\dot{J}_{\rm spin}$ is the rate of change of the rotational angular momentum and the equilibrium tide with convective damping is given by: 
\begin{eqnarray}
  \dot{J}_{\rm spin} = [k'_2(M_1-M_{\rm c})R_1^2+k'_3M_{\rm c}R^2_{\rm c}]\dot{\Omega}_{\rm spin}.
  \label{eq:15}
\end{eqnarray}
The constants $k'_2$ and $k'_3$ depends on the internal structure of a star \citep{hurley2000}, and $\dot{\Omega}_{\rm spin}$ is the rate of change of the spin angular frequency, which is given by \citep{hut1981}: 
\begin{eqnarray}
  \dot{\Omega}_{\rm spin} = 3\left(\frac{k}{T}\right)_{\rm conv}\frac{q^2}{r^2_{\rm g}}\left(\frac{R_1}{a}\right)^6\Omega_{\rm orb}\left(1-\frac{\Omega_{\rm spin}}{\Omega_{\rm orb}}\right), %check the fns value.
  \label{eq:16}
\end{eqnarray}
for convective damping, where $k$ is the apsidal motion constant, $T$ is the damping time-scale, $q=M_1/M_2$, and $r_g\equiv I/(M_1R_1^2)$ is the radius of gyration with momentum of inertia $I$ \citep{hurley2002}. The change rate of spin angular frequency is positively associated with $(R_1/a)^6$, and the tidal torque will have a non-negligible effect on the orbital evolution for a star at the giant stage (RGB or AGB; see \citealt{tauris2001}).}

%We do not consider the effects of spin-orbit coupling and tidal dissipation, and the eccentricity is set to be 0.

\subsection{Common envelope evolution}
\label{subsec:3.3}

Two mass transfer stability criteria are considered in this work. In the first model, the critical mass ratio $q_{\rm c}$ is set to 3 for donors on the MS, He MS, and in the He-core burning phase, and to 4 during the HG \citep{hurley2002}. For the donors on the giant branch, $q_{\rm c}$ is adopted from Eq. \ref{eq:3}. In the second model, $q_{\rm c}$ is derived from {realistic stellar structures with standard mixing-length convective envelopes}\footnote{{For mass loss in the superadiabatic surface convection regions, the adiabatic approximation fails. To quantify the possible impact of the superadiabatic mass loss on the thresholds of mass transfer stability, \citet{geh2010} constructed the model of mass-loss sequences with isentropic envelope. The critical mass ratios with the standard mixing-length convective envelope models are smaller than that with the assumption of isentropic envelopes, and would give results more conservatively. In fact, the critical mass ratios are affected by many factors, e.g., the mass transfer efficiency, angular momentum loss, superadiabatic layer of the giant, etc, where the isentropic envelopes are used to solve the problems of superadiabatic layer. In this work, we intend to study the influence of mass transfer stability on the DWD populations with simple assumptions as a first step. Therefore, we adopted the $q_{\rm c}$ derived from the standard mixing-length models.}}, as done in \citet{geh2010,geh2015,geh2020}. The criteria depend on the treatments of mass and angular momentum loss\footnote{In \citet{geh2020}, only the results of fully conservation case are published, the grids and codes of non-conservative cases can be requested via email to the authors.}. To keep the consistency in this paper, we adopt the case of the non-conservative mass transfer with $\beta_{\rm rlof}=0.5$\footnote{{The accretion efficiency given in Eq. \ref{eq:10} are not a constant. It is a simplified approach that we adopt a flat value of $\beta_{\rm rlof}=0.5$ to derive the critical mass ratios from the adiabatic mass-loss sequences. Nevertheless, we checked the accretion efficiencies of the simulated binaries that experience the stable mass transfer phase in the fiducial model (see more details in Section 4) and found that the accretion efficiencies for most binaries are close to 0.5. A high value of $\beta_{\rm rlof}$ inferred from Eq. \ref{eq:10} is because the thermal time scale of the accretor is not too much longer than the mass transfer time scale of the donor for most cases (see also \citealt{willems2004}).}}. If the mass transfer occurs at AGB, we set an upper limit of critical mass ratio of $5$, since the thermal timescale mass transfer at the tip of AGB is comparable with that of the dynamical timescale mass transfer \citep{geh2020}. %\citet{geh2020} obtained the $q_{\rm c}$ in three cases, i.e. fully conservation case of $\beta_{\rm rlof} = 1$, non-conservative case of $\beta_{\rm rlof} = 0.5$, and fully non-conservative case $\beta_{\rm rlof} = 0$. To keep the consistency in this paper, we adopt the case of the non-conservative mass transfer with $\beta_{\rm rlof}=0.5$. 

As the mass ratio at the onset of mass transfer is higher than $q_{\rm c}$, the binary enters in the CE phase. We adopt the standard energy mechanism to describe the CE ejection process \citep{webbink1984}, in which the orbital energy is released to eject the envelope during the orbit shrinkage. It can be expressed as \citep{webbink1984,livio1988,dekool1990}:
\begin{eqnarray}
E_{\rm bind} = \alpha_{\rm ce}(E_{\rm orb,f}-E_{\rm orb,i}),
\label{eq:17}
\end{eqnarray}
where $E_{\rm orb,i}$ and $E_{\rm orb,f}$ are the orbital energy before and after the CE phase, respectively; $\alpha_{\rm CE}$ is the CE efficiency parameter and $E_{\rm bind}$ is the binding energy, defined as
\begin{eqnarray}
E_{\rm bind} = -G\left(\frac{M_1M_{\rm env}}{\lambda R_1}\right),
\label{eq:18}
\end{eqnarray}
where $\lambda$ is the envelope structure parameter and depends on the relative mass distribution of the envelope. In this work, $\lambda$ is adopted from the results of \citet{dewi2000,claeys2014}. The value of $\lambda$ is in the range of $0.25-0.75$ for HG stars, in the range of $1.0-2.0$ for GB and AGB stars, and $\lambda=0.5$ for helium stars. 
{The CE efficiency is a poorly constrained parameter, and the choice of $\alpha_{\rm CE}$ has a large effect on the simulation results. In this work, we mainly focus on the influences of mass transfer stability on the DWD populations. For convenience, we artificially chose a value of $\alpha_{\rm CE}=1$ as the fiducial model to provide a more detailed investigation. Nevertheless, some other values of $\alpha_{\rm CE}$ in the range of $0.25-3.0$ are also calculated and the effects of $\alpha_{\rm CE}$ are carefully discussed. }
The accretion during the CE phase is neglected due to its short timescale.

\subsection{Population synthesis parameters}
\label{subsec:3.4}

We generated $5\times 10^6$ primordial binaries for the Monte Carlo simulation\footnote{{The number of primordial binaries is dozens of times larger than that in some previous works, e.g., \citet{han1998,nelemans2000}, and it is sufficient to model the DWD populations in the Galaxy. We carried out a test with sampling $1\times 10^7$ binaries and the results display almost no change.}}. The main input distributions are introduced as follows.

First, the primary mass, $M_{\rm 1,i}$, is given by the following initial mass function (IMF, \citealt{miller1979,eggleton1989}):
\begin{eqnarray}
  M_{\rm 1,i} = \frac{0.19X}{(1-X)^{0.75}+0.032(1-X)^{0.25}},
\label{eq:19}
\end{eqnarray}
where $X$ is a random number between 0 and 1, which gives the mass ranging from $0.1$ to $100M_\odot$. Second, the secondary mass is given according to the initial mass ratio distribution \citep{mazeh1992}, that is,
\begin{eqnarray}
n(q') = 1,\;\;q'<1,
\label{eq:20}
\end{eqnarray}
where $q'$ is the mass ratio of the primordial binary. Third, the initial binary separation distribution is given by \citep{han1998}
\begin{eqnarray}
    an(a)=
    \begin{cases}
    0.07(a/a_0)^{1.2},\qquad a\le a_0 \\
    0.07, \qquad \qquad \quad a_0 \le a \le a_1,
    \end{cases}
    \label{eq:21}
\end{eqnarray}
where $a_0=10\;\rm{R}_{\odot}$, $a_1=5.75\times 10^6\;\rm{R}_{\odot}=0.13\;\rm pc$. {This distribution implies that approximately 50 percent of systems are binary systems with orbital periods less than 100 yrs, namely, the initial binary fraction is assumed to be 50 percent (see also \citealt{han1995}).} Fourth, the star formation rate is adopted to be $5\;M_\odot\rm yr^{-1}$ within 14 Gyr. It gives the total mass of the Galaxy of $7\times 10^{10}M_\odot$ , which is consistent with the observation estimation \citep{binney2008}.

\section{Binary population synthesis results}
\label{sec:4}

\begin{figure*}
        \centering
        \includegraphics[width=\textwidth]{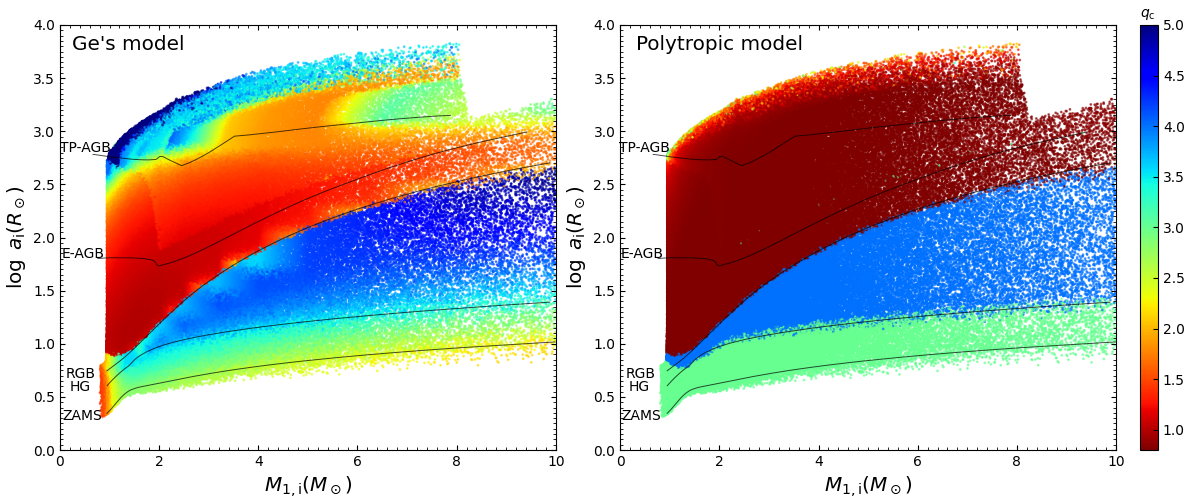}
    \caption{Critical mass ratios for stars with different masses and in different evolutionary stages. Left and right panels are for Ge's and polytropic models, respectively. The black solid lines represent the minimum separation ($q=1$) for which the primary fills its Roche lobe at a certain evolutionary stage, as indicated in the panels. {A star with a mass $\gtrsim 8M_\odot$ has no TP-AGB phase because the CO core will ignite carbon in the AGB and results in the SN explosion \citep{hurley2000}. We note that the results are obtained following the BPS simulations and all of the simulated binaries are considered. We find about $40.5\%$ and $74.0\%$ binaries would enter into CE phase for Ge's and polytropic models, respectively. Abbreviations are as follows:\ ZAMS—zero-age main sequence, HG-Hertzsprung gap, RGB-red giant branch, AGB—asymptotic giant branch, TP-AGB—thermal pulse AGB.}}
    \label{fig:1}
\end{figure*}

\subsection{Critical mass ratio for the two models}
To give an impression of the differences of $q_{\rm c}$ 
between the two models adopted in this work, 
we show values of $q_{\rm c}$ for stars with various masses and in different evolutionary stages in Figure \ref{fig:1}, 
where the left and right panels are for Ge's and the polytropic models, respectively. {We note that the results presented in Figure \ref{fig:1} include all of the simulated binaries and the distributions of the initial binary parameters follow the input parameters in BPS simulations of Section 3.4.} 
We see that for Ge's model, $q_{\rm c}$ is about $3$ for MS donors and it is $4$ for donors during HG stages, which is consistent with that in the polytropic model. A significant difference appears when the donors are giant stars, namely, the donor is an RGB star or an AGB star. The value of $q_{\rm c}$ is less than 1 in most cases for the polytropic model and it is in the range of $1-2$ in most cases based on Ge's model. This means that the mass transfer tends to be more stable for donors on the RGB and AGB stages in Ge's model. 
By comparing the mass ratio of binary systems and $q_{\rm c}$, we find that about $40.5\%$ of binaries would enter into CE phase in Ge's model, while that proportion rises to $74.0\%$ for the polytropic model. Besides the DWD populations, the mass transfer stability criterion should also exert a meaningful impact on the post-CE binaries, such as WD+MS binary and cataclysmic variables (CVs, e.g., \citealt{zorotovic2010,schreiber2016}), which are worthy of investigation in further studies.

\subsection{Evolution examples for DWDs}
\label{subsec:4.1}

\begin{figure*}
        \begin{minipage}[t]{0.5\textwidth}
                \centering
                \includegraphics[width=\textwidth]{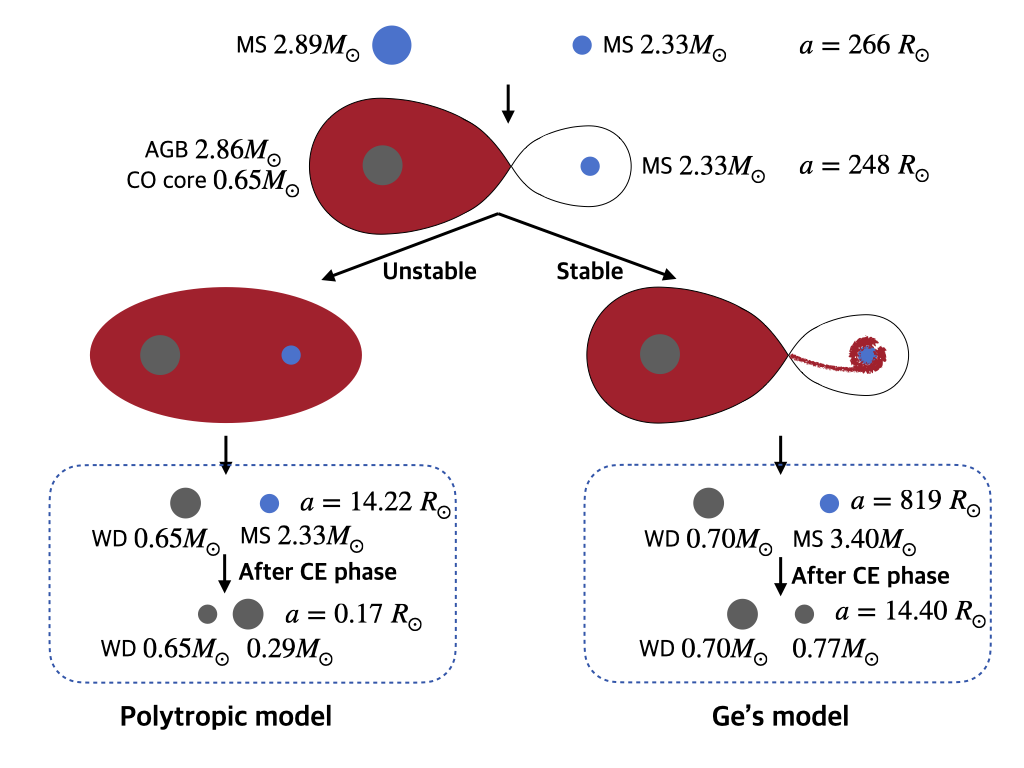}
        \end{minipage}
        \begin{minipage}[t]{0.5\textwidth}
                \centering
                \includegraphics[width=\textwidth]{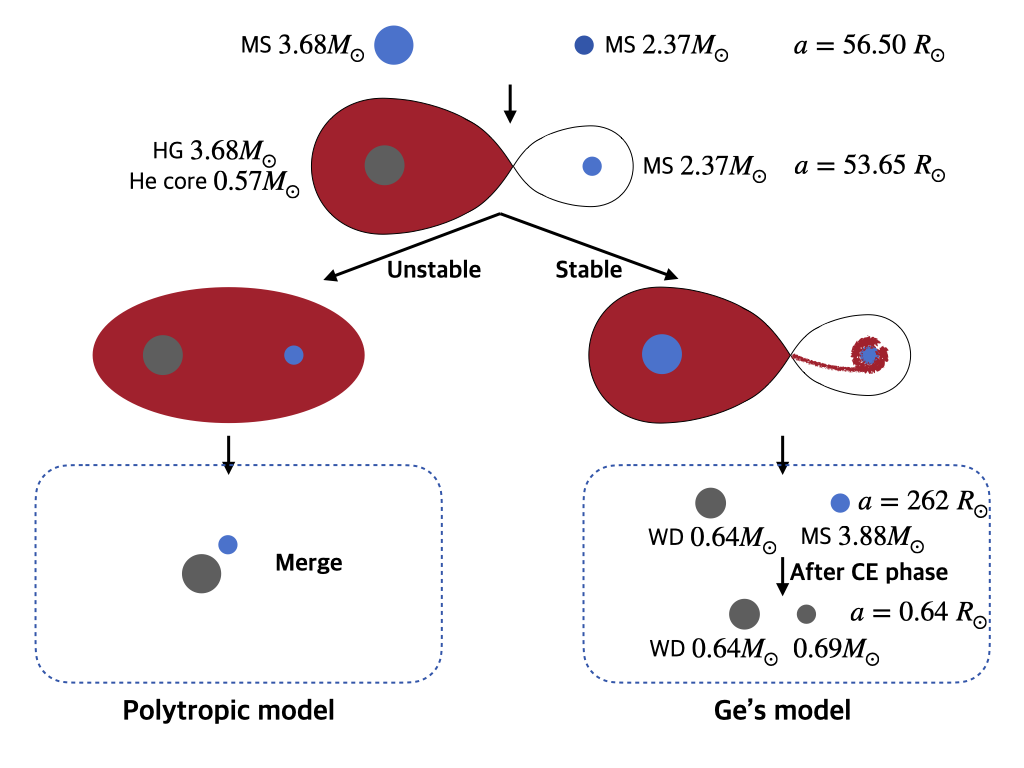}
        \end{minipage}
        \caption{Examples for the formation of DWDs from different mass transfer stability criteria ($\alpha_{\rm CE}=1$). Left panel shows that the DWDs produced from different models have different binary parameters. In the right panel, the DWD is formed in Ge's model, while the binary merges in the polytropic model (see more details in the text). Abbreviations are the same as in Fig. 1. }
    \label{fig:2}
\end{figure*}

We present two examples to illustrate the influence of mass transfer stability on the formation of DWDs, as shown in Figure \ref{fig:2}, {where $\alpha_{\rm CE}=1$ for all of the CE ejection phases. }

{\bf Example 1}: The initial binary consists of two MS stars with masses of {2.89}$M_\odot$ and {2.33}$M_\odot$, respectively, and has a separation of {266}$R_\odot$. The more massive one evolves faster and fills its Roche lobe at the AGB stage. At this moment, the star develops a CO core with a mass of {0.65}$M_\odot$. {The orbital shrinkage by about {7}$\%$ {(from $266R_\odot$ to $248R_\odot$)} is mainly due to tidal friction. As shown in Eqs. \ref{eq:14}-\ref{eq:16}, the change of orbital angular momentum is positively associated with $(R_1/a)^6$. Even though the binary separation is large, the stellar radius at the tip of AGB is larger than {64$R_\odot$}. So, the value of $R/a$ is about $\sim 0.3-0.4$. The tidal torque would exert a non-negligible effect on the orbital evolution. The result is consistent with that by \citet{tauris2001}, who found a $\sim 10\%$ orbital shrinkage for the pre-RLOF spin-orbit couplings in low-mass X-ray binaries.}
{According to the polytropic model, $q_{\rm c}=0.79$, the mass transfer thus is supposed to be unstable, and the binary enters into the CE phase}. After the ejection of CE, the binary encompasses a CO WD and an evolved MS star with an orbital separation of {14.22}$R_\odot$. {The binary subsequently enters into the second CE phase, resulting in the formation of DWD finally}. The WD formed later has a mass of {0.29}$M_\odot$, which is significantly less than that of the first-formed WD --
{the reason for this is that after the first phase of CE ejection, the orbital period is relatively short and the secondary consequently cannot contain a large core before the onset of the second mass transfer phase.} 

{Based on Ge's model, {$q_{\rm c}=$ 1.26} in the first mass transfer phase, and the binary thus experiences stable RLOF.} Different from that of the CE phase, the binary orbit would not shrink dramatically at the end of this stable mass transfer phase, and the secondary could accrete a part of the material from the donor. The secondary would have a large core mass at the onset of the second CE process, and a DWD is produced after the subsequent CE ejection event. Besides, the separation of the DWD formed in this way is relatively wider than that of the polytropic model. %{Here the orbital shrinkage during the stable RLOF phase is mainly caused by the wind accretion processes. Detailed discussions about orbital evolution due to the wind-fed mass transfer are given in Appendix \ref{app:A}}.  
From this example, we see that the new mass transfer stability would have significant effects on the masses and binary orbital separations of DWDs.

{\bf Example 2}: {The binary has two MS components with masses of $3.68M_\odot$ and $2.37M_\odot$, and an initial orbital separation of $56.50R_\odot$. The more massive one fills its Roche lobe close to the end of HG. The mass transfer at the beginning is stable since the critical mass ratio of an HG star for both models is very large (about $4$, as shown in Figure \ref{fig:1}). However, when about $0.4M_\odot$ material is lost, the donor climbs upward along the giant branch in the later evolutionary phase due to the residue hydrogen-rich envelope. At this moment, the binary encompasses a donor of $3.29M_\odot$ and an accretor of $3.49M_\odot$ (this epoch is not shown in the plot for clarity). For the polytropic model, the binary would enter into the CE phase on account of the relatively small $q_{\rm c}$ in the RGB (about $0.76$). The orbital energy is not enough to eject the CE (with $\lambda\simeq 1.6$) and the binary merge into a single star. In contrast, Ge's model gives a prediction that the mass transfer is still stable this time ($q_{\rm c} \simeq 1.2$) and a DWD is produced finally.} It suggests that more DWDs would be produced by adopting Ge's model.

\begin{figure*}
                \centering
                \includegraphics[width=\textwidth]{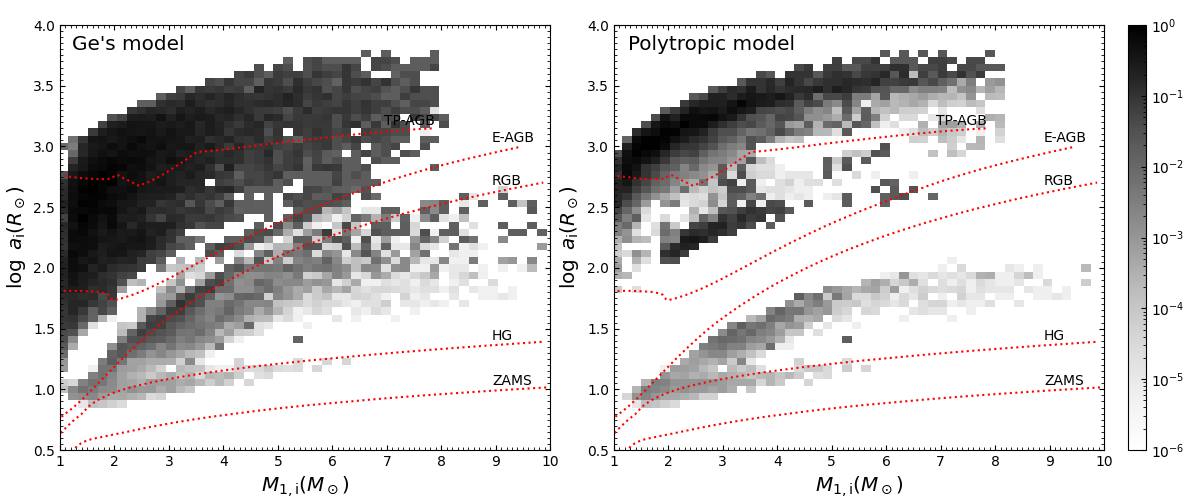}
        \caption{Distribution of the initial primary mass (initial massive one) and {initial} binary separation {of binaries that produce DWDs}. Left and right panels are for Ge's and polytropic models, respectively. The dotted lines represent the minimum separation ($q = 1$) for which the primary fills its Roche lobe at a certain evolutionary stage, as indicated in the panels.  Abbreviations are the same as in Fig. 1.}
    \label{fig:3}
\end{figure*}

\subsection{The progenitors}
\label{subsec:4.2}

In this section, we provide a discussion of the progenitor evolution of the DWD populations. The number density distributions of the initial primary mass (the initial, more massive one) and initial separation that form DWDs in the current epoch are shown in Figure \ref{fig:3}, where the left and right panels refer to Ge's and polytropic models, respectively. In the polytropic model, the density distribution is mainly divided into two parts, one is below the RGB line, and the other is above the early-AGB (E-AGB) line. For the first part, most primary stars start mass transfer during the HG stage, and the first mass transfer phase is stable due to the radiative envelopes of the primary stars. Eventually, most DWDs from this part are produced by the RL+CE channel. {For the part above the E-AGB, but below the tip of AGB, the first mass transfer phase are almost {always} unstable and the final DWDs are mainly produced by the CE+CE channel. However, a significant fraction of the E-AGB part for stars close to the tip of AGB could also produce DWDs from the RL+CE channel due to relatively large $q_{\rm c}$ at the tip of AGB and low primary masses resulting from the non-negligible effect of stellar wind on the AGB stage.}

For Ge's model, the progenitors are also divided into two parts, and {much more DWDs would be produced in comparison to the polytropic model.} Meanwhile, {the first mass transfer phase proceeds on a stable way for most stars filling their Roche lobes on the giant branches.} We find that more than $96\%$ binaries that form DWDs would experience a stable RLOF process in this mass transfer phase. We also note that many stars that start mass transfer in the HG stage can eventually produce DWDs in Ge's model. {These binaries follow similar physical processes as described in Example 2 of Figure \ref{fig:2}, that is, the binaries enter into the CE phase after a short phase of stable mass transfer based on the polytropic model, while the mass transfer would be always stable for Ge's model, leaving binaries with relatively long orbital periods at the termination of this mass transfer phase.}

\begin{figure*}
        \begin{minipage}[t]{0.5\textwidth}
                \centering
                \includegraphics[width=\textwidth]{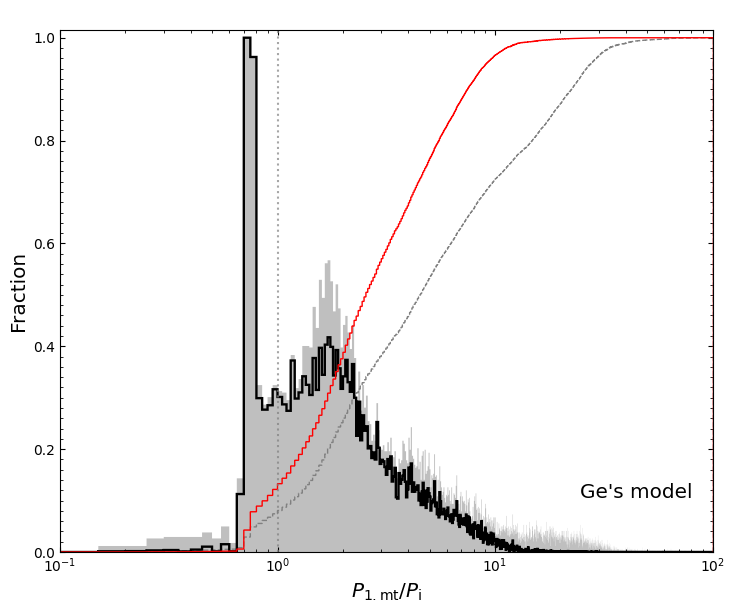}
        \end{minipage}
        \begin{minipage}[t]{0.5\textwidth}
                \centering
                \includegraphics[width=\textwidth]{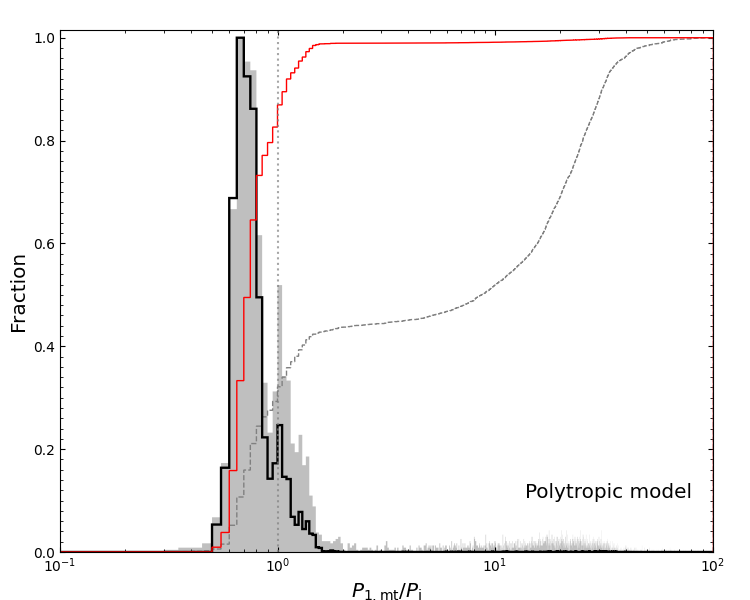}
        \end{minipage}
    \caption{{Properties of the period change after the first stable mass transfer phase. The period change distributions for binaries that produce DWDs in the current Galaxy are shown in the black histogram, and the corresponding cumulative distributions are shown in red solid lines. Left and right panels are for Ge's and polytropic models, respectively. Here, $P_{\rm 1,mt}$ refers to the orbital period after the termination of the first mass transfer phase and $P_{\rm i}$ is the initial orbital period of primordial binary. We note that most binaries with $P_{\rm 1,mt}/P_{\rm i}>1$ would enter into the CE processes in the subsequent evolution and some of the resulted DWDs have very short orbital periods (merge within the Hubble timescale). The DWD mergers lead to the decrease of the number for binaries with $P_{\rm 1,mt}/P_{\rm i}>1$. The results without considering the DWD mergers are shown in grey histograms and grey dashed lines. It is clear that the number of DWDs with very short orbital periods decreases a lot due to the DWD mergers. For binaries experiencing CE processes in the first mass transfer phase, the period changes of $P_{\rm 1,mt}/P_{\rm i}$ is less than 0.05, which are not shown for concision. }}
    \label{fig:4}
\end{figure*}

\subsection{Properties after the first mass-transfer phase}
\label{subsec:4.3}

{The change of orbital period after the first mass transfer phase is important to understand the subsequent evolution of a binary. Figure \ref{fig:4} shows the orbital periods $P_{\rm 1,mt}$ (related to initial periods of $P_{\rm i}$) of the DWD progenitors after the first mass transfer phase, where the left and right panels are for Ge's model and for the polytropic model, respectively. The period change distributions for binaries that produce DWDs in the current Galaxy are shown in the black histogram and the corresponding cumulative distributions are shown in red solid lines. Binaries experiencing the CE processes in the first mass transfer phase would lead to a remarkable orbit shrinkage, namely,  $P_{\rm 1,mt}/P_{\rm i}<0.05$, in general. We have not shown this part in the figure for concision. }

{During the stable RLOF phase, the change of the orbital period is mainly determined by the mass loss and wind accretion process. If the mass transfer occurs at the tip of AGB, the mass loss due to the wind dominates the orbital evolution. The wind accretion processes may lead to the shrinkage of orbital separation. In Appendix \ref{app:A}, we give a brief discussion about the effect of wind accretion on the orbit. In Figure \ref{fig:3}, we see that many systems enter the mass transfer stage at the tip of AGB in both models, and the shrinkage of the orbit due to wind accretion eventually results in the peak around $P_{\rm 1,mt}/P_{\rm i}=0.7-0.8$.}

{For a star filling the Roche lobe below the AGB, the mass loss due to the wind is generally weaker than that of the stable mass transfer. Then the orbital evolution of a binary is mainly determined by the stable RLOF. If the mass ratio of an initial binary is around $1-2$, the binary generally has a widened orbit after the stable mass transfer phase, as also shown in \citet{woods2012}. 
We note that most binaries with $P_{\rm 1,mt}/P_{\rm i}>1$ would enter into the CE processes in the subsequent evolution and some of the resulted DWDs have very short orbital periods (merge within the Hubble timescale). The DWD mergers lead to the decrease of the number for binaries with $P_{\rm 1,mt}/P_{\rm i}>1$. The results without considering the DWD mergers are shown in grey histograms and grey dashed lines for comparison, and we see that most of the binaries have widened orbits ($P_{\rm 1,mt}/P_{\rm i}>1$) for both models ($\sim 90\%$ for Ge's model and $\sim 70 \%$ for the Polytropic model). Since much more binaries experience the stable mass transfer phase in Ge's model, the fraction of binaries with $P_{\rm 1,mt}/P_{\rm i}>1$ is larger than that in the polytropic model. In addition, we note that there are some binaries with $P_{\rm 1,mt}/P_{\rm i}<0.5$ for both models. These binaries have relatively large mass ratios (of about 4), and the donors fill the Roche lobe at the HG stage. The orbital separations during the stable mass transfer phase tend to shrink since the massive components lose material \citep{tauris2006,postnov2014}.}

\subsection{Birthrate and number}
\label{subsec:4.4}

\begin{table*}
  \caption{Statistical properties of DWDs with different $\alpha_{\rm CE}$ values in the Galaxy.}
  \label{tab:1}
\centering
\begin{tabular}{lcccccccccc}
\hline\hline
& $N_{\rm sim}$ & $N_{\rm Gal}$ & $f_{\rm RL+CE}$ & $f_{\rm CE+CE}$ & $f_{\rm CE+RL}$ & $\nu$ & $N_{g\rm <21}$ & $\rho$ & $N_{<60\;\rm min}$ & $N_{\rm LISA}$\\
& & $\times 10^8$ & & & & $\rm yr^{-1}$ & & $\rm kpc^{-3}$ & &\\
\hline
$\alpha_{\rm CE}=0.25$&&&&& \\ 
Pol model & 12156 & $0.76$ &$79.1\%$ &$0.2\%$ & $20.7\%$& 0.009 & $1.08\times 10^5$ & 939.1 & $0.22\times 10^5$ & $1452$\\
Ge's model & 60025 & $3.53$ & $98.59\%$ & $0.04\%$ & $1.37\%$& 0.042 & $5.05\times 10^4 $ & 439.1 & $2.55\times 10^5$ & 14098\\ %sss
\hline
$\alpha_{\rm CE}=0.5$&&&&& \\ 
Pol model & 23010 & 0.83 &$75.8\%$ &$5.7\%$ & $18.5\%$& 0.016 & $1.29\times 10^5$& 1121.7 & $2.21\times 10^5$ & 11096\\
Ge's model & 71714 & $4.34$ & $96.80\%$ & $0.05\%$ & $3.15\%$& 0.051 & $9.81\times 10^4$& 853.0 & $2.46\times 10^5$ & 12940\\ %sss
\hline
$\alpha_{\rm CE}=0.75$&&&&& \\ 
Pol model & 42890 & $1.26$ &$71.43\%$ &$15.71\%$ & $12.86\%$& 0.030 & $1.93\times 10^5$ & 1678.3 & $6.43\times 10^5$ & 25427\\
Ge's model & 87035 & $4.55$ &$96.63\%$ &$0.16\%$ & $3.21\%$& 0.060 & $1.34\times 10^5$ & 1165.2 & $4.03\times 10^5$ & 19504\\
\hline
$\alpha_{\rm CE}=1.0$&&&&& \\ 
Pol model & 58789& $1.27$ &$49.8\%$ &$38.3\%$ & $11.9\%$& 0.041 & $3.08\times 10^5$ & 2678.3 & $1.04\times 10^6$ & 36487\\
Ge's model & 95045 &  $4.70$ & $96.4\%$ & $0.6\%$ & $3.0\%$& 0.067 & $1.55\times 10^5$ & 1347.8 & $7.16\times 10^5$ & 27137\\  %sss
\hline %4.82e8 gamma=-1
$\alpha_{\rm CE}=1.25$&&&&& \\ 
Pol model & 72021 & $1.74$ &$39.10\%$ &$52.61\%$ & $8.29\%$& 0.050 & $4.03\times 10^5$ & $3504.3$ & $1.4\times 10^6$ & 42479\\
Ge's model & 103222 & $4.96$ & $95.95\%$ & $1.32\%$ & $2.73\%$& 0.072 & $1.47\times 10^5$ & 1278.3 &$9.8\times10^{5}$ & 31147\\ %sss
\hline
$\alpha_{\rm CE}=1.5$&&&&& \\ 
Pol model & 81155 & $2.07$ &$33.0\%$ &$62.7\%$ & $4.3\%$& 0.057 & $4.87\times 10^5$ &4234.8 & $1.70\times 10^6$ & $45384$ \\
Ge's model & 109009 &  $5.11$ & $96.0\%$ & $1.7\%$ & $2.3\%$& 0.077 & $1.59\times 10^5$ & 1382.6 & $1.34\times 10^6$ & 37003\\ %sss
\hline
$\alpha_{\rm CE}=1.75$&&&&& \\ 
Pol model & 89422 & $2.60$ &$29.47\%$ &$67.73\%$ & $2.80\%$& 0.063 & $5.91\times 10^{5}$ & 5139.1 & $1.92\times 10^6$ & $49010$ \\
Ge's model & 112830 & $5.39$ & $95.15\%$ & $3.60\%$ & $1.25\%$& 0.079 & $1.99\times 10^5$& 1730.4 & $1.58\times 10^6$ & 41173\\ %sss
\hline
$\alpha_{\rm CE}=2.0$&&&&& \\ 
Pol model & 95738 & $3.13$ &$27.60\%$ &$70.43\%$ & $1.97\%$& 0.067 & $6.99\times 10^5$ & 6078.3 & $2.04\times 10^{6}$ & 49632\\
Ge's model & 116307 & $5.75$ & $93.91\%$ & $5.17\%$ & $0.92\%$& 0.082 & $2.44\times 10^5$ & $2121.7$ & $1.73\times 10^6$ & 43255\\ %sss
\hline

$\alpha_{\rm CE}=2.25$&&&&& \\ 
Pol model & 101575 & $3.66$ &$26.60\%$ &$71.57\%$ & $1.83\%$& 0.071 & $8.03\times 10^5$ &6982.6 & $2.14\times 10^6$ & 49679 \\
Ge's model & 119860 & $6.21$ & $92.43\%$ & $6.70\%$ & $0.87\%$& 0.084 & $3.00\times 10^5$ &2608.7 & $1.82\times 10^6$ & 43521\\ %sss
\hline

$\alpha_{\rm CE}=2.5$&&&&& \\ 
Pol model & 106727 & $4.18$ &$26.66\%$ &$71.66\%$ & $1.68\%$& 0.075 & $8.98\times 10^{5}$ &7808.7 & $2.17\times 10^6$ & 48720\\
Ge's model & 123088 & $6.70$ & $90.94\%$ & $8.24\%$ & $0.82\%$& 0.086 & $3.54\times 10^5$ & 3078.3 & $1.84\times 10^6$ & 43006\\ %sss
\hline

$\alpha_{\rm CE}=2.75$&&&&& \\ 
Pol model & 110871& $4.66$ &$27.30\%$ &$71.21\%$ & $1.49\%$& 0.078 & $1.00\times10^6$ & 8695.7 & $2.17\times 10^6$ & 47370\\
Ge's model & 125615 & $7.54$ & $89.73\%$ & $9.55\%$ & $0.72\%$& 0.088 & $4.08\times 10^5$ & 3547.8 & $1.81\times 10^6$ & 47936\\ %sss
\hline

$\alpha_{\rm CE}=3.0$&&&&& \\ 
Pol model & 114485 & $5.12$ &$28.23\%$ &$70.44\%$ & $1.33\%$& 0.080 & $1.10\times 10^6$ & 9565.2& $2.14\times 10^6$ & 45498\\
Ge's model & 127869 & $7.60$ & $88.68\%$ & $10.68\%$ & $0.64\%$& 0.090 & $4.5\times 10^5$ & 3913.0 & $1.71\times 10^6$ & 46312\\ %sss
\hline
\hline
\end{tabular}
\tablefoot{{$N_{\rm sim}$ is the total number of DWDs from $5\times 10^6$ primordial binaries. $N_{\rm Gal}$ is the total number of DWDs in the current Galaxy; $f$ means the percentage of the number of DWDs from the corresponding channels; $\nu$ is the DWD birth rate; $N_{\rm g<21}$ is the total number of magnitude-limit samples in the simulations; $\rho \equiv N_{\rm g<21}/V_{\rm eff}$ is the space density (see Section 5.1); $N_{\rm <60\;\rm min}$ is the number of DWDs with orbital periods $<60\;\rm min$; $N_{\rm LISA}$ is the number of LISA detectable sources (see Section 5.4). }
}
\end{table*}

The statistical results of DWD populations are shown in Table \ref{tab:1}, where the total number, the percentage contribution of each evolutionary channel to the total number, the birth rate, {and so on,} are listed. For both models, the total numbers and percentage of CE+CE channel increase with the CE efficiencies. {However, it should be noted that the CE+CE channel is always subdominant in Ge's model regardless of $\alpha_{\rm CE}$. } For the case of $\alpha_{\rm CE} = 1$, about $96.4\%$ of DWDs are from the RL+CE channel in Ge’s model, while the number becomes $49.8\%$ for the polytropic model. As expected, the total number of DWDs in Ge’s model is much larger than (about four times) that of the polytropic model, since many objects that can produce DWDs from the RL+CE channel based on Ge's model may merge into single stars during the first CE evolution based on the polytropic model (see example 2 in Section \ref{subsec:4.1}). {The characteristics of DWDs in the two models, such as the DWD mass distributions and the orbital period distributions, will be addressed in the following sections.}

\subsection{Properties of DWD populations}
\label{subsec:4.5}

\begin{figure*}
        \begin{minipage}[t]{0.5\textwidth}
                \centering
                \includegraphics[width=\textwidth]{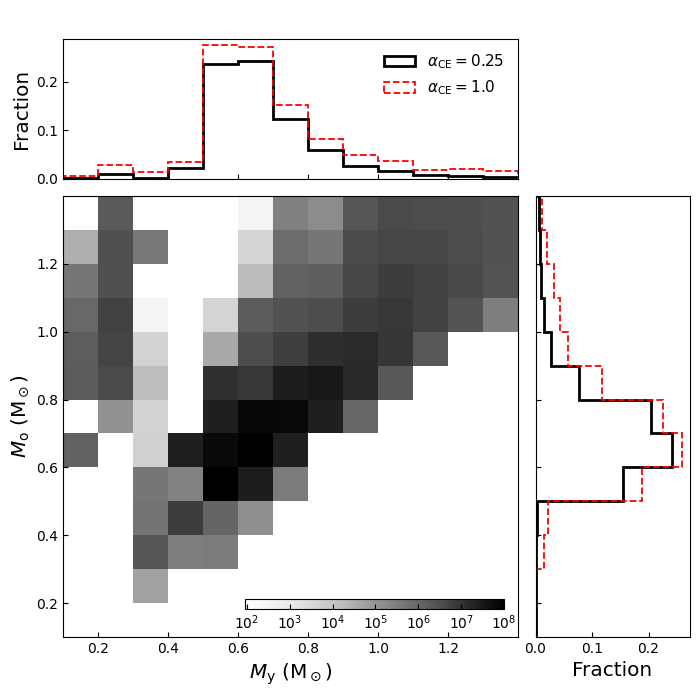}
        \end{minipage}
        \begin{minipage}[t]{0.5\textwidth}
                \centering
                \includegraphics[width=\textwidth]{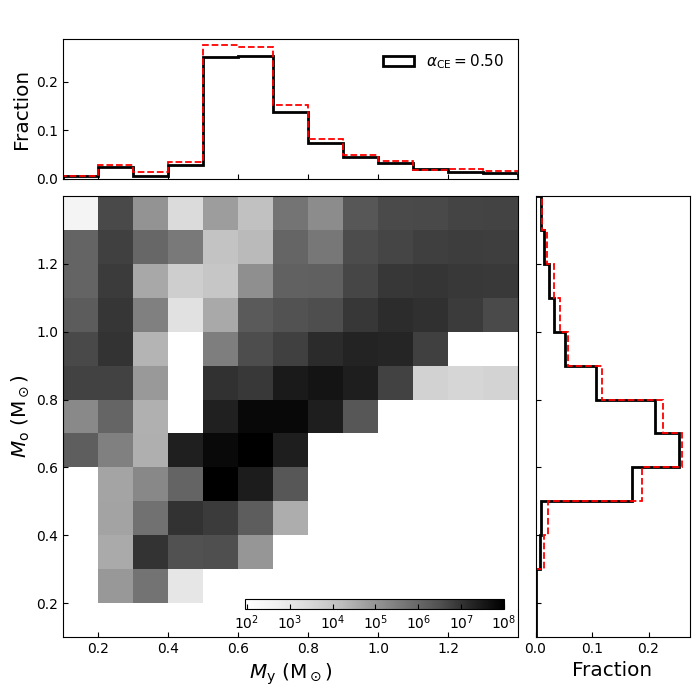}
        \end{minipage}
    \\
        \begin{minipage}[t]{0.5\textwidth}
                \centering
                \includegraphics[width=\textwidth]{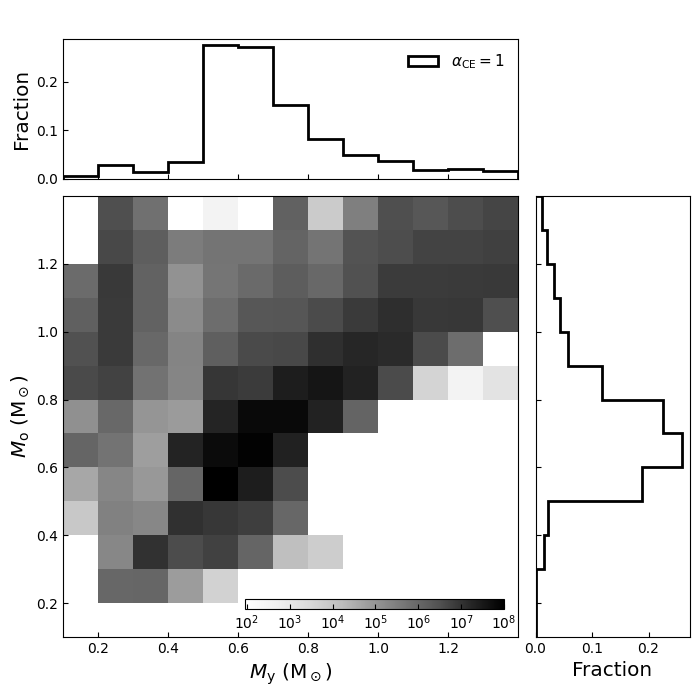}
        \end{minipage}
        \begin{minipage}[t]{0.5\textwidth}
                \centering
                \includegraphics[width=\textwidth]{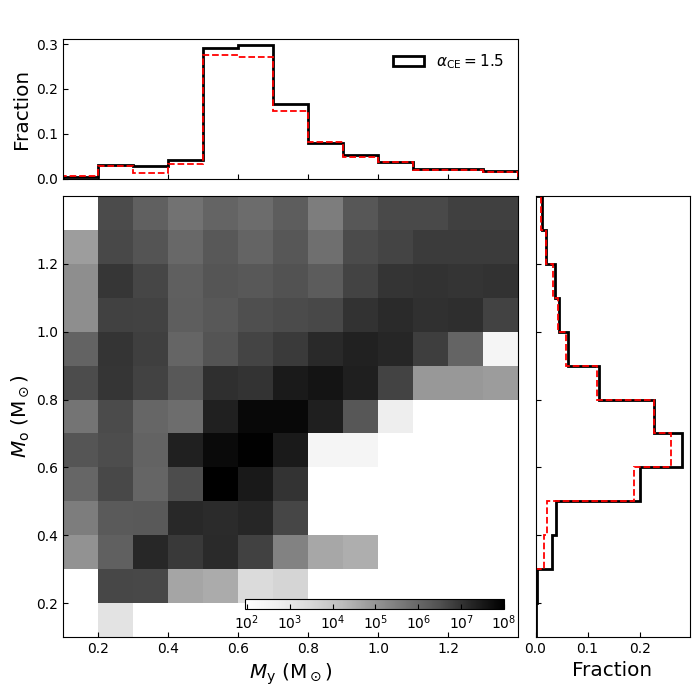}
        \end{minipage}
    \caption{{Number density distribution of DWD masses with various values of $\alpha_{\rm CE}$ in Ge's model, where $M_{\rm y}$ and $M_{\rm o}$ represent the mass of the younger and older (the later and former formed) WDs, respectively. The histogram of $\alpha_{\rm CE} = 1$ is also shown in other panels for comparisons.} 
}
    \label{fig:5}
\end{figure*}

The DWD is produced after the second mass transfer phase. As introduced in Section \ref{sec:2}, we mainly considered three evolution channels for the DWDs: RL + CE channel, CE+CE channel, and CE + RL channel. The CE+RL channel is one of two formation channels of DWDs with ELM WD companions and will be discussed in Section \ref{sec:5}. For the first two channels, the main difference is the first mass transfer phase. If the first mass transfer phase is stable, the less evolved secondary could increase its mass by accreting some material from the donor and would have a large core mass due to the relatively long orbital period. The WD born from the secondary then may have a relatively large mass. On the other hand, if the DWD is produced from the CE+CE channel, the mass of the latter-born WD should be less than that of the firstborn one. {In the following, we give a detailed investigation of the properties of DWD populations. We note the DWD populations refer in particular to the DWD populations in the current Galaxy, unless otherwise stated.} 

\subsubsection{Mass distributions}

\begin{figure*}
        \begin{minipage}[t]{0.5\textwidth}
                \centering
                \includegraphics[width=\textwidth]{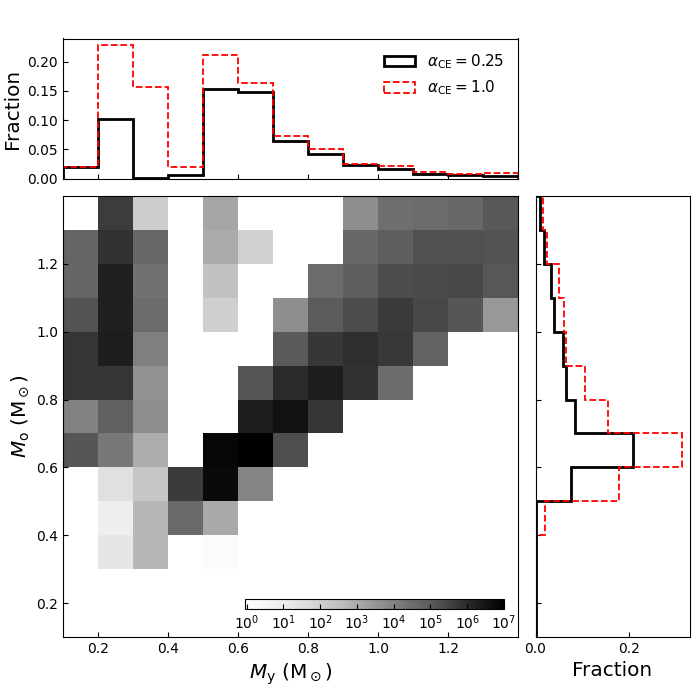}
        \end{minipage}
        \begin{minipage}[t]{0.5\textwidth}
                \centering
                \includegraphics[width=\textwidth]{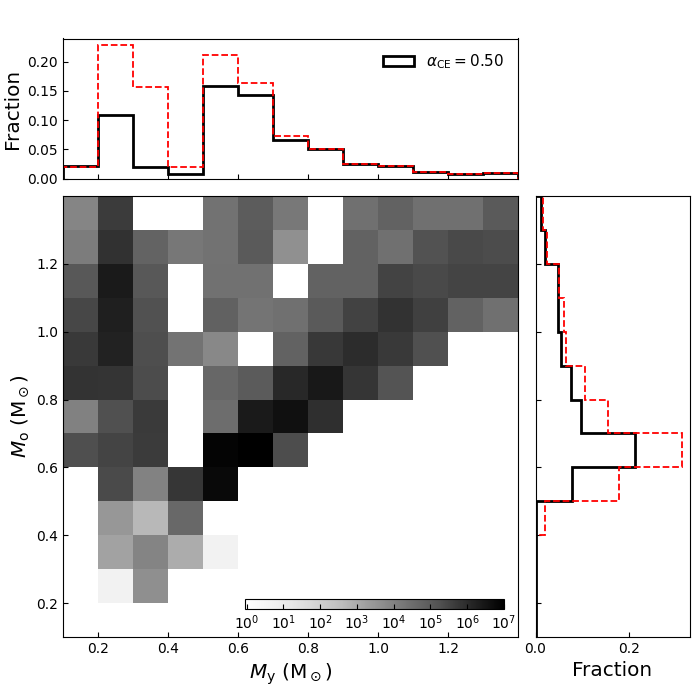}
        \end{minipage}
    \\
        \begin{minipage}[t]{0.5\textwidth}
                \centering
                \includegraphics[width=\textwidth]{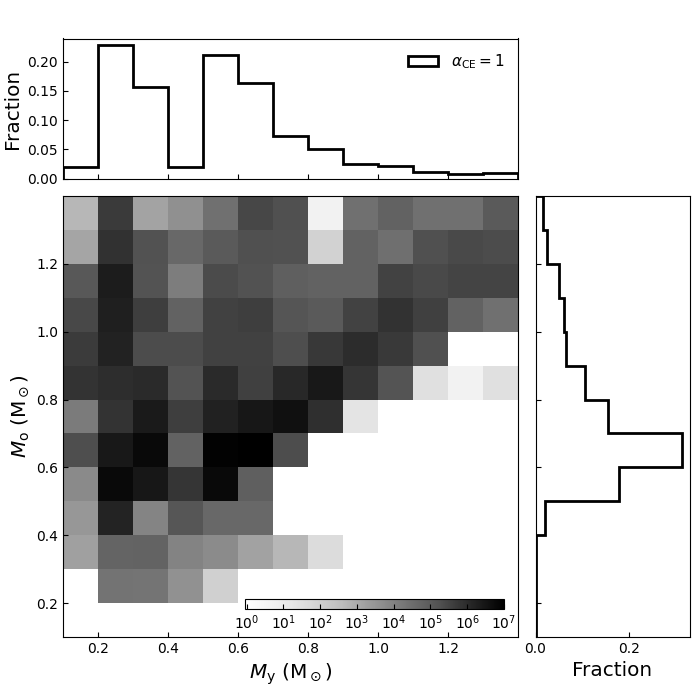}
        \end{minipage}
        \begin{minipage}[t]{0.5\textwidth}
                \centering
                \includegraphics[width=\textwidth]{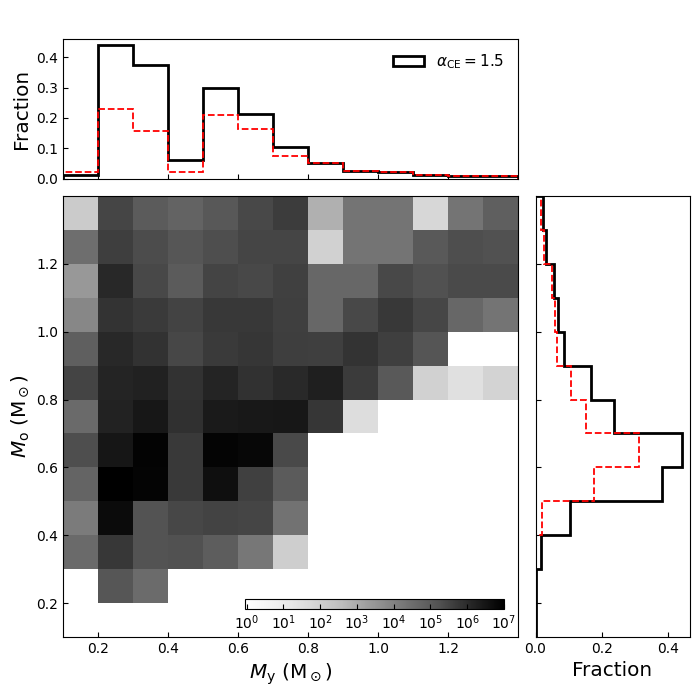}
        \end{minipage}
        \caption{{Number density distribution of DWD masses with various values of $\alpha_{\rm CE}$ in the polytropic model}. Details are similar to Figure \ref{fig:5} but for the polytropic model, where $M_{\rm y}$ and $M_{\rm o}$ represent the mass of the younger and older (the later and former formed) WDs, respectively. We note that the scale of the color bars differs from that of Figure \ref{fig:5}.}
    \label{fig:6}
\end{figure*}

{The DWD mass distributions for several selected values of $\alpha_{\rm CE}$ in Ge's model are shown in Figure \ref{fig:5}. With the increase in the value of $\alpha_{\rm CE}$, the total number of DWDs in the Galaxy changes slightly, as also shown in Table \ref{tab:1}. Moreover, the mass distributions for DWDs show similar statistical properties, namely, both of the younger WD mass ($M_{\rm y}$) and the older WD mass ($M_{\rm o}$) have peak values around $0.6M_\odot$. This suggests that the results in Ge's model are largely insensitive to the assumed $\alpha_{\rm CE}$. For the polytropic model, as shown in Figure \ref{fig:6}, the $\alpha_{\rm CE}$ becomes a determined parameter of the DWD populations. For $\alpha_{\rm CE}\lesssim 0.5$, the RL+CE channel is the dominant formation channel of DWDs, similar to that in Ge's model. For $\alpha_{\rm CE}\gtrsim 1$, the CE+CE channel becomes the dominant channel of DWDs. We see that there are two peaks, $\sim0.25M_\odot$ and $\sim 0.6M_\odot$, respectively, for $M_{\rm y}$ in the polytropic model. The peak of $\sim 0.25M_\odot$ is attributed to the ELM WDs, which are mainly produced from the CE+RL channel for $\alpha_{\rm CE}\lesssim 0.5$ and from CE+CE channel for $\alpha_{\rm CE}\gtrsim 1$. For the same value of $\alpha_{\rm CE}$, }we see that the peak values of $M_{\rm o}$ for the two models are similar (around $0.6M_\odot$), but it is obvious that the $M_{\rm o}$ from Ge's model is averagely larger than that of the polytropic model, for instance, for the fiducial model of $\alpha_{\rm CE}=1$, the mean and median values of $M_{\rm o}$ are $0.74M_\odot$ and $0.70M_\odot$ for Ge's model, respectively, and $0.53M_\odot$ and $0.54M_\odot$ for the polytropic model, respectively. This could be understood as follows:\ for the CE process, {the core mass does not increase due to the short timescale of the CE ejection processes, then the WD mass is mainly determined by the position of its progenitor on the giant branches and the distribution of $M_{\rm o}$ is only related to initial parameters of binary populations (e.g., initial mass function and orbital period distribution).} {However, there is a long time for the core to grow in mass if the mass transfer is stable}. Thus, for the RL process, the mass of the produced WD ($M_{\rm o}$) has also been affected by the detailed mass transfer process except for initial parameters, and should be larger than that from the CE process in general.

\begin{figure*}
        \begin{minipage}[t]{0.5\textwidth}
                \centering
                \includegraphics[width=\textwidth]{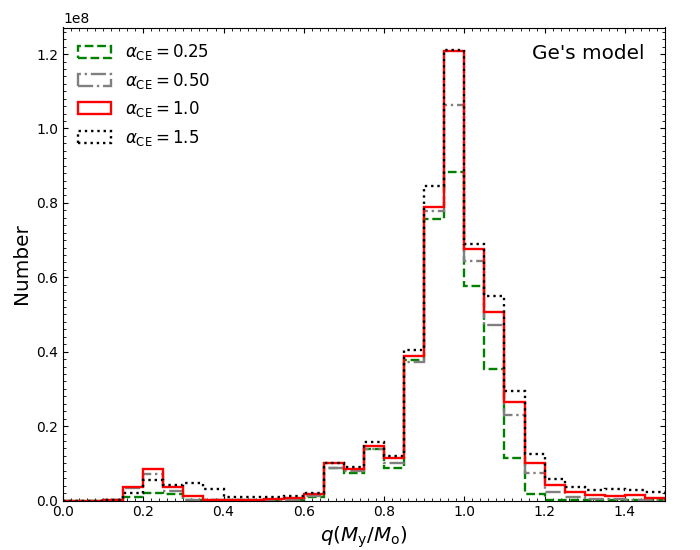}
        \end{minipage}
        \begin{minipage}[t]{0.5\textwidth}
                \centering
                \includegraphics[width=\textwidth]{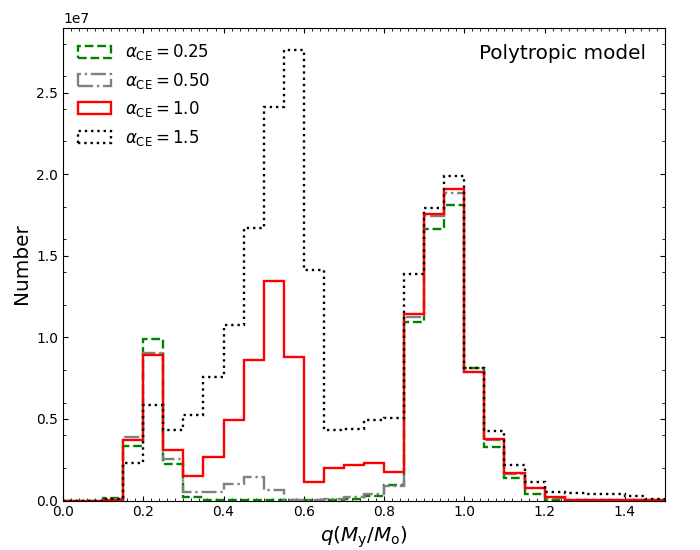}
        \end{minipage}
    \caption{{Mass ratio distribution for DWDs with selected values of $\alpha_{\rm CE}=0.25,0.5,1.0,1.5$, where $M_{\rm y}$ and $M_{\rm o}$ represent the mass of the younger and older (the later and former formed) WDs, respectively. Left and right panels are for Ge's and polytropic models, respectively.}
    }
    \label{fig:7}
\end{figure*}

\begin{figure*}
        \begin{minipage}[t]{0.5\textwidth}
                \centering
                \includegraphics[width=\textwidth]{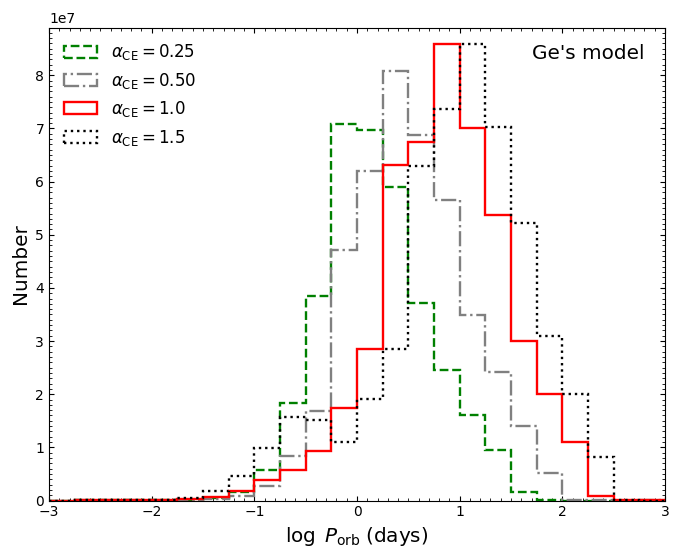}
        \end{minipage}
        \begin{minipage}[t]{0.5\textwidth}
                \centering
                \includegraphics[width=\textwidth]{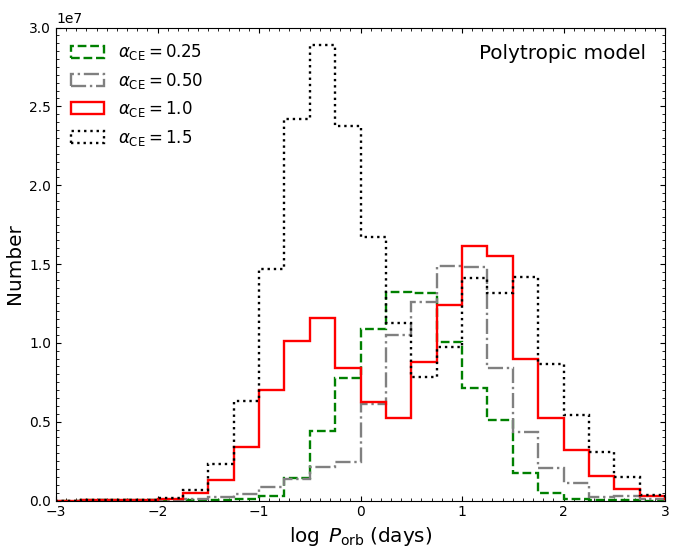}
        \end{minipage}
    \\
        \begin{minipage}[t]{0.5\textwidth}
                \centering
                \includegraphics[width=\textwidth]{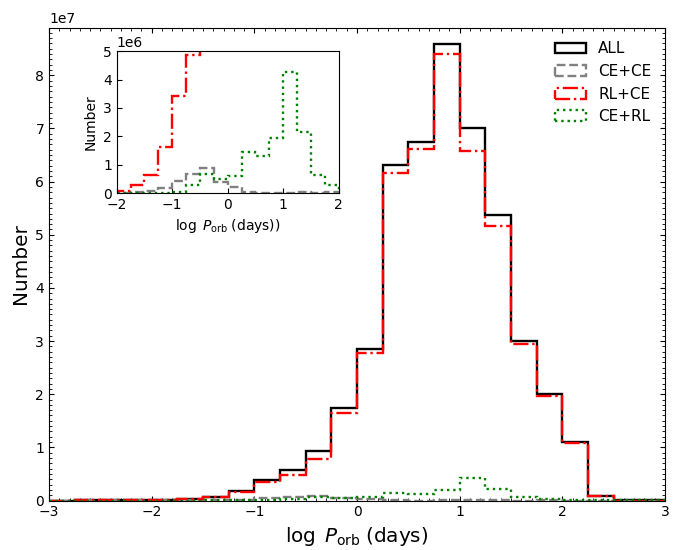}
        \end{minipage}
        \begin{minipage}[t]{0.5\textwidth}
                \centering
                \includegraphics[width=\textwidth]{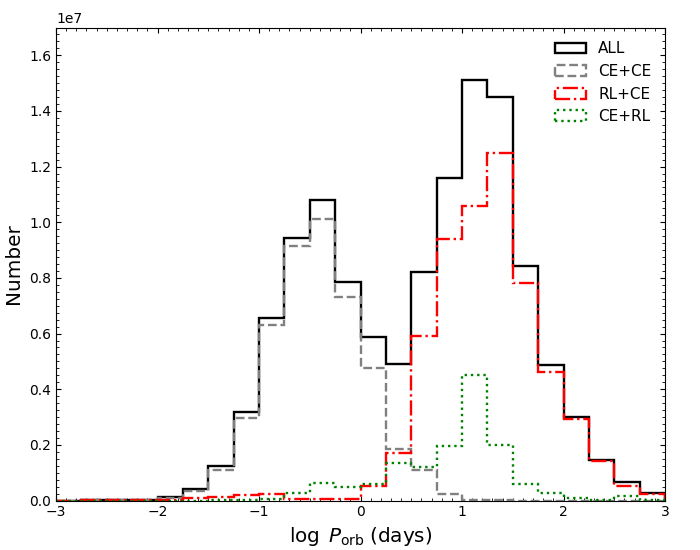}
        \end{minipage}
    \caption{{ Orbital period distributions for DWDs with $\alpha_{\rm CE}=0.25,0.5,1.0,1.5,    $ shown in the upper panels, orbital period distributions for different formation channels of $\alpha_{\rm CE}=1,$ shown in the lower panels. Left and right panels: Ge's and polytropic models, respectively. We note that the scale of the vertical axis is different for the two models.}}
    \label{fig:8}
\end{figure*}

\subsubsection{Mass ratio distributions}

{The mass ratio distributions for selected values of $\alpha_{\rm CE}$ are presented in Figure \ref{fig:7}, where the left and right panels are for Ge's and polytropic models, respectively.} For Ge's model, we see a main peak of $q$ around $1.0$ and a minor peak around 0.25 {regardless of the $\alpha_{\rm CE}$ value}. The former is for those from the RL+CE channel, while the latter is for the DWDs with ELM WDs, {which are mainly produced from the CE+RL channel}. However, the third peak (0.4-0.6) appears for the polytropic model ($\alpha_{\rm CE}\geq 0.5$), which is for the products from the CE+CE channel. {In addition, the contributions of CE+CE channel become increasingly important towards large values of $\alpha_{\rm CE}$.} The fact that the two WDs have comparable masses is a typical property for DWDs from the RL+CE channel and can be understood from an analytical estimation. For donor stars with degenerate cores, i.e. He cores with primary masses initially $\lesssim2.2M_\odot$ or CO cores with masses initially $\lesssim 6-8M_\odot$, there is a unique relation between the WD mass and the orbital period at the end of stable mass transfer (known as the WD mass-orbital period relation; e.g., \citealt{rappaport1983,rappaport1995,tauris1999,nelson2004,chenx2013}). This orbital period further determines the stellar radius (then the core mass) that the secondary can reach before the onset of the second mass transfer phase. The core mass is similar to that of the WD mass already formed since they are determined by the same orbital period. The second mass transfer phase is unstable and the resulted DWDs after the ejection of the CE would have comparable masses consequently. As we see, the masses of the two components are not strictly equal due to the details of binary evolution, for example, from the end of the first mass transfer phase (RL) to the onset of the second mass transfer phase (CE), the period could change (e.g., due to wind) but it should be narrow. 

\subsubsection{Period distributions}
{In Figure \ref{fig:8}, the orbital period distributions with different $\alpha_{\rm CE}$ for the two models are presented (upper panels), where the DWDs from different formation channels with $\alpha_{\rm CE}=1$ are shown in the lower panels. In Ge's model, most DWDs are produced from the RL+CE channel, as can be seen from the lower left panel, and the peak values of period distributions move to the right with the increase of $\alpha_{\rm CE}$. In the polytropic model, we see that there is only one peak for the case of $\alpha_{\rm CE}=0.25,0.5$, since the dominant formation channel of DWDs is RL+CE channel, similar to that in Ge's model. For the cases of $\alpha_{\rm CE}\geq 1$, the period distributions show two peaks, corresponding to the CE+CE channel and RL+CE channel, respectively, as can be seen from the lower right panel. Besides, DWDs from the CE+CE channel generally have short orbital periods ($\lesssim 1\;\rm d$) and we see that the proportion of close DWDs in the polytropic model is larger than that in Ge's model for $\alpha_{\rm CE}\geq 1$. However, it should be noted that the scale of the vertical coordinate is different for the left and right panels and the RL+CE channel is able to produce a lot of DWDs, with short orbital periods. In Table \ref{tab:1}, we count the number of DWDs with $P_{\rm orb}<60\;\rm min$, and the numbers of DWDs with very short orbital periods in both models are comparable for $\alpha_{\rm CE} \geq 1$ (see Section 5.4 for more details).}

\section{Implications for observations}
\label{sec:5}

\subsection{{Magnitude-limited sample}}
\label{subsec:5.1}
In this section, we make a comparison between the simulation results and the observations. To judge whether the binary can be detected by the telescope, we first constructed the {magnitude-limited sample} in the simulations. The cooling tracks for DWDs are taken from \citet{fontaine2001,lauffer2018,camisassa2019,lizw2019} with pure hydrogen atmospheres\footnote{The transformation of absolute magnitude to visual magnitude is done with the publicly available code \url{https://github.com/ SihaoCheng/WD_models}.}. 
For convenience, the SDSS $g-$band magnitude was adopted and the limited magnitude for the telescope is set to be 21 mag\footnote{In fact, the limiting magnitude of $g-$band in SDSS is about 23 mag \citep{york2000}. However, since the WD is an intrinsic faint object, the signal-to-noise of the spectroscopic observations is closely related to the observed magnitudes \citep{tremblay2011}. { And DWDs should be bright enough to allow for a precise determination of the orbital parameters. Therefore, the limiting magnitude is assumed to be 21 mag in this work. In table~\ref{tab:C1}, we see that most of the observed DWD samples are below this limit.}}.  
If any of the DWD has a $g-$band magnitude less than 21 mag, the binary is deemed to have been detected.

{We assume that the spatial density distribution of DWD population follows the spatial density distribution of all the stars in the disk, which we model as \citep{binney2008}:}
\begin{eqnarray}
  \rho_{\rm D}(R,z) = \frac{\sum_{\rm t}}{2z_{\rm t}}\exp\left(-\frac{R}{R_{\rm t}}-\frac{|z|}{z_{\rm t}}\right),
  \label{eq:22}
\end{eqnarray}
where $(R, z)$ is the Galactocentric distance and height in cylindrical coordinates, $\sum_{\rm t}= 970.294M_\odot \rm pc^{-2}$ and $z_{\rm t}=0.3$ kpc are the central surface density and the scale height of the disk {\citep{bland2016}}, respectively. The position of the Sun is adopted as $(R_{\rm sun},z_{\rm sun})$ = (8.5 kpc, 16.5 pc) \citep{freudenreich1998}. The extinction is simply adopted as 1 mag $\rm kpc^{-1}$ \citep{carrasco2014}. Finally, we obtained the magnitude-limited sample of DWDs with given location and visual magnitude of the DWDs.

\subsubsection{DWD mass distribution}
\label{subsubsec:5.1.1}

\begin{figure*}
        \begin{minipage}[t]{0.5\textwidth}
                \centering
                \includegraphics[width=\textwidth]{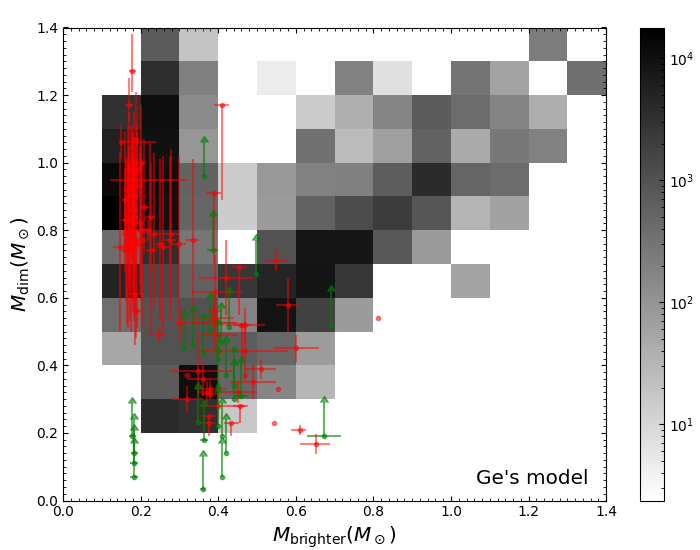}
        \end{minipage}
        \begin{minipage}[t]{0.5\textwidth}
                \centering
                \includegraphics[width=\textwidth]{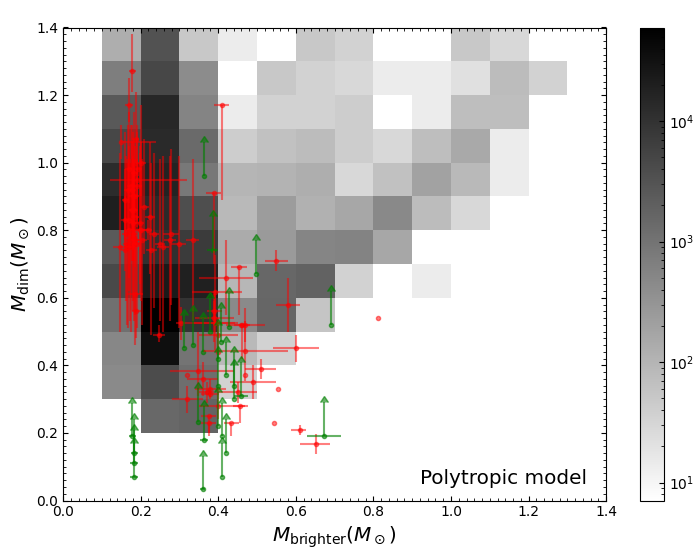}
        \end{minipage}
    \caption{Number density distributions of visible DWDs and the comparison with the observations. The red dots are for the observed samples with determined component masses {and the green dots are for those with only minimum companion (dimmer) masses determined \citep{kruckow2021}. The "brighter" and "dimmer" are defined as the WDs with higher and lower effective temperatures. The catalog of the observed samples is given in appendix Table \ref{tab:C1}.} Left and right panels are for Ge's and polytropic models, respectively.}
    \label{fig:9}
\end{figure*}

\begin{figure*}
        \begin{minipage}[t]{0.5\textwidth}
                \centering
                \includegraphics[width=\textwidth]{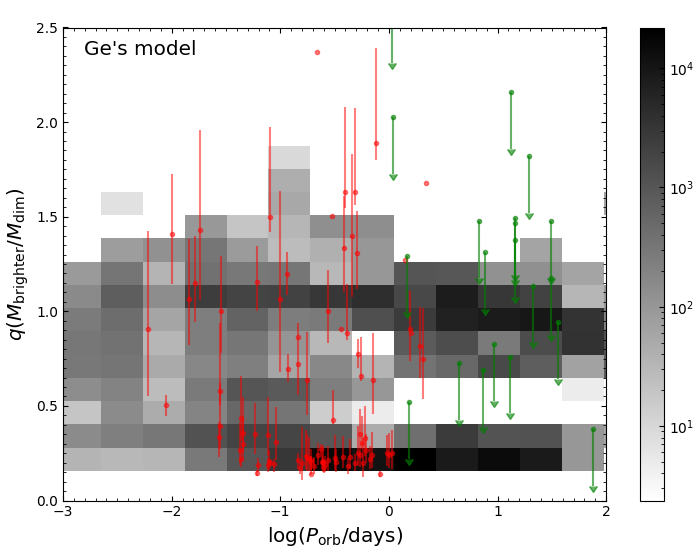}
        \end{minipage}
        \begin{minipage}[t]{0.5\textwidth}
                \centering
                \includegraphics[width=\textwidth]{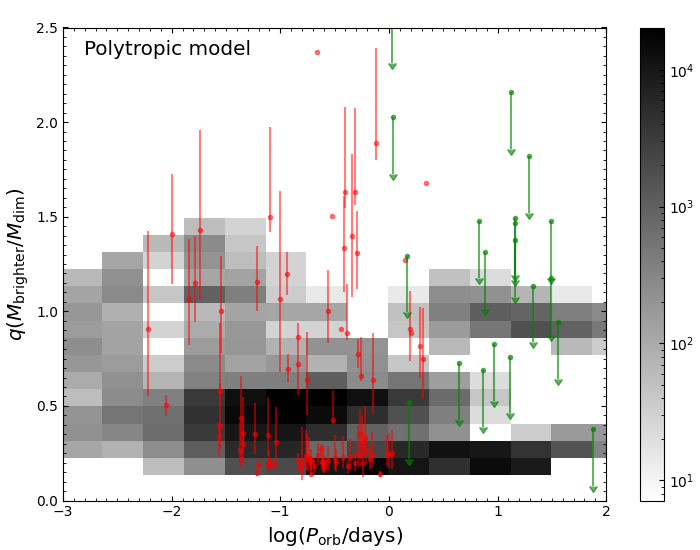}
        \end{minipage}
    \caption{Number density distribution of visible DWDs as a function of mass ratio and orbital period, where the mass ratio is defined as the ratio of the brighter WD mass to the dimmer WD mass. {The red dots are for the observed samples with determined component masses and the green dots are for those with only minimum companion (dimmer) masses determined \citep{kruckow2021}.} Left and right panels are for Ge's and polytropic models, respectively.}
    \label{fig:10}
\end{figure*}

The DWD mass distributions with $\alpha_{\rm CE}=1$ are shown in Figure \ref{fig:9}, where the left and right panels are related to Ge's and the polytropic model, respectively. {Even though the total number of DWDs is a few times higher in Ge's model, the number of detectable DWDs is actually larger in the polytropic model since more ELM WDs have been predicted. We find that the numbers are $3.08\times 10^5$ and $1.55\times 10^5 $ for the polytropic and Ge's models, respectively. The results with other values of $\alpha_{\rm CE}$ are presented in Table \ref{tab:1}. The observational DWDs taken from \citet{kruckow2021} are also shown for comparison and are listed in Appendix \ref{tab:C1}. It should be noted that the observation samples are collected from multiple observational projects and the observational biases are very different. The overall number distribution of the simulations and the observations cannot be directly compared \citep{kilic2011a,toonen2012}. Therefore, we only give a qualitative analysis of the properties of DWD populations here.}  

For Ge's model, we see that the proportion of {DWDs with ELM WD companions} increases with comparing to the results without selection effects (lower left panel of Figure \ref{fig:5}). The reason is that the ELM WD generally has a massive hydrogen-rich envelope and can sustain the high luminosity for a long time \citep{istrate2014b,chenx2017,lizw2019}. For the polytropic model, we see that the simulations concentrate on the DWDs with ELM WD companions, and DWDs with massive companions decrease a lot in comparison with the lower left panel of Figure \ref{fig:6}. {It is clear that the samples in the observations are weighted towards the low-mass WDs due to the large radial velocity variation. Our results may suggest that there are many massive DWDs expected to be detected for both models. Moreover, the simulations in Ge's model predict more DWDs with component masses $\gtrsim 0.5M_\odot$ than that in the polytropic model. There are several observational DWDs with $M_{\rm dim}\lesssim 0.2M_\odot$ that cannot be explained by both models. The possible reason can be understood as follows. In our simulations, the brighter WDs are younger in most cases. However, the extra heating mechanism, such as tidal heating, may increase the surface temperature of the older WD \citep{burdge2019}, thus, the older WD may be observed as the brighter one. Nevertheless, most of the observations can be covered by the simulations of both models. Presently, it is hard to determine which model supports the observation better due to the existence of the observational bias. A quantitative comparison between the simulations and the volume-limited DWD samples, therefore, is necessary and we will address this issue in Section 5.1.3.}

\subsubsection{Mass ratio-orbital period distribution}
\label{subsubsec:5.1.2}
Figure \ref{fig:10} shows the density distribution of visible DWD populations as a function of mass ratio and orbital period for Ge's and polytropic models with $\alpha_{\rm CE} = 1$. In Ge's model, the DWDs cluster around mass ratios of $q\sim 0.25$ and $\sim 1.0$. While the polytropic model shows $q\lesssim 0.5$ for most DWDs and it underestimates the contribution of DWDs with $q\sim 1$. The peak of mass ratio of $\sim 0.25$ is attributed to DWDs with ELM WD companions, since the ELM WDs can sustain high luminosity for a long time and are more easily to be discovered. In the early works, \citet{iben1997,han1998} studied the properties of DWD populations and found that the mass ratio distribution has a dominant peak of $\sim 0.5$ and a second peak of $\sim 1.8$. The former peak is consistent with the results in the polytropic model. However, the later peak of mass ratio is not found in this work. The possible reason is that the different stellar evolutionary tracks\footnote{\citet{toonen2012} calculated the DWD populations with stellar evolutionary tracks adopted from \citet{eggleton1973} and \citet{hurley2000}, respectively. They found more He WDs are produced by using the former tracks due to the difference in the stellar tracks related to helium ignition under degenerate conditions. Moreover, \citet{toonen2012} also found that the DWDs cluster around a mass ratio of $\sim 0.5$ with the stellar tracks, similarly to what is shown in this work, consistent with the results of the polytropic model.} we adopted. By using the $\gamma-$mechanism, \citet{nelemans2001a} found that the DWDs concentrate to $q\sim 1$. The result is similar to that in Ge's model. {It should be noted that the populations of DWDs with ELM WD companions are obtained with only rapid population synthesis method in these works, while we also considered the ELM WDs from the CE+RL channel by adopting the detailed binary evolution code. Then the contributions of DWDs with ELM WD companions may be underestimated in the previous works and, thus, the peak of mass ratio around $\sim 0.25$ is not presented.}

\subsubsection{Space density of DWDs}
\label{subsubsec:5.1.3}

We then calculate the space density of detectable DWDs and make a comparison with the observations, {where the results with $\alpha_{\rm CE}=1$ are discussed in detail (see table~\ref{tab:1} for the results of other $\alpha_{\rm CE}$}). {In the simulations, the detectable number of DWDs in 5 kpc for the polytropic model is about two times greater than that of Ge's model, where the numbers are $3.08\times 10^5$ and $1.55\times 10^5 $ for the polytropic and Ge's models, respectively. }The space density $\rho$ can be estimated by $\rho = N/V_{\rm eff}$, where $N$ is the number of the sources, and $V_{\rm eff}$ is the effective volume, which can be calculated by numerical integration of space density in Eq. \ref{eq:22} for a given distance (see also \citealt{inight2021} for more details). For $d = 5$ kpc, the $V_{\rm eff}$ is about $115\;\rm kpc^{3}$. Then, $\rho =1347 \;\rm kpc^{-3}$ and $2513\;\rm kpc^{-3}$ for Ge's and polytropic models, respectively (see Table \ref{tab:1} for other $\alpha_{\rm CE}$). {In the observations, the space density should be given with a nearly completeness sample. \citep{holberg2016} estimated the local space density of DWDs to be $>620\;\rm kpc^{-3}$ from a volume-limited survey of 25 pc. Recently, \citet{inight2021} extended the DWD samples to $300\;\rm pc$, and found the space density is about $730-1350\;\rm kpc^{-3}$, where the uncertainties come from the assumption of scale height of disk. We see that the result of Ge's model supports the observation estimation better than that of the polytropic model.}

Furthermore, the samples of ELM WD binaries in the ELM Survey are fairly complete \citep{brown2022}, and the local space density is estimated to be $160-300\;\rm kpc^{-3}$, depending on the disk models (see also \citealt{pelisoli2019b}, who find a space density of $275 \;\rm kpc^{-3}$). \citet{lizw2019} studied the populations of DWD with ELM WD companions, where the ELM WD is defined as He WD with mass $\leq 0.3M_\odot$, and its companion is assumed to be CO WD with mass $\geq 0.45M_\odot$, which is the most common companion type of ELM WDs in the observations \citep{brown2020}. They found the space density of ELM WD binaries in the simulations is about $1500 \;\rm kpc^{-3}$, which is significantly larger than that of the observations. Here we also calculate the space density of DWD with ELM WD companions. We chose the ELM WD binaries with $P_{\rm orb}<2\;\rm d$ according to the selection effects in the ELM Survey\footnote{To promise a high completeness of the observations, several selection effects are considered, including semi-amplitude $k>75 \;\rm km\; s^{-1}$, orbital period $P_{\rm orb}<2\;\rm  d$; surface gravity in the range of $4.85<\log g(\rm cm s^{-2})<7.15$; and color selection of $8000<T_{\rm eff}<22000\;\rm  K$, where $T_{\rm eff}$ is the effective temperature \citep{brown2016a,brown2020}. To keep the consistency in this work, we only consider the orbital period and magnitude limits rather than the selection effects in ELM Survey.} \citep{brown2016b}. The space densities of DWDs with ELM WD companions in this study are $473 \;\rm kpc^{-3}$ and $1443\;\rm {kpc^{-3}}$ for Ge's and polytropic models, respectively. The space density of DWDs with ELM WD companions in Ge's model is also close to the estimation in the observations. {Altogether, the simulations with the new mass transfer stability criterion support the volume-limited sample (or a sample with high completeness) in the observations better.}

\subsection{DWD Merger rate}
\label{subsec:5.2}

\begin{figure}
                \centering
                \includegraphics[width=\columnwidth]{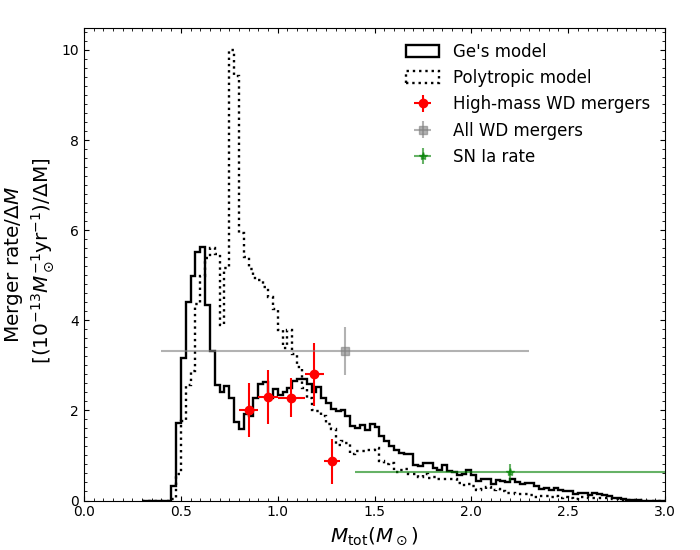}
                \centering
                \includegraphics[width=\columnwidth]{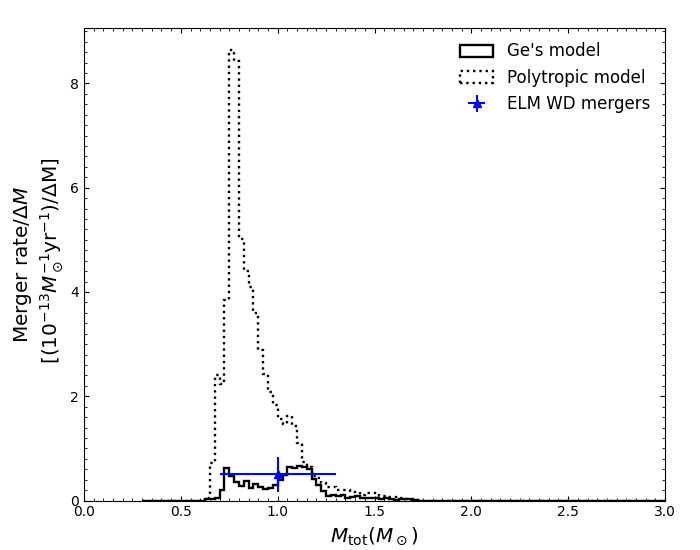}
        \caption{ {Merger rate distribution as a function of the merger mass for all DWD mergers (upper panel).} Black solid and dotted lines are for Ge’s and polytropic models, respectively. The red circles represent the DWD mergers that produce $0.8-1.32M_\odot$ WDs \citep{chengs2020}, the blue one is for the ELM WD mergers \citep{brown2020}, the grey one is for all DWD mergers \citep{maoz2018}, and the blue stars show the observed SN Ia rate \citep{liw2011}. {We note that the merger rate is divided by $\Delta M$, where $\Delta M$ equals the bin size ($0.025M_\odot$) in the simulations and equals the horizontal error bar of the data points. Comparisons between a data point and the histogram should be made in terms of the area under the horizontal error bar of the data point and the area under the histogram in the same mass range.}  The ELM WD is defined as He WD with mass $\leq0.30M_\odot$, and its companion is assumed to be CO WD ($\geq0.45M_\odot$), which is the most common companion type found in ELM Survey (shown in the lower panel).}
    \label{fig:11}
\end{figure}

After the birth of DWDs, the GWR is the only angular momentum loss mechanism that shrinks the orbit. For a DWD with typical WD masses of $0.6M_\odot$ and orbital period of $0.42\;\rm d$, the GW merger timescale is approximately equal to Hubble timescale. We then could expect many DWD mergers in the Galaxy as seen from period distribution in Figure \ref{fig:8}. The DWD merger products are rich, for instance,\ R CrB stars, single massive WDs, or SNe Ia \citep{paczynski1976,webbink1984,maoz2014}, depending on the WD structure and mass. In this section, we calculate the DWD merge rate and compare the simulated results with the observations.

Figure \ref{fig:11} presents the {DWD merger rate\footnote{{The merger rate has been divided by the total mass of stars in the Galaxy ($7\times 10^{10}M_\odot$), it has an unit of ${\rm yr}^{-1}M_\odot^{-1}$.}} distribution as a function of the merger mass, }where we do not consider the possible mass loss during the merger processes and the masses of the merger products simply equal to the sum of the DWD masses. \citet{chengs2020} analyzed the transverse-velocity distribution of high-mass WDs and obtained the relation between DWD merger rate and merger mass, as shown in red circles. Some other observations from DWD mergers and SN Ia rate are also shown for comparison \citep{maoz2018,brown2020,liw2011}. {To make a direct comparison between the simulations and the observations, the merger rate in Figure \ref{fig:11} is divided by $\Delta M$, where $\Delta M$ equals the bin size ($0.025M_\odot$) in the simulations and equals the horizontal error bar of the data points. Thus, the merger rate should be understood as the area under the histogram or the horizontal error bar (see \citealt{chengs2020} for more details). }

The simulation results of Ge's model are close to the high-mass WD mergers, while the results of the polytropic model predict more high-mass WD mergers with masses of $0.8-1.1M_\odot$. The merger rate of DWD mergers that produce $0.8-1.32M_\odot$ WDs in the observations is about $1.1 \pm0.3 \times 10^{-13}M_\odot^{-1}\rm yr^{-1}$ \citep{chengs2020}. In our simulations, the merger rates for those DWD mergers are $1.26\times 10^{-13}M_\odot^{-1}\rm yr^{-1}$ and $1.80\times 10^{-13}M_\odot^{-1}\rm yr^{-1}$ for Ge's and polytropic models, respectively. It is clear that the simulation results from Ge’s model support the observations better. We also note that more mergers with a mass larger than 1.4$M_\odot$ (Chandrasekhar mass) are produced in Ge's model, and it may indicate that DWD mergers in Ge's model have a relatively larger contribution to the SN Ia. We will address this issue in Section \ref{subsec:5.3}. Besides, \citet{chengs2020} also simulated the merger rate distribution of DWDs but with $\gamma-$mechanism. They find a dominant peak around $0.6M_\odot$. The results are similar to that of the new criterion in this work. It may suggest that the effect of $\gamma-$mechanism is similar to that of non-conservative stable RLOF in the first mass transfer phase {(\citealt{woods2012}; see Section \ref{sec:6} for a discussion.)}.

In the lower panel of Figure \ref{fig:11}, we only consider the merger rate of ELM WD binaries. In the observations, the total merger rate {(the merger rate in the lower panel multiply the total mass of the Galaxy)} of DWDs with ELM WD companions in the Galaxy is about $2\times10^{-3}\;\rm yr^{-1}$ \citep{brown2020}. The corresponding values in our simulations are $11.4\times10^{-3}\;\rm  yr^{-1}$ and $1.8\times10^{-3}\;\rm  yr^{-1}$ for the polytropic and Ge's models. The results show that the synthesized merge rate of ELM WD binaries in Ge's model supports the observations well without any tuning of parameters, {while the polytropic model overpredicts the merger rate of ELM WD binaries by a factor of $\sim 6$.}

\subsection{SN Ia birthrate}
\label{subsec:5.3}

\begin{figure}
                \centering
                \includegraphics[width=\columnwidth]{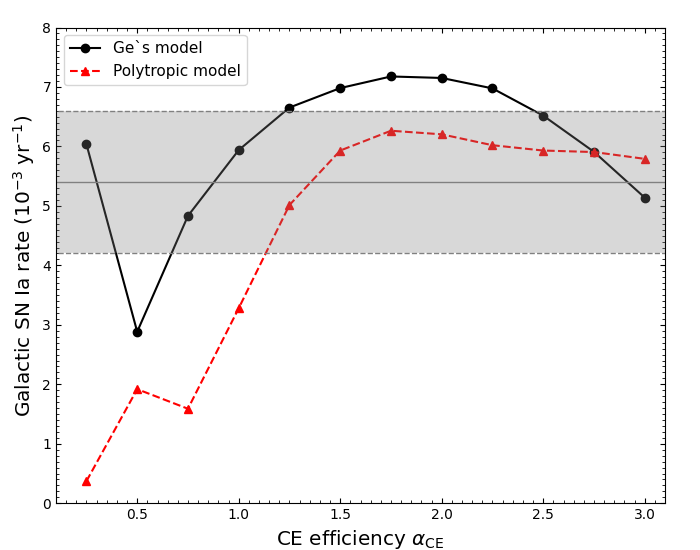}
        \caption{Influence of $\alpha_{\rm CE}$ on the Galactic SN Ia rate. Black solid and red dashed lines are for Ge's and polytropic model, respectively. The Galactic SN Ia rate in the observations is $(5.4 \pm 1.2) \times 10^{-3}\;{\rm yr^{-1}}$ \citep{liw2011}, as shown in grey shade region.}
    \label{fig:12}
\end{figure}

\begin{figure}
                \centering
                \includegraphics[width=\columnwidth]{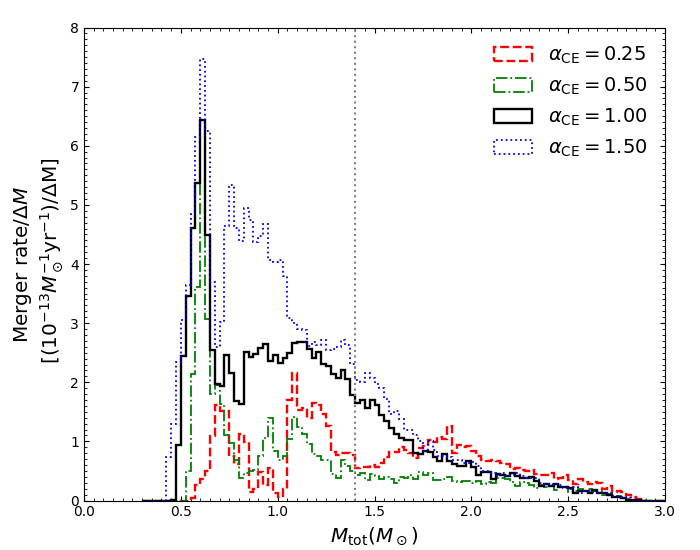}
        \caption{Merge rate distributions with different $\alpha_{\rm CE}$ in Ge's model. The vertical dotted line represents the Chandrasekhar mass ($1.4M_\odot$), beyond which the merger product is supposed to explode as SN Ia.}
    \label{fig:13}
\end{figure}

SNe Ia are thought to be thermonuclear explosions of degenerate WD, and are possessed of an important position in astrophysics \citep{perlmutter1999}. One of the main progenitor models of SN Ia is the single-degenerate model, and the WD accretes material from the non-degenerate companion star {\citep{whelan1973,nomoto1982}}. Another progenitor model is the double degenerate model, in which the merger of DWD has a mass larger than the Chandrasekhar mass {(\citealt{tutukov1981,iben1984,webbink1984}; see also \citep{wangb2012} and references therein)}. In this section, we discuss the contribution of the DWD mergers to the SNe Ia. For convenience, we consider as SNe Ia results of mergers with a total mass larger than $1.4M_\odot$, which gives an upper limit for the SN Ia rate under the double-degenerate scenario \citep{ablimit2016}.

\begin{figure*}
                \centering
                \includegraphics[width=\textwidth]{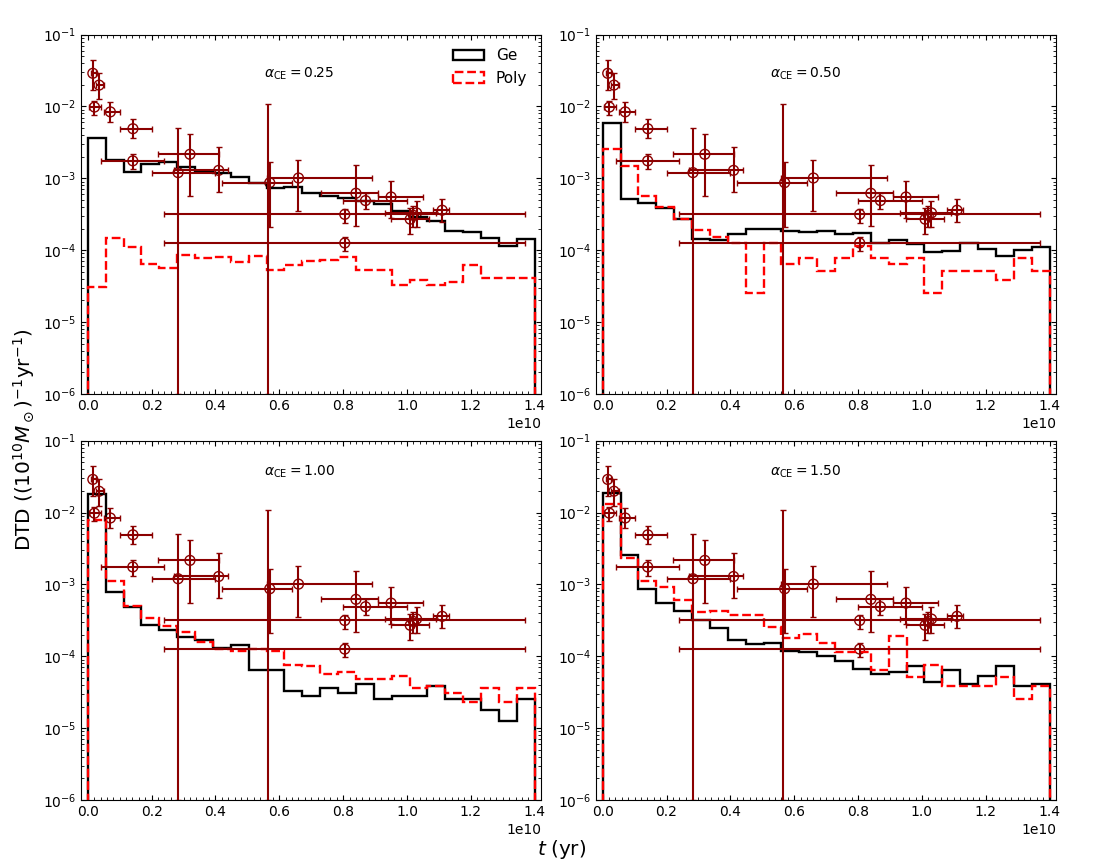}
        \caption{DTDs of SNe Ia with different $\alpha_{\rm CE}$. The open circles with error bars represent the observation SNe Ia rate from elliptical galaxies and galaxy clusters \citep{totani2008,maoz2010,maoz2012,graur2013}. Black solid and red dashed lines are for Ge’s and polytropic models, respectively.}
    \label{fig:14}
\end{figure*}

\subsubsection{Galactic SN Ia rate}
The current Galactic birth rate of SNe Ia with different CE efficiencies for the two models is presented in Figure \ref{fig:12}, where the SN Ia rate for a Milky Way-like galaxy in the observation is about $(5.4\pm1.2)\times 10^{-3}\;\rm yr^{-1}$ (with a systematic uncertainty of a factor of $\sim 2$, \citealt{liw2011}), as shown in grey shaded region. It is clear that the SN Ia rate is not a monotone function of $\alpha_{\rm CE}$. The reason can be understood as follows. With the increase of $\alpha_{\rm CE}$, the total number of DWDs also increases, since the CE ejection is efficient, as shown in Table~\ref{tab:1}. However, the CE efficiency also has an important effect on the final binary separation. For the CE process with a small $\alpha_{\rm CE}$, the binary separation will be short after the ejection of CE. Then the proportion of DWD mergers in the Hubble time may increase for a low $\alpha_{\rm CE}$ (see also \citealt{claeys2014}). In addition, the CE for binaries with massive components is more easily to be ejected for a small $\alpha_{\rm CE}$. 

In Fig. \ref{fig:12}, we note that the SN Ia rate for $\alpha_{\rm CE}=0.25$ in Ge's model is surprisedly large.
We then present in Figure \ref{fig:13} 
the merger rate variation with total mass of DWDs
within various $\alpha_{\rm CE}$ values in Ge's model.
We clearly see that the merger rate distribution versus on total mass for $\alpha_{\rm CE} = 0.25$ is totally different from that of 
other $\alpha_{\rm CE}$ values. 
An obvious feature is that the merger rate with $M_{\rm tot}\geq1.4M_\odot$ increases in comparison with that of $\alpha_{\rm CE} = 0.5$, and becomes comparable with that of $\alpha_{\rm CE}=1.0,1.5$.
{We can understand this according to the following. Only DWDs with very short orbital periods, generally $\lesssim 0.1\;\rm d$, would merge in the current epoch due to gravitational wave radiation. For the case of $\alpha_{\rm CE} = 0.25$, the binary components should be massive enough to eject the CE due to the extremely small CE efficiency, and the produced DWDs have very short orbital periods as well. The proportion of massive DWDs with short orbital periods is relatively large consequently, leading to more SNe Ia being produced. With increasing of $\alpha_{\rm CE}$ values, more and more DWDs with small component masses are produced, while those with massive components have relatively long orbital periods and cannot come to merge presently. We see in figure that the case of $\alpha_{\rm CE}=0.5$ has a minimum value for the SN Ia rate. 
}

The observation of SN Ia rate can be covered by both models. 
However, compared with the polytropic model, the results in Ge's model predict more SNe Ia for $\alpha_{\rm CE}\lesssim2.75$. 
In the fiducial model, Ge's model predicts a Galactic supernova Ia rate about $6\times 10^{-3}\;\rm yr^{-1}$, which supports the observations of $(5.4\pm1.2)\times 10^{-3}\;\rm yr^{-1}$. Besides, we also find the SN Ia rate is significantly affected by the $\alpha_{\rm CE}$ in the polytropic model, where the values of SN Ia rate spreads a wide range from $0.3\times 10^{-3}$ to $6.2\times 10^{-3}\;\rm yr^{-1}$ with the change of $\alpha_{\rm CE}$. While for Ge's model, the SN Ia rates are in a relatively narrow range of $\sim 3-7\times 10^{-3}\;\rm yr^{-1}$. The results suggest that the current SN Ia rate in the observations may put a limit on the CE efficiency.  

\subsubsection{Delay time distribution}
Figure \ref{fig:14} presents the delay time distributions (DTDs) of SNe Ia for a $10^{10}M_\odot$ single burst, where four typical examples of $\alpha_{\rm CE} =0.25,0.5,1,1.5$ are shown for comparison. The results of Ge's and polytropic models are shown in the black solid and red dashed lines, respectively. For the simulation with large CE efficiencies of $1.0,1.5$, the DTDs for the two models are comparable. And the results for both models are less than the observations as a whole, similarly to previous studies (e.g., \citealt{yungelson2017,liud2018}). With the decrease of $\alpha_{\rm CE}$, the total number of DWDs falls down significantly for the polytropic model, as shown in Table \ref{tab:1}, resulting in smaller SNe Ia rates.
For Ge's model, however, the total number of DWDs changes little with $\alpha_{\rm CE}$ as mentioned before,  
the CE efficiency therefore has a limited effect on the DTD of SN Ia.
In particular, for the case of $\alpha_{\rm CE}=0.25$, 
the results in the polytropic model are about $1$ order of magnitude less than that of the observations, 
while the DTD in Ge's model supports the observations well for delay time large than $\sim 1\;\rm Gyr$. 

{Unfortunately, none of the simulations are equipped to explain all of the observational constraints of the DTDs. In Ge's model, there is a large discrepancy between the simulations and the observations for short delay time ($\lesssim 1\;\rm Gyr$) in the case of $\alpha_{\rm CE}=0.25$, and for long delay time ($\gtrsim 1\;\rm Gyr$) in the cases of $\alpha_{\rm CE} = 1.0,1.5$. The results suggest that there should be other channels contributing to the SNe Ia. For example, \citet{wangb2009} found that the single-degenerate model of WD + He star channel has a significant contribution to the SNe Ia with short delay times (see also \citealt{claeys2014}, and references therein), and WD+MS and WD+RGB channels can produce SNe Ia with long delay times \citep{wangb2010}. In addition, some variants of single-degenerate and double degenerate models may also have a contribution to the SNe Ia, such as the sub-Chandrasekhar mass model, the single-star model, and the delayed dynamical instability model, etc \citep{wangb2012}. We expect that the future observations of large samples of SNe Ia could set further constraints on SN Ia progenitor models.}

\subsection{GW signal of DWDs}
\label{subsec:5.4}

Overall, DWDs are meant to be the dominant GW sources for future space-based GW detectors. According to BPS simulations, it has been suggested that about $(\sim1-3) \times10^4$ sources are expected to be individually detected by LISA or TianQin \citep{evans1987,webbink1998,nelemans2001a,liu2009,liu2010,ruiter2010,yus2010,nissanke2012,korol2017,korol2019,lamberts2019,huangs2020,korol2022a} and several tens of them are to be potentially observed with the combination of GW and electromagnetic observations \citep{korol2017,korol2019,kupfer2018,lizw2020}. Multi-messenger studies of DWDs provide more precise information \citep{shah2012,shah2013} and shed light on the physics of tidal interaction and mass transfer between DWDs \citep{fuller2013,baker2019}.

As discussed above, the mass transfer stability criterion has a significant effect on the formation of DWDs. Compared with the polytropic model, more DWDs are produced in Ge's model. However, the proportion of DWDs with short orbital periods decreases in Ge's model, since most binaries experience the stable RLOF process in the first mass transfer phase. {In Table \ref{tab:1}, we present the number of DWDs with $P_{\rm orb} < 60 \;\rm min$ for both models}. For the polytropic model, the close DWDs (here defined as DWD with $P_{\rm orb} < 60\;\rm  min$) are mainly produced from the CE+CE channel, while from the RL+CE channel in Ge's model {(CE+RL channel can also produce a small number of close DWDs; see \citet{chenh2022} for more details)}. Therefore, the CE efficiency would lead to a more substantial effect on the polytropic model. We see that for the cases of $\alpha_{\rm CE}\leq 0.5$, there are more close DWDs in Ge's model. With the increase of $\alpha_{\rm CE}$, the numbers of close DWDs in the polytropic model are larger than that of Ge's model.

We then calculated the numbers of DWDs that can be detected by LISA. The specific methods to obtain the signal-noise-ratios (S/N) of the DWDs can be found in \citet{lizw2020}. The main input parameters are described as follows. The distance of each simulated binary to the Sun is got from the spatial distribution of stars in the Galaxy, where the Galaxy model with disk and bulge components are adopted from \citet{binney2008}. The sensitivity curve of LISA is taken from \citet{robson2018}, where we adopted the current LISA configuration with an arm length of $2.5 \times 10^6$ km for the three detector arms, and a total observation time of 4 yr \citep{LISA2017}. A source could be detected by LISA if its S/N was higher than 7. 

The numbers of LISA detectable sources for different models are shown in Table \ref{tab:1}. 
For a relatively small $\alpha_{\rm CE}$ value of $0.25$, the number of LISA detectable sources is $1452$ in the polytropic model, which is only about 10\%\ of that of Ge's model. With the increase of $\alpha_{\rm CE}$, the polytropic models predict more LISA detectable sources. The results are comparable to the previous works on the GW studies (e.g., \citealt{nelemans2001b,yus2010,korol2017}). In the fiducial model of $\alpha_{\rm CE} = 1$,  the number of
detectable GW sources in the polytropic model is larger than that in Ge's model by about $35\%$. It should be noted that the DWDs with a orbital period $\lesssim 1\;\rm  d$ may contribute to the foreground noise, which has an important effect on the GW detections (e.g., \citealt{evans1987,nelemans2001b,nissanke2012,korol2022a,georgousi2022}). The orbital period distributions of the two models (Figures \ref{fig:8}) in this work show a large difference and whether it affects the foreground noise should be addressed in further studies.

\section{{Discussion: Stable RLOF versus $\gamma-$mechanism}}
\label{sec:6}

{The $\gamma-$mechanism suggested by \citet{nelemans2000} has been studied in many previous works (e.g., \citealt{toonen2012,toonen2017}) and the observed properties of DWDs can be well reproduced in $\gamma-$mechanism. 
The main argument for this mechanism is that there is no clear physical explanation (as mentioned previously). 
An alternative solution for the formation of DWDs is that substituting the $\gamma-$mechanism to stable but non-conservative mass transfer \citep{woods2012}, as investigated in this work. 
As pointed out by \citet{toonen2012}, the stable non-conservative mass transfer may have an effect alike the $\gamma-$description. 
Here, we offer a brief discussion on the similarity between the two mechanisms. }

{The governing equation of $\gamma-$mechanism can be expressed as: 
\begin{eqnarray}
\frac{\Delta J_{\rm lost}}{J_{\rm i}} = -\gamma\frac{\triangle M_{1}}{M}, 
\label{eq:23}
\end{eqnarray}
where $\Delta J_{\rm lost}$ is the difference value between the initial and final orbital angular momentum, and the dimensionless parameter of $\gamma$ is adopted around $1.5-1.75$ \citep{nelemans2005}. Then, the change of orbital separation determined by Eq. \ref{eq:23} becomes a function of the initial and final masses of the binary components. We note that it may give a widened orbit after the ejection of CE. On the other hand, the angular momentum loss caused by stable non-conservative mass transfer in Eq. \ref{eq:11} can also be expressed as \citep{postnov2014}
\begin{eqnarray}
  \frac{\triangle J_{\rm lost}}{J_{\rm i}} = -(1-\beta_{\rm rlof})\frac{M_1}{M_2}\frac{\triangle M_1}{M},
  \label{eq:24}
\end{eqnarray}
Comparing Eqs. \ref{eq:23} and \ref{eq:24}, we see that the orbital angular momentum loss of the $\gamma-$mechanism and non-conservative mass transfer is within one order of magnitude. This may suggest that the $\gamma-$mechanism and stable non-conservative mass transfer have similar effects on the change of orbital separation.} 

{In this work, we carry out BPS simulations with the new mass transfer stability criterion from Ge's model and the results show that most of DWDs have experienced stable RLOF in the first mass transfer phase regardless of the adopted value of $\alpha_{\rm CE}$ (see Table \ref{tab:1}). It means that we have succeeded in constructing the models that substitute the $\gamma-$mechanism with a stable non-conservative mass transfer for the first mass-transfer phase. Our results confirm that the stable non-conservative RLOF + CE channel can produce abundant DWDs and some of the observational properties of DWDs can be adequately reproduced in Ge's model. Comparing the results with that of $\gamma-$mechanism (e.g., \citealt{nelemans2001b,toonen2012,toonen2017,chengs2020}), both of these models give a mass ratio peak value of $\sim 1$ and quite similar profiles for the merger rate distribution, as shown in Section \ref{sec:5}. It supports the aforementioned viewpoints that the stable non-conservative mass transfer may have an effect akin to the $\gamma-$description. However, it should be noted that the accretor can accumulate a significant part of masses during the non-conservative mass transfer, which is contrary to the $\gamma-$mechanism. Therefore, these two models should lead to some differences in the DWD populations and this issue is worthy of investigation in future studies.}

\section{Conclusions}
\label{sec:7}

In this paper, we systemically investigate the influence of mass transfer stability on the DWD populations. Two criteria of stability are considered for comparison. In the two models, the critical mass ratios are obtained from {the adiabatic mass-loss model with polytropic stellar structures (polytropic model) and with realistic stellar structures (Ge's model), respectively}. We performed Monte Carlo simulations to obtain the Galactic DWD populations and to make a comparison with the observations. The implications for the SN Ia rate and GW observations are also discussed. Our main conclusions are summarized as follows.
\begin{enumerate}
  \item In the model featuring a new mass transfer stability (Ge's model), the mass transfer tends to be stable for stars filling the Roche lobe at RGB or AGB phase. {Most binaries experience stable RLOF in the first mass transfer phase and more than $\sim 90\%$ DWDs are produced from the RL+CE channel regardless of the values of $\alpha_{\rm CE}$. For the polytropic model, $\alpha_{\rm CE}$ is a determined parameter of the DWD populations, that is, the RL+CE channel is the dominant formation channel of DWDs for $\alpha_{\rm CE}<1$, and the CE+CE channel becomes the dominant channel of DWDs for high value of the $\alpha_{\rm CE}$. In the fiducial model where $\alpha_{\rm CE} = 1$, the number of DWDs in Ge's model is about four times greater than that of the polytropic model because certain CE mergers are avoided.}
  \item We constructed the magnitude-limited sample for the two models. {Due to the incompleteness of the observed samples, we give a qualitative discussion about the properties of DWDs in the DWD mass distribution diagram and the mass ratio-orbital period diagram. Most of the observed DWDs can be explained in our results. Then we calculate the space density of the detectable DWDs and those with ELM WD companions in Ge's model, which are $1347 \;\rm kpc^{-3}$ and $473\;\rm kpc^{-3}$, respectively. The values are close to the observation estimations. While the polytropic model overpredicts the space density of DWDs by a factor of about 2-3.}
  \item We calculated the merger rate of DWDs and their mass dependence, and the results of Ge's model reproduce the observations better than the polytropic model. For the case of DWDs with ELM WD companions, we find the merger rate is about $1.8\times10^{-3}\;\rm  yr^{-1}$, which is comparable to the observation estimation of $2 \times 10^{-3}\;\rm yr^{-1}$.
  \item {The results in Ge's model predict relatively more DWD mergers contributing to the Galactic SNe Ia than that in polytropic model. 
For the DTDs of SNe Ia, none of the simulations could explain all of the observational constraints. It suggests that there should be other channels contributing to the SNe Ia, such as the single-degenerate model.}
    \item We offer a discussion about the influence of mass transfer stability on the GW observations. The numbers of DWDs that can be detected by LISA for both models have been calculated. The number of LISA detectable sources is significantly affected by $\alpha_{\rm CE}$. In the fiducial model of $\alpha_{\rm CE} = 1$, the number of detectable GW sources in the polytropic model is larger than that in Ge's model by about $35\%$.
\end{enumerate}
We confirm that mass transfer stability criterion plays an important role in the formation of DWDs. The results from Ge's model support the observations better than what can be derived from the polytropic model.

\begin{acknowledgements}
  {We are grateful to the anonymous referee for the helpful comments and suggested improvements.} We thank Huaju Li, Chuanxi Liu, Yangyang Zhang, {Chengyuan Wu} for the helpful discussions. This work is supported by National Key R$\&$D Program of China (Grant Nos. 2021YFA1600403, 2021YFA1600400), National Natural Science Foundation of China (Nos. 11733008,12103086,12090040/12090043,12125303,12073071,12173081,12288102), the Key Research Program of Frontier Sciences of CAS (Grant No. ZDBS-LY-7005), the Western Light Youth Project of CAS and the Yunnan Fundamental Research Projects (Nos. 202101AU070276, 202101AV070001, 202201AU070234, 202001AT070058, 202101AW070003). We also acknowledge the science research grant from the China Manned Space Project with No.CMS-CSST-2021-A10. 
  
\end{acknowledgements}

\bibliographystyle{a_a}
\bibliography{aa}

\begin{thebibliography}{198}
\expandafter\ifx\csname natexlab\endcsname\relax\def\natexlab#1{#1}\fi

\bibitem[{{Ablimit} {et~al.}(2016){Ablimit}, {Maeda}, \& {Li}}]{ablimit2016}
{Ablimit}, I., {Maeda}, K., \& {Li}, X.-D. 2016, \apj, 826, 53

\bibitem[{{Althaus} {et~al.}(2010){Althaus}, {C{\'o}rsico}, {Isern}, \&
  {Garc{\'\i}a-Berro}}]{althaus2010}
{Althaus}, L.~G., {C{\'o}rsico}, A.~H., {Isern}, J., \& {Garc{\'\i}a-Berro}, E.
  2010, \aapr, 18, 471

\bibitem[{{Amaro-Seoane} {et~al.}(2017){Amaro-Seoane}, {Audley}, {Babak},
  {Baker}, {Barausse}, {Bender}, {Berti}, {Binetruy}, {Born}, {Bortoluzzi},
  {Camp}, {Caprini}, {Cardoso}, {Colpi}, {Conklin}, {Cornish}, {Cutler},
  {Danzmann}, {Dolesi}, {Ferraioli}, {Ferroni}, {Fitzsimons}, {Gair}, {Gesa
  Bote}, {Giardini}, {Gibert}, {Grimani}, {Halloin}, {Heinzel}, {Hertog},
  {Hewitson}, {Holley-Bockelmann}, {Hollington}, {Hueller}, {Inchauspe},
  {Jetzer}, {Karnesis}, {Killow}, {Klein}, {Klipstein}, {Korsakova}, {Larson},
  {Livas}, {Lloro}, {Man}, {Mance}, {Martino}, {Mateos}, {McKenzie},
  {McWilliams}, {Miller}, {Mueller}, {Nardini}, {Nelemans}, {Nofrarias},
  {Petiteau}, {Pivato}, {Plagnol}, {Porter}, {Reiche}, {Robertson},
  {Robertson}, {Rossi}, {Russano}, {Schutz}, {Sesana}, {Shoemaker}, {Slutsky},
  {Sopuerta}, {Sumner}, {Tamanini}, {Thorpe}, {Troebs}, {Vallisneri},
  {Vecchio}, {Vetrugno}, {Vitale}, {Volonteri}, {Wanner}, {Ward}, {Wass},
  {Weber}, {Ziemer}, \& {Zweifel}}]{LISA2017}
{Amaro-Seoane}, P., {Audley}, H., {Babak}, S., {et~al.} 2017, arXiv e-prints
  [\eprint[arXiv]{1702.00786}]

\bibitem[{{Badenes} \& {Maoz}(2012)}]{badenes2012}
{Badenes}, C. \& {Maoz}, D. 2012, \apjl, 749, L11

\bibitem[{{Badenes} {et~al.}(2009){Badenes}, {Mullally}, {Thompson}, \&
  {Lupton}}]{badenes2009}
{Badenes}, C., {Mullally}, F., {Thompson}, S.~E., \& {Lupton}, R.~H. 2009,
  \apj, 707, 971

\bibitem[{{Baker} {et~al.}(2019){Baker}, {Haiman}, {Rossi}, {Berger}, {Brandt},
  {Breedt}, {Breivik}, {Charisi}, {Derdzinski}, {D'Orazio}, {Ford}, {Greene},
  {Hill}, {Holley-Bockelmann}, {Shapiro Key}, {Kocsis}, {Kupfer}, {Larson},
  {Madau}, {Marsh}, {McKernan}, {McWilliams}, {Natarajan}, {Nissanke}, {Noble},
  {Phinney}, {Ramsay}, {Schnittman}, {Sesana}, {Shoemaker}, {Stone}, {Toonen},
  {Trakhtenbrot}, {Vikhlinin}, \& {Volonteri}}]{baker2019}
{Baker}, J., {Haiman}, Z., {Rossi}, E.~M., {et~al.} 2019, arXiv e-prints
  [\eprint[arXiv]{1903.04417}]

\bibitem[{{Bellm} {et~al.}(2019){Bellm}, {Kulkarni}, {Graham}, {Dekany},
  {Smith}, {Riddle}, {Masci}, {Helou}, \& {Prince}}]{ZTF2019}
{Bellm}, E.~C., {Kulkarni}, S.~R., {Graham}, M.~J., {et~al.} 2019, \pasp, 131,
  018002

\bibitem[{{Bergeron} {et~al.}(1989){Bergeron}, {Wesemael}, {Liebert}, \&
  {Fontaine}}]{bergeron1989}
{Bergeron}, P., {Wesemael}, F., {Liebert}, J., \& {Fontaine}, G. 1989, \apjl,
  345, L91

\bibitem[{{Binney} \& {Tremaine}(2008)}]{binney2008}
{Binney}, J. \& {Tremaine}, S. 2008, {Galactic Dynamics: Second Edition}
  (Princeton University Press)

\bibitem[{{Bland-Hawthorn} \& {Gerhard}(2016)}]{bland2016}
{Bland-Hawthorn}, J. \& {Gerhard}, O. 2016, \araa, 54, 529

\bibitem[{{Boffin} \& {Jorissen}(1988)}]{boffin1988}
{Boffin}, H.~M.~J. \& {Jorissen}, A. 1988, \aap, 205, 155

\bibitem[{{Bondi} \& {Hoyle}(1944)}]{bondi1944}
{Bondi}, H. \& {Hoyle}, F. 1944, \mnras, 104, 273

\bibitem[{{Bours} {et~al.}(2015){Bours}, {Marsh}, {G{\"a}nsicke}, {Tauris},
  {Istrate}, {Badenes}, {Dhillon}, {Gal-Yam}, {Hermes}, {Kengkriangkrai},
  {Kilic}, {Koester}, {Mullally}, {Prasert}, {Steeghs}, {Thompson}, \&
  {Thorstensen}}]{bours2015}
{Bours}, M.~C.~P., {Marsh}, T.~R., {G{\"a}nsicke}, B.~T., {et~al.} 2015,
  \mnras, 450, 3966

\bibitem[{{Bours} {et~al.}(2014){Bours}, {Marsh}, {Parsons}, {Copperwheat},
  {Dhillon}, {Littlefair}, {G{\"a}nsicke}, {Gianninas}, \&
  {Tremblay}}]{bours2014}
{Bours}, M.~C.~P., {Marsh}, T.~R., {Parsons}, S.~G., {et~al.} 2014, \mnras,
  438, 3399

\bibitem[{{Breedt} {et~al.}(2017){Breedt}, {Steeghs}, {Marsh}, {Gentile
  Fusillo}, {Tremblay}, {Green}, {De Pasquale}, {Hermes}, {G{\"a}nsicke},
  {Parsons}, {Bours}, {Longa-Pe{\~n}a}, \& {Rebassa-Mansergas}}]{breedt2017}
{Breedt}, E., {Steeghs}, D., {Marsh}, T.~R., {et~al.} 2017, \mnras, 468, 2910

\bibitem[{{Brown} {et~al.}(2011{\natexlab{a}}){Brown}, {Kilic}, {Brown}, \&
  {Kenyon}}]{brown2011a}
{Brown}, J.~M., {Kilic}, M., {Brown}, W.~R., \& {Kenyon}, S.~J.
  2011{\natexlab{a}}, \apj, 730, 67

\bibitem[{{Brown} {et~al.}(2016{\natexlab{a}}){Brown}, {Gianninas}, {Kilic},
  {Kenyon}, \& {Allende Prieto}}]{brown2016a}
{Brown}, W.~R., {Gianninas}, A., {Kilic}, M., {Kenyon}, S.~J., \& {Allende
  Prieto}, C. 2016{\natexlab{a}}, \apj, 818, 155

\bibitem[{{Brown} {et~al.}(2013){Brown}, {Kilic}, {Allende Prieto},
  {Gianninas}, \& {Kenyon}}]{brown2013}
{Brown}, W.~R., {Kilic}, M., {Allende Prieto}, C., {Gianninas}, A., \&
  {Kenyon}, S.~J. 2013, \apj, 769, 66

\bibitem[{{Brown} {et~al.}(2010){Brown}, {Kilic}, {Allende Prieto}, \&
  {Kenyon}}]{brown2010}
{Brown}, W.~R., {Kilic}, M., {Allende Prieto}, C., \& {Kenyon}, S.~J. 2010,
  \apj, 723, 1072

\bibitem[{{Brown} {et~al.}(2012){Brown}, {Kilic}, {Allende Prieto}, \&
  {Kenyon}}]{brown2012}
{Brown}, W.~R., {Kilic}, M., {Allende Prieto}, C., \& {Kenyon}, S.~J. 2012,
  \apj, 744, 142

\bibitem[{{Brown} {et~al.}(2011{\natexlab{b}}){Brown}, {Kilic}, {Hermes},
  {Allende Prieto}, {Kenyon}, \& {Winget}}]{brown2011b}
{Brown}, W.~R., {Kilic}, M., {Hermes}, J.~J., {et~al.} 2011{\natexlab{b}},
  \apjl, 737, L23

\bibitem[{{Brown} {et~al.}(2016{\natexlab{b}}){Brown}, {Kilic}, {Kenyon}, \&
  {Gianninas}}]{brown2016b}
{Brown}, W.~R., {Kilic}, M., {Kenyon}, S.~J., \& {Gianninas}, A.
  2016{\natexlab{b}}, \apj, 824, 46

\bibitem[{{Brown} {et~al.}(2020){Brown}, {Kilic}, {Kosakowski}, {Andrews},
  {Heinke}, {Ag{\"u}eros}, {Camilo}, {Gianninas}, {Hermes}, \&
  {Kenyon}}]{brown2020}
{Brown}, W.~R., {Kilic}, M., {Kosakowski}, A., {et~al.} 2020, \apj, 889, 49

\bibitem[{{Brown} {et~al.}(2017){Brown}, {Kilic}, {Kosakowski}, \&
  {Gianninas}}]{brown2017}
{Brown}, W.~R., {Kilic}, M., {Kosakowski}, A., \& {Gianninas}, A. 2017, \apj,
  847, 10

\bibitem[{{Brown} {et~al.}(2022){Brown}, {Kilic}, {Kosakowski}, \&
  {Gianninas}}]{brown2022}
{Brown}, W.~R., {Kilic}, M., {Kosakowski}, A., \& {Gianninas}, A. 2022, \apj,
  933, 94

\bibitem[{{Burdge} {et~al.}(2020{\natexlab{a}}){Burdge}, {Coughlin}, {Fuller},
  {Kaplan}, {Kulkarni}, {Marsh}, {Bellm}, {Dekany}, {Duev}, {Graham},
  {Mahabal}, {Masci}, {Laher}, {Riddle}, {Soumagnac}, \&
  {Prince}}]{burdge2020a}
{Burdge}, K.~B., {Coughlin}, M.~W., {Fuller}, J., {et~al.} 2020{\natexlab{a}},
  \apjl, 905, L7

\bibitem[{{Burdge} {et~al.}(2019{\natexlab{a}}){Burdge}, {Fuller}, {Phinney},
  {van Roestel}, {Claret}, {Cukanovaite}, {Gentile Fusillo}, {Coughlin},
  {Kaplan}, {Kupfer}, {Tremblay}, {Dekany}, {Duev}, {Feeney}, {Riddle},
  {Kulkarni}, \& {Prince}}]{burdge2019}
{Burdge}, K.~B., {Fuller}, J., {Phinney}, E.~S., {et~al.} 2019{\natexlab{a}},
  \apjl, 886, L12

\bibitem[{{Burdge} {et~al.}(2019{\natexlab{b}}){Burdge}, {Fuller}, {Phinney},
  {van Roestel}, {Claret}, {Cukanovaite}, {Gentile Fusillo}, {Coughlin},
  {Kaplan}, {Kupfer}, {Tremblay}, {Dekany}, {Duev}, {Feeney}, {Riddle},
  {Kulkarni}, \& {Prince}}]{burdge2019b}
{Burdge}, K.~B., {Fuller}, J., {Phinney}, E.~S., {et~al.} 2019{\natexlab{b}},
  \apjl, 886, L12

\bibitem[{{Burdge} {et~al.}(2020{\natexlab{b}}){Burdge}, {Prince}, {Fuller},
  {Kaplan}, {Marsh}, {Tremblay}, {Zhuang}, {Bellm}, {Caiazzo}, {Coughlin},
  {Dhillon}, {Gaensicke}, {Rodr{\'\i}guez-Gil}, {Graham}, {Hermes}, {Kupfer},
  {Littlefair}, {Mr{\'o}z}, {Phinney}, {van Roestel}, {Yao}, {Dekany}, {Drake},
  {Duev}, {Hale}, {Feeney}, {Helou}, {Kaye}, {Mahabal}, {Masci}, {Riddle},
  {Smith}, {Soumagnac}, \& {Kulkarni}}]{burdge2020b}
{Burdge}, K.~B., {Prince}, T.~A., {Fuller}, J., {et~al.} 2020{\natexlab{b}},
  \apj, 905, 32

\bibitem[{{Camisassa} {et~al.}(2019){Camisassa}, {Althaus}, {C{\'o}rsico}, {De
  Ger{\'o}nimo}, {Miller Bertolami}, {Novarino}, {Rohrmann}, {Wachlin}, \&
  {Garc{\'\i}a-Berro}}]{camisassa2019}
{Camisassa}, M.~E., {Althaus}, L.~G., {C{\'o}rsico}, A.~H., {et~al.} 2019,
  \aap, 625, A87

\bibitem[{{Carrasco} {et~al.}(2014){Carrasco}, {Catal{\'a}n}, {Jordi},
  {Tremblay}, {Napiwotzki}, {Luri}, {Robin}, \& {Kowalski}}]{carrasco2014}
{Carrasco}, J.~M., {Catal{\'a}n}, S., {Jordi}, C., {et~al.} 2014, \aap, 565,
  A11

\bibitem[{{Chandra} {et~al.}(2021){Chandra}, {Hwang}, {Zakamska},
  {G{\"a}nsicke}, {Hermes}, {Schwope}, {Badenes}, {Tovmassian}, {Bauer},
  {Maoz}, {Schreiber}, {Toloza}, {Inight}, {Rix}, \& {Brown}}]{chandra2021}
{Chandra}, V., {Hwang}, H.-C., {Zakamska}, N.~L., {et~al.} 2021, \apj, 921, 160

\bibitem[{{Chen} {et~al.}(2022){Chen}, {Tauris}, {Chen}, \& {Han}}]{chenh2022}
{Chen}, H.-L., {Tauris}, T.~M., {Chen}, X., \& {Han}, Z. 2022, \apj, 925, 89

\bibitem[{{Chen} \& {Han}(2003)}]{chen2003}
{Chen}, X. \& {Han}, Z. 2003, \mnras, 341, 662

\bibitem[{{Chen} \& {Han}(2008)}]{chenx2008}
{Chen}, X. \& {Han}, Z. 2008, \mnras, 387, 1416

\bibitem[{{Chen} {et~al.}(2013){Chen}, {Han}, {Deca}, \&
  {Podsiadlowski}}]{chenx2013}
{Chen}, X., {Han}, Z., {Deca}, J., \& {Podsiadlowski}, P. 2013, \mnras, 434,
  186

\bibitem[{{Chen} {et~al.}(2017){Chen}, {Maxted}, {Li}, \& {Han}}]{chenx2017}
{Chen}, X., {Maxted}, P.~F.~L., {Li}, J., \& {Han}, Z. 2017, \mnras, 467, 1874

\bibitem[{{Chen} {et~al.}(2020){Chen}, {Ivanova}, \&
  {Carroll-Nellenback}}]{chenz2020}
{Chen}, Z., {Ivanova}, N., \& {Carroll-Nellenback}, J. 2020, \apj, 892, 110

\bibitem[{{Cheng} {et~al.}(2020){Cheng}, {Cummings}, {M{\'e}nard}, \&
  {Toonen}}]{chengs2020}
{Cheng}, S., {Cummings}, J.~D., {M{\'e}nard}, B., \& {Toonen}, S. 2020, \apj,
  891, 160

\bibitem[{{Claeys} {et~al.}(2014){Claeys}, {Pols}, {Izzard}, {Vink}, \&
  {Verbunt}}]{claeys2014}
{Claeys}, J.~S.~W., {Pols}, O.~R., {Izzard}, R.~G., {Vink}, J., \& {Verbunt},
  F.~W.~M. 2014, \aap, 563, A83

\bibitem[{{Coughlin} {et~al.}(2020){Coughlin}, {Burdge}, {Phinney}, {van
  Roestel}, {Bellm}, {Dekany}, {Delacroix}, {Duev}, {Feeney}, {Graham},
  {Kulkarni}, {Kupfer}, {Laher}, {Masci}, {Prince}, {Riddle}, {Rosnet},
  {Smith}, {Serabyn}, \& {Walters}}]{coughlin2020}
{Coughlin}, M.~W., {Burdge}, K., {Phinney}, E.~S., {et~al.} 2020, \mnras, 494,
  L91

\bibitem[{{Davis} {et~al.}(2010){Davis}, {Kolb}, \& {Willems}}]{davis2010}
{Davis}, P.~J., {Kolb}, U., \& {Willems}, B. 2010, \mnras, 403, 179

\bibitem[{{De Kool}(1990)}]{dekool1990}
{De Kool}, M. 1990, \apj, 358, 189

\bibitem[{{Debes} {et~al.}(2015){Debes}, {Kilic}, {Tremblay},
  {L{\'o}pez-Morales}, {Anglada-Escude}, {Napiwotzki}, {Osip}, \&
  {Weinberger}}]{debes2015}
{Debes}, J.~H., {Kilic}, M., {Tremblay}, P.-E., {et~al.} 2015, \aj, 149, 176

\bibitem[{{Dewi} \& {Tauris}(2000)}]{dewi2000}
{Dewi}, J.~D.~M. \& {Tauris}, T.~M. 2000, \aap, 360, 1043

\bibitem[{{Eggleton} {et~al.}(1973){Eggleton}, {Faulkner}, \&
  {Flannery}}]{eggleton1973}
{Eggleton}, P.~P., {Faulkner}, J., \& {Flannery}, B.~P. 1973, \aap, 23, 325

\bibitem[{{Eggleton} {et~al.}(1989){Eggleton}, {Fitchett}, \&
  {Tout}}]{eggleton1989}
{Eggleton}, P.~P., {Fitchett}, M.~J., \& {Tout}, C.~A. 1989, \apj, 347, 998

\bibitem[{{El-Badry} \& {Rix}(2018)}]{elbadry2018}
{El-Badry}, K. \& {Rix}, H.-W. 2018, \mnras, 480, 4884

\bibitem[{{El-Badry} {et~al.}(2021){El-Badry}, {Rix}, \&
  {Heintz}}]{elbadry2021}
{El-Badry}, K., {Rix}, H.-W., \& {Heintz}, T.~M. 2021, \mnras, 506, 2269

\bibitem[{{Evans} {et~al.}(1987){Evans}, {Iben}, \& {Smarr}}]{evans1987}
{Evans}, C.~R., {Iben}, Icko, J., \& {Smarr}, L. 1987, \apj, 323, 129

\bibitem[{{Fontaine} {et~al.}(2001){Fontaine}, {Brassard}, \&
  {Bergeron}}]{fontaine2001}
{Fontaine}, G., {Brassard}, P., \& {Bergeron}, P. 2001, \pasp, 113, 409

\bibitem[{{Freudenreich}(1998)}]{freudenreich1998}
{Freudenreich}, H.~T. 1998, \apj, 492, 495

\bibitem[{{Fuller} \& {Lai}(2013)}]{fuller2013}
{Fuller}, J. \& {Lai}, D. 2013, \mnras, 430, 274

\bibitem[{{Gaia Collaboration} {et~al.}(2016){Gaia Collaboration}, {Prusti},
  {de Bruijne}, {Brown}, {Vallenari}, {Babusiaux}, {Bailer-Jones}, {Bastian},
  {Biermann}, {Evans}, {Eyer}, {Jansen}, {Jordi}, {Klioner}, {Lammers},
  {Lindegren}, {Luri}, {Mignard}, {Milligan}, {Panem}, {Poinsignon},
  {Pourbaix}, {Randich}, {Sarri}, {Sartoretti}, {Siddiqui}, {Soubiran},
  {Valette}, {van Leeuwen}, {Walton}, {Aerts}, {Arenou}, {Cropper}, {Drimmel},
  {H{\o}g}, {Katz}, {Lattanzi}, {O'Mullane}, {Grebel}, {Holland}, {Huc},
  {Passot}, {Bramante}, {Cacciari}, {Casta{\~n}eda}, {Chaoul}, {Cheek}, {De
  Angeli}, {Fabricius}, {Guerra}, {Hern{\'a}ndez}, {Jean-Antoine-Piccolo},
  {Masana}, {Messineo}, {Mowlavi}, {Nienartowicz}, {Ord{\'o}{\~n}ez-Blanco},
  {Panuzzo}, {Portell}, {Richards}, {Riello}, {Seabroke}, {Tanga},
  {Th{\'e}venin}, {Torra}, {Els}, {Gracia-Abril}, {Comoretto},
  {Garcia-Reinaldos}, {Lock}, {Mercier}, {Altmann}, {Andrae}, {Astraatmadja},
  {Bellas-Velidis}, {Benson}, {Berthier}, {Blomme}, {Busso}, {Carry},
  {Cellino}, {Clementini}, {Cowell}, {Creevey}, {Cuypers}, {Davidson}, {De
  Ridder}, {de Torres}, {Delchambre}, {Dell'Oro}, {Ducourant}, {Fr{\'e}mat},
  {Garc{\'\i}a-Torres}, {Gosset}, {Halbwachs}, {Hambly}, {Harrison}, {Hauser},
  {Hestroffer}, {Hodgkin}, {Huckle}, {Hutton}, {Jasniewicz}, {Jordan},
  {Kontizas}, {Korn}, {Lanzafame}, {Manteiga}, {Moitinho}, {Muinonen},
  {Osinde}, {Pancino}, {Pauwels}, {Petit}, {Recio-Blanco}, {Robin}, {Sarro},
  {Siopis}, {Smith}, {Smith}, {Sozzetti}, {Thuillot}, {van Reeven}, {Viala},
  {Abbas}, {Abreu Aramburu}, {Accart}, {Aguado}, {Allan}, {Allasia},
  {Altavilla}, {{\'A}lvarez}, {Alves}, {Anderson}, {Andrei}, {Anglada Varela},
  {Antiche}, {Antoja}, {Ant{\'o}n}, {Arcay}, {Atzei}, {Ayache}, {Bach},
  {Baker}, {Balaguer-N{\'u}{\~n}ez}, {Barache}, {Barata}, {Barbier}, {Barblan},
  {Baroni}, {Barrado y Navascu{\'e}s}, {Barros}, {Barstow}, {Becciani},
  {Bellazzini}, {Bellei}, {Bello Garc{\'\i}a}, {Belokurov}, {Bendjoya},
  {Berihuete}, {Bianchi}, {Bienaym{\'e}}, {Billebaud}, {Blagorodnova},
  {Blanco-Cuaresma}, {Boch}, {Bombrun}, {Borrachero}, {Bouquillon}, {Bourda},
  {Bouy}, {Bragaglia}, {Breddels}, {Brouillet}, {Br{\"u}semeister},
  {Bucciarelli}, {Budnik}, {Burgess}, {Burgon}, {Burlacu}, {Busonero}, {Buzzi},
  {Caffau}, {Cambras}, {Campbell}, {Cancelliere}, {Cantat-Gaudin}, {Carlucci},
  {Carrasco}, {Castellani}, {Charlot}, {Charnas}, {Charvet}, {Chassat},
  {Chiavassa}, {Clotet}, {Cocozza}, {Collins}, {Collins}, {Costigan}, {Crifo},
  {Cross}, {Crosta}, {Crowley}, {Dafonte}, {Damerdji}, {Dapergolas}, {David},
  {David}, {De Cat}, {de Felice}, {de Laverny}, {De Luise}, {De March}, {de
  Martino}, {de Souza}, {Debosscher}, {del Pozo}, {Delbo}, {Delgado},
  {Delgado}, {di Marco}, {Di Matteo}, {Diakite}, {Distefano}, {Dolding}, {Dos
  Anjos}, {Drazinos}, {Dur{\'a}n}, {Dzigan}, {Ecale}, {Edvardsson}, {Enke},
  {Erdmann}, {Escolar}, {Espina}, {Evans}, {Eynard Bontemps}, {Fabre},
  {Fabrizio}, {Faigler}, {Falc{\~a}o}, {Farr{\`a}s Casas}, {Faye}, {Federici},
  {Fedorets}, {Fern{\'a}ndez-Hern{\'a}ndez}, {Fernique}, {Fienga}, {Figueras},
  {Filippi}, {Findeisen}, {Fonti}, {Fouesneau}, {Fraile}, {Fraser}, {Fuchs},
  {Furnell}, {Gai}, {Galleti}, {Galluccio}, {Garabato}, {Garc{\'\i}a-Sedano},
  {Gar{\'e}}, {Garofalo}, {Garralda}, {Gavras}, {Gerssen}, {Geyer}, {Gilmore},
  {Girona}, {Giuffrida}, {Gomes}, {Gonz{\'a}lez-Marcos},
  {Gonz{\'a}lez-N{\'u}{\~n}ez}, {Gonz{\'a}lez-Vidal}, {Granvik}, {Guerrier},
  {Guillout}, {Guiraud}, {G{\'u}rpide}, {Guti{\'e}rrez-S{\'a}nchez}, {Guy},
  {Haigron}, {Hatzidimitriou}, {Haywood}, {Heiter}, {Helmi}, {Hobbs},
  {Hofmann}, {Holl}, {Holland }, {Hunt}, {Hypki}, {Icardi}, {Irwin}, {Jevardat
  de Fombelle}, {Jofr{\'e}}, {Jonker}, {Jorissen}, {Julbe}, {Karampelas},
  {Kochoska}, {Kohley}, {Kolenberg}, {Kontizas}, {Koposov}, {Kordopatis},
  {Koubsky}, {Kowalczyk}, {Krone-Martins}, {Kudryashova}, {Kull}, {Bachchan},
  {Lacoste-Seris}, {Lanza}, {Lavigne}, {Le Poncin-Lafitte}, {Lebreton},
  {Lebzelter}, {Leccia}, {Leclerc}, {Lecoeur-Taibi}, {Lemaitre}, {Lenhardt},
  {Leroux}, {Liao}, {Licata}, {Lindstr{\o}m}, {Lister}, {Livanou}, {Lobel},
  {L{\"o}ffler}, {L{\'o}pez}, {Lopez-Lozano}, {Lorenz}, {Loureiro},
  {MacDonald}, {Magalh{\~a}es Fernandes}, {Managau}, {Mann}, {Mantelet},
  {Marchal}, {Marchant}, {Marconi}, {Marie}, {Marinoni}, {Marrese},
  {Marschalk{\'o}}, {Marshall}, {Mart{\'\i}n-Fleitas}, {Martino}, {Mary},
  {Matijevi{\v{c}}}, {Mazeh}, {McMillan}, {Messina}, {Mestre}, {Michalik},
  {Millar}, {Miranda}, {Molina}, {Molinaro}, {Molinaro}, {Moln{\'a}r},
  {Moniez}, {Montegriffo}, {Monteiro}, {Mor}, {Mora}, {Morbidelli}, {Morel},
  {Morgenthaler}, {Morley}, {Morris}, {Mulone}, {Muraveva}, {Musella},
  {Narbonne}, {Nelemans}, {Nicastro}, {Noval}, {Ord{\'e}novic},
  {Ordieres-Mer{\'e}}, {Osborne}, {Pagani}, {Pagano}, {Pailler}, {Palacin},
  {Palaversa}, {Parsons}, {Paulsen}, {Pecoraro}, {Pedrosa}, {Pentik{\"a}inen},
  {Pereira}, {Pichon}, {Piersimoni}, {Pineau}, {Plachy}, {Plum}, {Poujoulet},
  {Pr{\v{s}}a}, {Pulone}, {Ragaini}, {Rago}, {Rambaux}, {Ramos-Lerate},
  {Ranalli}, {Rauw}, {Read}, {Regibo}, {Renk}, {Reyl{\'e}}, {Ribeiro},
  {Rimoldini}, {Ripepi}, {Riva}, {Rixon}, {Roelens}, {Romero-G{\'o}mez},
  {Rowell}, {Royer}, {Rudolph}, {Ruiz-Dern}, {Sadowski}, {Sagrist{\`a}
  Sell{\'e}s}, {Sahlmann}, {Salgado}, {Salguero}, {Sarasso}, {Savietto},
  {Schnorhk}, {Schultheis}, {Sciacca}, {Segol}, {Segovia}, {Segransan},
  {Serpell}, {Shih}, {Smareglia}, {Smart}, {Smith}, {Solano}, {Solitro},
  {Sordo}, {Soria Nieto}, {Souchay}, {Spagna}, {Spoto}, {Stampa}, {Steele},
  {Steidelm{\"u}ller}, {Stephenson}, {Stoev}, {Suess}, {S{\"u}veges}, {Surdej},
  {Szabados}, {Szegedi-Elek}, {Tapiador}, {Taris}, {Tauran}, {Taylor},
  {Teixeira}, {Terrett}, {Tingley}, {Trager}, {Turon}, {Ulla}, {Utrilla},
  {Valentini}, {van Elteren}, {Van Hemelryck}, {van Leeuwen}, {Varadi},
  {Vecchiato}, {Veljanoski}, {Via}, {Vicente}, {Vogt}, {Voss}, {Votruba},
  {Voutsinas}, {Walmsley}, {Weiler}, {Weingrill}, {Werner}, {Wevers},
  {Whitehead}, {Wyrzykowski}, {Yoldas}, {{\v{Z}}erjal}, {Zucker}, {Zurbach},
  {Zwitter}, {Alecu}, {Allen}, {Allende Prieto}, {Amorim},
  {Anglada-Escud{\'e}}, {Arsenijevic}, {Azaz}, {Balm}, {Beck}, {Bernstein},
  {Bigot}, {Bijaoui}, {Blasco}, {Bonfigli}, {Bono}, {Boudreault}, {Bressan},
  {Brown}, {Brunet}, {Bunclark}, {Buonanno}, {Butkevich}, {Carret}, {Carrion},
  {Chemin}, {Ch{\'e}reau}, {Corcione}, {Darmigny}, {de Boer}, {de Teodoro}, {de
  Zeeuw}, {Delle Luche}, {Domingues}, {Dubath}, {Fodor}, {Fr{\'e}zouls},
  {Fries}, {Fustes}, {Fyfe}, {Gallardo}, {Gallegos}, {Gardiol}, {Gebran},
  {Gomboc}, {G{\'o}mez}, {Grux}, {Gueguen}, {Heyrovsky}, {Hoar}, {Iannicola},
  {Isasi Parache}, {Janotto}, {Joliet}, {Jonckheere}, {Keil}, {Kim},
  {Klagyivik}, {Klar}, {Knude}, {Kochukhov}, {Kolka}, {Kos}, {Kutka}, {Lainey},
  {LeBouquin}, {Liu}, {Loreggia}, {Makarov}, {Marseille}, {Martayan},
  {Martinez-Rubi}, {Massart}, {Meynadier}, {Mignot}, {Munari}, {Nguyen},
  {Nordlander}, {Ocvirk}, {O'Flaherty}, {Olias Sanz}, {Ortiz}, {Osorio},
  {Oszkiewicz}, {Ouzounis}, {Palmer}, {Park}, {Pasquato}, {Peltzer}, {Peralta},
  {P{\'e}turaud}, {Pieniluoma}, {Pigozzi}, {Poels}, {Prat}, {Prod'homme},
  {Raison}, {Rebordao}, {Risquez}, {Rocca-Volmerange}, {Rosen}, {Ruiz-Fuertes},
  {Russo}, {Sembay}, {Serraller Vizcaino}, {Short}, {Siebert}, {Silva},
  {Sinachopoulos}, {Slezak}, {Soffel}, {Sosnowska}, {Strai{\v{z}}ys}, {ter
  Linden}, {Terrell}, {Theil}, {Tiede}, {Troisi}, {Tsalmantza}, {Tur},
  {Vaccari}, {Vachier}, {Valles}, {Van Hamme}, {Veltz}, {Virtanen}, {Wallut},
  {Wichmann}, {Wilkinson}, {Ziaeepour}, \& {Zschocke}}]{GAIA2016}
{Gaia Collaboration}, {Prusti}, T., {de Bruijne}, J.~H.~J., {et~al.} 2016,
  \aap, 595, A1

\bibitem[{{Ge} {et~al.}(2010){Ge}, {Hjellming}, {Webbink}, {Chen}, \&
  {Han}}]{geh2010}
{Ge}, H., {Hjellming}, M.~S., {Webbink}, R.~F., {Chen}, X., \& {Han}, Z. 2010,
  \apj, 717, 724

\bibitem[{{Ge} {et~al.}(2015){Ge}, {Webbink}, {Chen}, \& {Han}}]{geh2015}
{Ge}, H., {Webbink}, R.~F., {Chen}, X., \& {Han}, Z. 2015, \apj, 812, 40

\bibitem[{{Ge} {et~al.}(2020){Ge}, {Webbink}, {Chen}, \& {Han}}]{geh2020}
{Ge}, H., {Webbink}, R.~F., {Chen}, X., \& {Han}, Z. 2020, \apj, 899, 132

\bibitem[{{Georgousi} {et~al.}(2022){Georgousi}, {Karnesis}, {Korol},
  {Pieroni}, \& {Stergioulas}}]{georgousi2022}
{Georgousi}, M., {Karnesis}, N., {Korol}, V., {Pieroni}, M., \& {Stergioulas},
  N. 2022, arXiv e-prints, arXiv:2204.07349

\bibitem[{{Gianninas} {et~al.}(2014){Gianninas}, {Dufour}, {Kilic}, {Brown},
  {Bergeron}, \& {Hermes}}]{gianninas2014}
{Gianninas}, A., {Dufour}, P., {Kilic}, M., {et~al.} 2014, \apj, 794, 35

\bibitem[{{Gianninas} {et~al.}(2015){Gianninas}, {Kilic}, {Brown}, {Canton}, \&
  {Kenyon}}]{gianninas2015}
{Gianninas}, A., {Kilic}, M., {Brown}, W.~R., {Canton}, P., \& {Kenyon}, S.~J.
  2015, \apj, 812, 167

\bibitem[{{Graur} \& {Maoz}(2013)}]{graur2013}
{Graur}, O. \& {Maoz}, D. 2013, \mnras, 430, 1746

\bibitem[{{Hachisu} {et~al.}(1999){Hachisu}, {Kato}, \& {Nomoto}}]{hachisu1999}
{Hachisu}, I., {Kato}, M., \& {Nomoto}, K. 1999, \apj, 522, 487

\bibitem[{{Hallakoun} {et~al.}(2016){Hallakoun}, {Maoz}, {Kilic}, {Mazeh},
  {Gianninas}, {Agol}, {Bell}, {Bloemen}, {Brown}, {Debes}, {Faigler}, {Kull},
  {Kupfer}, {Loeb}, {Morris}, \& {Mullally}}]{hallakoun2016}
{Hallakoun}, N., {Maoz}, D., {Kilic}, M., {et~al.} 2016, \mnras, 458, 845

\bibitem[{{Hamann} \& {Koesterke}(1998)}]{hamann1998}
{Hamann}, W.~R. \& {Koesterke}, L. 1998, \aap, 335, 1003

\bibitem[{{Hamann} {et~al.}(1995){Hamann}, {Koesterke}, \&
  {Wessolowski}}]{hamann1995}
{Hamann}, W.~R., {Koesterke}, L., \& {Wessolowski}, U. 1995, \aap, 299, 151

\bibitem[{{Han}(1998)}]{han1998}
{Han}, Z. 1998, \mnras, 296, 1019

\bibitem[{{Han} {et~al.}(1995){Han}, {Podsiadlowski}, \& {Eggleton}}]{han1995}
{Han}, Z., {Podsiadlowski}, P., \& {Eggleton}, P.~P. 1995, \mnras, 272, 800

\bibitem[{{Han} {et~al.}(2020){Han}, {Ge}, {Chen}, \& {Chen}}]{han2020}
{Han}, Z.-W., {Ge}, H.-W., {Chen}, X.-F., \& {Chen}, H.-L. 2020, Research in
  Astronomy and Astrophysics, 20, 161

\bibitem[{{Hermes} {et~al.}(2014){Hermes}, {Brown}, {Kilic}, {Gianninas},
  {Chote}, {Sullivan}, {Winget}, {Bell}, {Falcon}, {Winget}, {Mason},
  {Harrold}, \& {Montgomery}}]{hermes2014}
{Hermes}, J.~J., {Brown}, W.~R., {Kilic}, M., {et~al.} 2014, \apj, 792, 39

\bibitem[{{Hermes} {et~al.}(2012){Hermes}, {Kilic}, {Brown}, {Winget}, {Allende
  Prieto}, {Gianninas}, {Mukadam}, {Cabrera-Lavers}, \& {Kenyon}}]{hermes2012}
{Hermes}, J.~J., {Kilic}, M., {Brown}, W.~R., {et~al.} 2012, \apjl, 757, L21

\bibitem[{{Hermes} {et~al.}(2013){Hermes}, {Montgomery}, {Gianninas}, {Winget},
  {Brown}, {Harrold}, {Bell}, {Kenyon}, {Kilic}, \& {Castanheira}}]{hermes2013}
{Hermes}, J.~J., {Montgomery}, M.~H., {Gianninas}, A., {et~al.} 2013, \mnras,
  436, 3573

\bibitem[{{Hjellming} \& {Webbink}(1987)}]{hjellming1987}
{Hjellming}, M.~S. \& {Webbink}, R.~F. 1987, \apj, 318, 794

\bibitem[{{Holberg} {et~al.}(2016){Holberg}, {Oswalt}, {Sion}, \&
  {McCook}}]{holberg2016}
{Holberg}, J.~B., {Oswalt}, T.~D., {Sion}, E.~M., \& {McCook}, G.~P. 2016,
  \mnras, 462, 2295

\bibitem[{{Huang} {et~al.}(2020){Huang}, {Hu}, {Korol}, {Li}, {Liang}, {Lu},
  {Wang}, {Yu}, \& {Mei}}]{huangs2020}
{Huang}, S.-J., {Hu}, Y.-M., {Korol}, V., {et~al.} 2020, \prd, 102, 063021

\bibitem[{{Hurley} {et~al.}(2000){Hurley}, {Pols}, \& {Tout}}]{hurley2000}
{Hurley}, J.~R., {Pols}, O.~R., \& {Tout}, C.~A. 2000, \mnras, 315, 543

\bibitem[{{Hurley} {et~al.}(2002){Hurley}, {Tout}, \& {Pols}}]{hurley2002}
{Hurley}, J.~R., {Tout}, C.~A., \& {Pols}, O.~R. 2002, \mnras, 329, 897

\bibitem[{{Hut}(1981)}]{hut1981}
{Hut}, P. 1981, \aap, 99, 126

\bibitem[{{Iben} \& {Renzini}(1983)}]{iben1983}
{Iben}, I., J. \& {Renzini}, A. 1983, \araa, 21, 271

\bibitem[{{Iben} \& {Tutukov}(1984)}]{iben1984}
{Iben}, I., J. \& {Tutukov}, A.~V. 1984, \apjs, 54, 335

\bibitem[{{Iben} {et~al.}(1997){Iben}, {Tutukov}, \& {Yungelson}}]{iben1997}
{Iben}, Icko, J., {Tutukov}, A.~V., \& {Yungelson}, L.~R. 1997, \apj, 475, 291

\bibitem[{{Inight} {et~al.}(2021){Inight}, {G{\"a}nsicke}, {Breedt}, {Marsh},
  {Pala}, \& {Raddi}}]{inight2021}
{Inight}, K., {G{\"a}nsicke}, B.~T., {Breedt}, E., {et~al.} 2021, \mnras, 504,
  2420

\bibitem[{{Istrate} {et~al.}(2014){Istrate}, {Tauris}, {Langer}, \&
  {Antoniadis}}]{istrate2014b}
{Istrate}, A.~G., {Tauris}, T.~M., {Langer}, N., \& {Antoniadis}, J. 2014,
  \aap, 571, L3

\bibitem[{{Karl} {et~al.}(2003){Karl}, {Napiwotzki}, {Nelemans}, {Christlieb},
  {Koester}, {Heber}, \& {Reimers}}]{karl2003}
{Karl}, C.~A., {Napiwotzki}, R., {Nelemans}, G., {et~al.} 2003, \aap, 410, 663

\bibitem[{{Kato} \& {Hachisu}(2004)}]{kato2004}
{Kato}, M. \& {Hachisu}, I. 2004, \apjl, 613, L129

\bibitem[{{Keller} {et~al.}(2022){Keller}, {Breedt}, {Hodgkin}, {Belokurov},
  {Wild}, {Garc{\'\i}a-Soriano}, \& {Wise}}]{keller2022}
{Keller}, P.~M., {Breedt}, E., {Hodgkin}, S., {et~al.} 2022, \mnras, 509, 4171

\bibitem[{{Kilic} {et~al.}(2021){Kilic}, {B{\'e}dard}, \&
  {Bergeron}}]{kilic2021}
{Kilic}, M., {B{\'e}dard}, A., \& {Bergeron}, P. 2021, \mnras, 502, 4972

\bibitem[{{Kilic} {et~al.}(2011{\natexlab{a}}){Kilic}, {Brown}, {Allende
  Prieto}, {Ag{\"u}eros}, {Heinke}, \& {Kenyon}}]{kilic2011a}
{Kilic}, M., {Brown}, W.~R., {Allende Prieto}, C., {et~al.} 2011{\natexlab{a}},
  \apj, 727, 3

\bibitem[{{Kilic} {et~al.}(2012){Kilic}, {Brown}, {Allende Prieto}, {Kenyon},
  {Heinke}, {Ag{\"u}eros}, \& {Kleinman}}]{kilic2012}
{Kilic}, M., {Brown}, W.~R., {Allende Prieto}, C., {et~al.} 2012, \apj, 751,
  141

\bibitem[{{Kilic} {et~al.}(2010){Kilic}, {Brown}, {Allende Prieto}, {Kenyon},
  \& {Panei}}]{kilic2010b}
{Kilic}, M., {Brown}, W.~R., {Allende Prieto}, C., {Kenyon}, S.~J., \& {Panei},
  J.~A. 2010, \apj, 716, 122

\bibitem[{{Kilic} {et~al.}(2009){Kilic}, {Brown}, {Allende Prieto}, {Swift},
  {Kenyon}, {Liebert}, \& {Ag{\"u}eros}}]{kilic2009}
{Kilic}, M., {Brown}, W.~R., {Allende Prieto}, C., {et~al.} 2009, \apjl, 695,
  L92

\bibitem[{{Kilic} {et~al.}(2014){Kilic}, {Brown}, {Gianninas}, {Hermes},
  {Allende Prieto}, \& {Kenyon}}]{kilic2014}
{Kilic}, M., {Brown}, W.~R., {Gianninas}, A., {et~al.} 2014, \mnras, 444, L1

\bibitem[{{Kilic} {et~al.}(2011{\natexlab{b}}){Kilic}, {Brown}, {Hermes},
  {Allende Prieto}, {Kenyon}, {Winget}, \& {Winget}}]{kilic2011b}
{Kilic}, M., {Brown}, W.~R., {Hermes}, J.~J., {et~al.} 2011{\natexlab{b}},
  \mnras, 418, L157

\bibitem[{{Kilic} {et~al.}(2011{\natexlab{c}}){Kilic}, {Brown}, {Kenyon},
  {Allende Prieto}, {Andrews}, {Kleinman}, {Winget}, {Winget}, \&
  {Hermes}}]{kilic2011c}
{Kilic}, M., {Brown}, W.~R., {Kenyon}, S.~J., {et~al.} 2011{\natexlab{c}},
  \mnras, 413, L101

\bibitem[{{Kilic} {et~al.}(2007){Kilic}, {Stanek}, \&
  {Pinsonneault}}]{kilic2007}
{Kilic}, M., {Stanek}, K.~Z., \& {Pinsonneault}, M.~H. 2007, \apj, 671, 761

\bibitem[{{Knapp} {et~al.}(1998){Knapp}, {Young}, {Lee}, \&
  {Jorissen}}]{knapp1998}
{Knapp}, G.~R., {Young}, K., {Lee}, E., \& {Jorissen}, A. 1998, \apjs, 117, 209

\bibitem[{{Korol} {et~al.}(2022{\natexlab{a}}){Korol}, {Belokurov}, \&
  {Toonen}}]{korol2022b}
{Korol}, V., {Belokurov}, V., \& {Toonen}, S. 2022{\natexlab{a}}, \mnras, 515,
  1228

\bibitem[{{Korol} {et~al.}(2022{\natexlab{b}}){Korol}, {Hallakoun}, {Toonen},
  \& {Karnesis}}]{korol2022a}
{Korol}, V., {Hallakoun}, N., {Toonen}, S., \& {Karnesis}, N.
  2022{\natexlab{b}}, \mnras, 511, 5936

\bibitem[{{Korol} {et~al.}(2019){Korol}, {Rossi}, \& {Barausse}}]{korol2019}
{Korol}, V., {Rossi}, E.~M., \& {Barausse}, E. 2019, \mnras, 483, 5518

\bibitem[{{Korol} {et~al.}(2017){Korol}, {Rossi}, {Groot}, {Nelemans},
  {Toonen}, \& {Brown}}]{korol2017}
{Korol}, V., {Rossi}, E.~M., {Groot}, P.~J., {et~al.} 2017, \mnras, 470, 1894

\bibitem[{{Kosakowski} {et~al.}(2021){Kosakowski}, {Kilic}, \&
  {Brown}}]{kosakowski2021}
{Kosakowski}, A., {Kilic}, M., \& {Brown}, W. 2021, \mnras, 500, 5098

\bibitem[{{Kosakowski} {et~al.}(2020){Kosakowski}, {Kilic}, {Brown}, \&
  {Gianninas}}]{kosakowski2020}
{Kosakowski}, A., {Kilic}, M., {Brown}, W.~R., \& {Gianninas}, A. 2020, \apj,
  894, 53

\bibitem[{{Kruckow} {et~al.}(2021){Kruckow}, {Neunteufel}, {Di Stefano}, {Gao},
  \& {Kobayashi}}]{kruckow2021}
{Kruckow}, M.~U., {Neunteufel}, P.~G., {Di Stefano}, R., {Gao}, Y., \&
  {Kobayashi}, C. 2021, \apj, 920, 86

\bibitem[{{Kupfer} {et~al.}(2018){Kupfer}, {Korol}, {Shah}, {Nelemans},
  {Marsh}, {Ramsay}, {Groot}, {Steeghs}, \& {Rossi}}]{kupfer2018}
{Kupfer}, T., {Korol}, V., {Shah}, S., {et~al.} 2018, \mnras, 480, 302

\bibitem[{{Lamberts} {et~al.}(2019){Lamberts}, {Blunt}, {Littenberg},
  {Garrison-Kimmel}, {Kupfer}, \& {Sanderson}}]{lamberts2019}
{Lamberts}, A., {Blunt}, S., {Littenberg}, T., {et~al.} 2019, arXiv e-prints,
  arXiv:1907.00014

\bibitem[{{Landau} \& {Lifshitz}(1975)}]{landau1975}
{Landau}, L.~D. \& {Lifshitz}, E.~M. 1975, {The classical theory of fields}
  (New York: Pergamon Press,~Oxford)

\bibitem[{{Lauffer} {et~al.}(2018){Lauffer}, {Romero}, \&
  {Kepler}}]{lauffer2018}
{Lauffer}, G.~R., {Romero}, A.~D., \& {Kepler}, S.~O. 2018, \mnras, 480, 1547

\bibitem[{{Li} {et~al.}(2011){Li}, {Chornock}, {Leaman}, {Filippenko},
  {Poznanski}, {Wang}, {Ganeshalingam}, \& {Mannucci}}]{liw2011}
{Li}, W., {Chornock}, R., {Leaman}, J., {et~al.} 2011, \mnras, 412, 1473

\bibitem[{{Li} {et~al.}(2019){Li}, {Chen}, {Chen}, \& {Han}}]{lizw2019}
{Li}, Z., {Chen}, X., {Chen}, H.-L., \& {Han}, Z. 2019, \apj, 871, 148

\bibitem[{{Li} {et~al.}(2020){Li}, {Chen}, {Chen}, {Li}, {Yu}, \&
  {Han}}]{lizw2020}
{Li}, Z., {Chen}, X., {Chen}, H.-L., {et~al.} 2020, \apj, 893, 2

\bibitem[{{Liu} {et~al.}(2018){Liu}, {Wang}, \& {Han}}]{liud2018}
{Liu}, D., {Wang}, B., \& {Han}, Z. 2018, \mnras, 473, 5352

\bibitem[{{Liu}(2009)}]{liu2009}
{Liu}, J. 2009, \mnras, 400, 1850

\bibitem[{{Liu} {et~al.}(2010){Liu}, {Han}, {Zhang}, \& {Zhang}}]{liu2010}
{Liu}, J., {Han}, Z., {Zhang}, F., \& {Zhang}, Y. 2010, \apj, 719, 1546

\bibitem[{{Liu} {et~al.}(2017){Liu}, {Stancliffe}, {Abate}, \&
  {Matrozis}}]{liuz2017}
{Liu}, Z.-W., {Stancliffe}, R.~J., {Abate}, C., \& {Matrozis}, E. 2017, \apj,
  846, 117

\bibitem[{{Livio} \& {Soker}(1988)}]{livio1988}
{Livio}, M. \& {Soker}, N. 1988, \apj, 329, 764

\bibitem[{{Luo} {et~al.}(2016){Luo}, {Chen}, {Duan}, {Gong}, {Hu}, {Ji}, {Liu},
  {Mei}, {Milyukov}, {Sazhin}, {Shao}, {Toth}, {Tu}, {Wang}, {Wang}, {Yeh},
  {Zhan}, {Zhang}, {Zharov}, \& {Zhou}}]{luo2015}
{Luo}, J., {Chen}, L.-S., {Duan}, H.-Z., {et~al.} 2016, Classical and Quantum
  Gravity, 33, 035010

\bibitem[{{Maoz} {et~al.}(2018){Maoz}, {Hallakoun}, \& {Badenes}}]{maoz2018}
{Maoz}, D., {Hallakoun}, N., \& {Badenes}, C. 2018, \mnras, 476, 2584

\bibitem[{{Maoz} {et~al.}(2012){Maoz}, {Mannucci}, \& {Brandt}}]{maoz2012}
{Maoz}, D., {Mannucci}, F., \& {Brandt}, T.~D. 2012, \mnras, 426, 3282

\bibitem[{{Maoz} {et~al.}(2014){Maoz}, {Mannucci}, \& {Nelemans}}]{maoz2014}
{Maoz}, D., {Mannucci}, F., \& {Nelemans}, G. 2014, \araa, 52, 107

\bibitem[{{Maoz} {et~al.}(2010){Maoz}, {Sharon}, \& {Gal-Yam}}]{maoz2010}
{Maoz}, D., {Sharon}, K., \& {Gal-Yam}, A. 2010, \apj, 722, 1879

\bibitem[{{Marsh}(2000)}]{marsh2000b}
{Marsh}, T.~R. 2000, \nar, 44, 119

\bibitem[{{Marsh} {et~al.}(1995){Marsh}, {Dhillon}, \& {Duck}}]{marsh1995}
{Marsh}, T.~R., {Dhillon}, V.~S., \& {Duck}, S.~R. 1995, \mnras, 275, 828

\bibitem[{{Marsh} {et~al.}(2011){Marsh}, {G{\"a}nsicke}, {Steeghs},
  {Southworth}, {Koester}, {Harris}, \& {Merry}}]{marsh2011}
{Marsh}, T.~R., {G{\"a}nsicke}, B.~T., {Steeghs}, D., {et~al.} 2011, \apj, 736,
  95

\bibitem[{{Maxted} {et~al.}(2002{\natexlab{a}}){Maxted}, {Burleigh}, {Marsh},
  \& {Bannister}}]{maxted2002a}
{Maxted}, P.~F.~L., {Burleigh}, M.~R., {Marsh}, T.~R., \& {Bannister}, N.~P.
  2002{\natexlab{a}}, \mnras, 334, 833

\bibitem[{{Maxted} \& {Marsh}(1999)}]{maxted1999}
{Maxted}, P.~F.~L. \& {Marsh}, T.~R. 1999, \mnras, 307, 122

\bibitem[{{Maxted} {et~al.}(2002{\natexlab{b}}){Maxted}, {Marsh}, \&
  {Moran}}]{maxted2002c}
{Maxted}, P.~F.~L., {Marsh}, T.~R., \& {Moran}, C.~K.~J. 2002{\natexlab{b}},
  \mnras, 332, 745

\bibitem[{{Maxted} {et~al.}(2000){Maxted}, {Marsh}, {Moran}, \&
  {Han}}]{maxted2000a}
{Maxted}, P.~F.~L., {Marsh}, T.~R., {Moran}, C.~K.~J., \& {Han}, Z. 2000,
  \mnras, 314, 334

\bibitem[{{Mazeh} {et~al.}(1992){Mazeh}, {Goldberg}, {Duquennoy}, \&
  {Mayor}}]{mazeh1992}
{Mazeh}, T., {Goldberg}, D., {Duquennoy}, A., \& {Mayor}, M. 1992, \apj, 401,
  265

\bibitem[{{Miller} \& {Scalo}(1979)}]{miller1979}
{Miller}, G.~E. \& {Scalo}, J.~M. 1979, \apjs, 41, 513

\bibitem[{{Morales-Rueda} {et~al.}(2005){Morales-Rueda}, {Marsh}, {Maxted},
  {Nelemans}, {Karl}, {Napiwotzki}, \& {Moran}}]{morales2005}
{Morales-Rueda}, L., {Marsh}, T.~R., {Maxted}, P.~F.~L., {et~al.} 2005, \mnras,
  359, 648

\bibitem[{{Moran} {et~al.}(1997){Moran}, {Marsh}, \& {Bragaglia}}]{moran1997}
{Moran}, C., {Marsh}, T.~R., \& {Bragaglia}, A. 1997, \mnras, 288, 538

\bibitem[{{Mullally} {et~al.}(2009){Mullally}, {Badenes}, {Thompson}, \&
  {Lupton}}]{mullally2009}
{Mullally}, F., {Badenes}, C., {Thompson}, S.~E., \& {Lupton}, R. 2009, \apjl,
  707, L51

\bibitem[{{Napiwotzki} {et~al.}(2020){Napiwotzki}, {Karl}, {Lisker},
  {Catal{\'a}n}, {Drechsel}, {Heber}, {Homeier}, {Koester}, {Leibundgut},
  {Marsh}, {Moehler}, {Nelemans}, {Reimers}, {Renzini}, {Str{\"o}er}, \&
  {Yungelson}}]{napiwotzki2020}
{Napiwotzki}, R., {Karl}, C.~A., {Lisker}, T., {et~al.} 2020, \aap, 638, A131

\bibitem[{{Napiwotzki} {et~al.}(2007){Napiwotzki}, {Karl}, {Nelemans},
  {Yungelson}, {Christlieb}, {Drechsel}, {Heber}, {Homeier}, {Koester},
  {Leibundgut}, {Marsh}, {Moehler}, {Renzini}, \& {Reimers}}]{napiwotzki2007}
{Napiwotzki}, R., {Karl}, C.~A., {Nelemans}, G., {et~al.} 2007, in Astronomical
  Society of the Pacific Conference Series, Vol. 372, 15th European Workshop on
  White Dwarfs, ed. R.~{Napiwotzki} \& M.~R. {Burleigh}, 387

\bibitem[{{Napiwotzki} {et~al.}(2002){Napiwotzki}, {Koester}, {Nelemans},
  {Yungelson}, {Christlieb}, {Renzini}, {Reimers}, {Drechsel}, \&
  {Leibundgut}}]{napiwotzki2002}
{Napiwotzki}, R., {Koester}, D., {Nelemans}, G., {et~al.} 2002, \aap, 386, 957

\bibitem[{{Nelemans} {et~al.}(2005){Nelemans}, {Napiwotzki}, {Karl}, {Marsh},
  {Voss}, {Roelofs}, {Izzard}, {Montgomery}, {Reerink}, {Christlieb}, \&
  {Reimers}}]{nelemans2005b}
{Nelemans}, G., {Napiwotzki}, R., {Karl}, C., {et~al.} 2005, \aap, 440, 1087

\bibitem[{{Nelemans} {et~al.}(2001{\natexlab{a}}){Nelemans}, {Portegies Zwart},
  {Verbunt}, \& {Yungelson}}]{nelemans2001b}
{Nelemans}, G., {Portegies Zwart}, S.~F., {Verbunt}, F., \& {Yungelson}, L.~R.
  2001{\natexlab{a}}, \aap, 368, 939

\bibitem[{{Nelemans} \& {Tout}(2005)}]{nelemans2005}
{Nelemans}, G. \& {Tout}, C.~A. 2005, \mnras, 356, 753

\bibitem[{{Nelemans} {et~al.}(2000){Nelemans}, {Verbunt}, {Yungelson}, \&
  {Portegies Zwart}}]{nelemans2000}
{Nelemans}, G., {Verbunt}, F., {Yungelson}, L.~R., \& {Portegies Zwart}, S.~F.
  2000, \aap, 360, 1011

\bibitem[{{Nelemans} {et~al.}(2001{\natexlab{b}}){Nelemans}, {Yungelson},
  {Portegies Zwart}, \& {Verbunt}}]{nelemans2001a}
{Nelemans}, G., {Yungelson}, L.~R., {Portegies Zwart}, S.~F., \& {Verbunt}, F.
  2001{\natexlab{b}}, \aap, 365, 491

\bibitem[{{Nelson} {et~al.}(2004){Nelson}, {Dubeau}, \&
  {MacCannell}}]{nelson2004}
{Nelson}, L.~A., {Dubeau}, E., \& {MacCannell}, K.~A. 2004, \apj, 616, 1124

\bibitem[{{Nissanke} {et~al.}(2012){Nissanke}, {Vallisneri}, {Nelemans}, \&
  {Prince}}]{nissanke2012}
{Nissanke}, S., {Vallisneri}, M., {Nelemans}, G., \& {Prince}, T.~A. 2012,
  \apj, 758, 131

\bibitem[{{Nomoto}(1982)}]{nomoto1982}
{Nomoto}, K. 1982, \apj, 253, 798

\bibitem[{{Paczynski}(1976)}]{paczynski1976}
{Paczynski}, B. 1976, in Structure and Evolution of Close Binary Systems, ed.
  P.~{Eggleton}, S.~{Mitton}, \& J.~{Whelan}, Vol.~73, 75

\bibitem[{{Parsons} {et~al.}(2013){Parsons}, {G{\"a}nsicke}, {Marsh}, {Drake},
  {Dhillon}, {Littlefair}, {Pyrzas}, {Rebassa-Mansergas}, \&
  {Schreiber}}]{parsons2013}
{Parsons}, S.~G., {G{\"a}nsicke}, B.~T., {Marsh}, T.~R., {et~al.} 2013, \mnras,
  429, 256

\bibitem[{{Pastetter} \& {Ritter}(1989)}]{pastetter1989}
{Pastetter}, L. \& {Ritter}, H. 1989, \aap, 214, 186

\bibitem[{{Pavlovskii} \& {Ivanova}(2015)}]{pavlovskii2015}
{Pavlovskii}, K. \& {Ivanova}, N. 2015, \mnras, 449, 4415

\bibitem[{{Paxton} {et~al.}(2011){Paxton}, {Bildsten}, {Dotter}, {Herwig},
  {Lesaffre}, \& {Timmes}}]{paxton2011}
{Paxton}, B., {Bildsten}, L., {Dotter}, A., {et~al.} 2011, \apjs, 192, 3

\bibitem[{{Paxton} {et~al.}(2013){Paxton}, {Cantiello}, {Arras}, {Bildsten},
  {Brown}, {Dotter}, {Mankovich}, {Montgomery}, {Stello}, {Timmes}, \&
  {Townsend}}]{paxton2013}
{Paxton}, B., {Cantiello}, M., {Arras}, P., {et~al.} 2013, \apjs, 208, 4

\bibitem[{{Paxton} {et~al.}(2015){Paxton}, {Marchant}, {Schwab}, {Bauer},
  {Bildsten}, {Cantiello}, {Dessart}, {Farmer}, {Hu}, {Langer}, {Townsend},
  {Townsley}, \& {Timmes}}]{paxton2015}
{Paxton}, B., {Marchant}, P., {Schwab}, J., {et~al.} 2015, \apjs, 220, 15

\bibitem[{{Pelisoli} \& {Vos}(2019)}]{pelisoli2019b}
{Pelisoli}, I. \& {Vos}, J. 2019, \mnras, 488, 2892

\bibitem[{{Perlmutter} {et~al.}(1999){Perlmutter}, {Aldering}, {Goldhaber},
  {Knop}, {Nugent}, {Castro}, {Deustua}, {Fabbro}, {Goobar}, {Groom}, {Hook},
  {Kim}, {Kim}, {Lee}, {Nunes}, {Pain}, {Pennypacker}, {Quimby}, {Lidman},
  {Ellis}, {Irwin}, {McMahon}, {Ruiz-Lapuente}, {Walton}, {Schaefer}, {Boyle},
  {Filippenko}, {Matheson}, {Fruchter}, {Panagia}, {Newberg}, {Couch}, \&
  {Project}}]{perlmutter1999}
{Perlmutter}, S., {Aldering}, G., {Goldhaber}, G., {et~al.} 1999, \apj, 517,
  565

\bibitem[{{Postnov} \& {Yungelson}(2014)}]{postnov2014}
{Postnov}, K.~A. \& {Yungelson}, L.~R. 2014, Living Reviews in Relativity, 17,
  3

\bibitem[{{Rappaport} {et~al.}(1995){Rappaport}, {Podsiadlowski}, {Joss}, {Di
  Stefano}, \& {Han}}]{rappaport1995}
{Rappaport}, S., {Podsiadlowski}, P., {Joss}, P.~C., {Di Stefano}, R., \&
  {Han}, Z. 1995, \mnras, 273, 731

\bibitem[{{Rappaport} {et~al.}(1983){Rappaport}, {Verbunt}, \&
  {Joss}}]{rappaport1983}
{Rappaport}, S., {Verbunt}, F., \& {Joss}, P.~C. 1983, \apj, 275, 713

\bibitem[{{Rebassa-Mansergas} {et~al.}(2017){Rebassa-Mansergas}, {Parsons},
  {Garc{\'\i}a-Berro}, {G{\"a}nsicke}, {Schreiber}, {Rybicka}, \&
  {Koester}}]{rebassa2017}
{Rebassa-Mansergas}, A., {Parsons}, S.~G., {Garc{\'\i}a-Berro}, E., {et~al.}
  2017, \mnras, 466, 1575

\bibitem[{{Reimers}(1975)}]{reimers1975}
{Reimers}, D. 1975, in Problems in stellar atmospheres and envelopes., 229--256

\bibitem[{{Reindl} {et~al.}(2020){Reindl}, {Schaffenroth}, {Miller Bertolami},
  {Geier}, {Finch}, {Barstow}, {Casewell}, \& {Taubenberger}}]{reindl2020}
{Reindl}, N., {Schaffenroth}, V., {Miller Bertolami}, M.~M., {et~al.} 2020,
  \aap, 638, A93

\bibitem[{{Riess} {et~al.}(1998){Riess}, {Filippenko}, {Challis},
  {Clocchiatti}, {Diercks}, {Garnavich}, {Gilliland}, {Hogan}, {Jha},
  {Kirshner}, {Leibundgut}, {Phillips}, {Reiss}, {Schmidt}, {Schommer},
  {Smith}, {Spyromilio}, {Stubbs}, {Suntzeff}, \& {Tonry}}]{riess1998}
{Riess}, A.~G., {Filippenko}, A.~V., {Challis}, P., {et~al.} 1998, \aj, 116,
  1009

\bibitem[{{Robson} {et~al.}(2018){Robson}, {Cornish}, \& {Liu}}]{robson2018}
{Robson}, T., {Cornish}, N., \& {Liu}, C. 2018, arXiv e-prints
  [\eprint[arXiv]{1803.01944}]

\bibitem[{{Ruan} {et~al.}(2018){Ruan}, {Guo}, {Cai}, \& {Zhang}}]{taiji2018}
{Ruan}, W.-H., {Guo}, Z.-K., {Cai}, R.-G., \& {Zhang}, Y.-Z. 2018, arXiv
  e-prints, arXiv:1807.09495

\bibitem[{{Ruiter} {et~al.}(2010){Ruiter}, {Belczynski}, {Benacquista},
  {Larson}, \& {Williams}}]{ruiter2010}
{Ruiter}, A.~J., {Belczynski}, K., {Benacquista}, M., {Larson}, S.~L., \&
  {Williams}, G. 2010, \apj, 717, 1006

\bibitem[{{Saffer} {et~al.}(1988){Saffer}, {Liebert}, \&
  {Olszewski}}]{saffer1988}
{Saffer}, R.~A., {Liebert}, J., \& {Olszewski}, E.~W. 1988, \apj, 334, 947

\bibitem[{{Saladino} {et~al.}(2018){Saladino}, {Pols}, {van der Helm},
  {Pelupessy}, \& {Portegies Zwart}}]{saladino2018}
{Saladino}, M.~I., {Pols}, O.~R., {van der Helm}, E., {Pelupessy}, I., \&
  {Portegies Zwart}, S. 2018, \aap, 618, A50

\bibitem[{{Santander-Garc{\'\i}a} {et~al.}(2015){Santander-Garc{\'\i}a},
  {Rodr{\'\i}guez-Gil}, {Corradi}, {Jones}, {Miszalski}, {Boffin},
  {Rubio-D{\'\i}ez}, \& {Kotze}}]{santander2015}
{Santander-Garc{\'\i}a}, M., {Rodr{\'\i}guez-Gil}, P., {Corradi}, R.~L.~M.,
  {et~al.} 2015, \nat, 519, 63

\bibitem[{{Schreiber} {et~al.}(2016){Schreiber}, {Zorotovic}, \&
  {Wijnen}}]{schreiber2016}
{Schreiber}, M.~R., {Zorotovic}, M., \& {Wijnen}, T.~P.~G. 2016, \mnras, 455,
  L16

\bibitem[{{Shah} {et~al.}(2013){Shah}, {Nelemans}, \& {van der
  Sluys}}]{shah2013}
{Shah}, S., {Nelemans}, G., \& {van der Sluys}, M. 2013, \aap, 553, A82

\bibitem[{{Shah} {et~al.}(2012){Shah}, {van der Sluys}, \&
  {Nelemans}}]{shah2012}
{Shah}, S., {van der Sluys}, M., \& {Nelemans}, G. 2012, \aap, 544, A153

\bibitem[{{Soberman} {et~al.}(1997){Soberman}, {Phinney}, \& {van den
  Heuvel}}]{soberman1997}
{Soberman}, G.~E., {Phinney}, E.~S., \& {van den Heuvel}, E.~P.~J. 1997, \aap,
  327, 620

\bibitem[{{Tauris}(2001)}]{tauris2001}
{Tauris}, T.~M. 2001, in Astronomical Society of the Pacific Conference Series,
  Vol. 229, Evolution of Binary and Multiple Star Systems, ed.
  P.~{Podsiadlowski}, S.~{Rappaport}, A.~R. {King}, F.~{D'Antona}, \&
  L.~{Burderi}, 145

\bibitem[{{Tauris} \& {Savonije}(1999)}]{tauris1999}
{Tauris}, T.~M. \& {Savonije}, G.~J. 1999, \aap, 350, 928

\bibitem[{{Tauris} \& {van den Heuvel}(2006)}]{tauris2006}
{Tauris}, T.~M. \& {van den Heuvel}, E.~P.~J. 2006, in Compact stellar X-ray
  sources, Vol.~39, 623--665

\bibitem[{{Toonen} {et~al.}(2017){Toonen}, {Hollands}, {G{\"a}nsicke}, \&
  {Boekholt}}]{toonen2017}
{Toonen}, S., {Hollands}, M., {G{\"a}nsicke}, B.~T., \& {Boekholt}, T. 2017,
  \aap, 602, A16

\bibitem[{{Toonen} {et~al.}(2012){Toonen}, {Nelemans}, \& {Portegies
  Zwart}}]{toonen2012}
{Toonen}, S., {Nelemans}, G., \& {Portegies Zwart}, S. 2012, \aap, 546, A70

\bibitem[{{Torres} {et~al.}(2022){Torres}, {Canals}, {Jim{\'e}nez-Esteban},
  {Rebassa-Mansergas}, \& {Solano}}]{torres2022}
{Torres}, S., {Canals}, P., {Jim{\'e}nez-Esteban}, F.~M., {Rebassa-Mansergas},
  A., \& {Solano}, E. 2022, \mnras, 511, 5462

\bibitem[{{Totani} {et~al.}(2008){Totani}, {Morokuma}, {Oda}, {Doi}, \&
  {Yasuda}}]{totani2008}
{Totani}, T., {Morokuma}, T., {Oda}, T., {Doi}, M., \& {Yasuda}, N. 2008,
  \pasj, 60, 1327

\bibitem[{{Tout} {et~al.}(1997){Tout}, {Aarseth}, {Pols}, \&
  {Eggleton}}]{tout1997}
{Tout}, C.~A., {Aarseth}, S.~J., {Pols}, O.~R., \& {Eggleton}, P.~P. 1997,
  \mnras, 291, 732

\bibitem[{{Tremblay} {et~al.}(2011){Tremblay}, {Bergeron}, \&
  {Gianninas}}]{tremblay2011}
{Tremblay}, P.~E., {Bergeron}, P., \& {Gianninas}, A. 2011, \apj, 730, 128

\bibitem[{{Tremblay} {et~al.}(2014){Tremblay}, {Kalirai}, {Soderblom},
  {Cignoni}, \& {Cummings}}]{tremblay2014}
{Tremblay}, P.~E., {Kalirai}, J.~S., {Soderblom}, D.~R., {Cignoni}, M., \&
  {Cummings}, J. 2014, \apj, 791, 92

\bibitem[{{Tutukov} \& {Yungelson}(1981)}]{tutukov1981}
{Tutukov}, A.~V. \& {Yungelson}, L.~R. 1981, Nauchnye Informatsii, 49, 3

\bibitem[{{Vassiliadis} \& {Wood}(1993)}]{vassiliadis1993}
{Vassiliadis}, E. \& {Wood}, P.~R. 1993, \apj, 413, 641

\bibitem[{{Vennes} {et~al.}(2011){Vennes}, {Thorstensen}, {Kawka},
  {N{\'e}meth}, {Skinner}, {Pigulski}, {Ste\&{\c{s}}acute}, {licki},
  {Ko{\l}aczkowski}, \& {{\'S}r{\'o}dka}}]{vennes2011}
{Vennes}, S., {Thorstensen}, J.~R., {Kawka}, A., {et~al.} 2011, \apjl, 737, L16

\bibitem[{{Vos} {et~al.}(2019){Vos}, {Vu{\v{c}}kovi{\'c}}, {Chen}, {Han},
  {Boudreaux}, {Barlow}, {{\O}stensen}, \& {N{\'e}meth}}]{vos2019}
{Vos}, J., {Vu{\v{c}}kovi{\'c}}, M., {Chen}, X., {et~al.} 2019, \mnras, 482,
  4592

\bibitem[{{Wang} {et~al.}(2009){Wang}, {Chen}, {Meng}, \& {Han}}]{wangb2009}
{Wang}, B., {Chen}, X., {Meng}, X., \& {Han}, Z. 2009, \apj, 701, 1540

\bibitem[{{Wang} \& {Han}(2012)}]{wangb2012}
{Wang}, B. \& {Han}, Z. 2012, \nar, 56, 122

\bibitem[{{Wang} {et~al.}(2010){Wang}, {Li}, \& {Han}}]{wangb2010}
{Wang}, B., {Li}, X.-D., \& {Han}, Z.-W. 2010, \mnras, 401, 2729

\bibitem[{{Webbink}(1979)}]{webbink1979}
{Webbink}, R.~F. 1979, in IAU Colloq. 53: White Dwarfs and Variable Degenerate
  Stars, ed. H.~M. {van Horn}, V.~{Weidemann}, \& M.~P. {Savedoff}, 426

\bibitem[{{Webbink}(1984)}]{webbink1984}
{Webbink}, R.~F. 1984, \apj, 277, 355

\bibitem[{{Webbink}(1988)}]{webbink1988}
{Webbink}, R.~F. 1988, in Critical Observations Versus Physical Models for
  Close Binary Systems, 403--446

\bibitem[{{Webbink}(2008)}]{webbink2008}
{Webbink}, R.~F. 2008, in Astrophysics and Space Science Library, Vol. 352,
  Astrophysics and Space Science Library, ed. E.~F. {Milone}, D.~A. {Leahy}, \&
  D.~W. {Hobill}, 233

\bibitem[{{Webbink} \& {Han}(1998)}]{webbink1998}
{Webbink}, R.~F. \& {Han}, Z. 1998, in American Institute of Physics Conference
  Series, Vol. 456, Laser Interferometer Space Antenna, Second International
  LISA Symposium on the Detection and Observation of Gravitational Waves in
  Space, ed. W.~M. {Folkner}, 61--67

\bibitem[{{Whelan} \& {Iben}(1973)}]{whelan1973}
{Whelan}, J. \& {Iben}, Icko, J. 1973, \apj, 186, 1007

\bibitem[{{Willems} \& {Kolb}(2004)}]{willems2004}
{Willems}, B. \& {Kolb}, U. 2004, \aap, 419, 1057

\bibitem[{{Woods} \& {Ivanova}(2011)}]{woods2011}
{Woods}, T.~E. \& {Ivanova}, N. 2011, \apjl, 739, L48

\bibitem[{{Woods} {et~al.}(2012){Woods}, {Ivanova}, {van der Sluys}, \&
  {Chaichenets}}]{woods2012}
{Woods}, T.~E., {Ivanova}, N., {van der Sluys}, M.~V., \& {Chaichenets}, S.
  2012, \apj, 744, 12

\bibitem[{{York} {et~al.}(2000){York}, {Adelman}, {Anderson}, {Anderson},
  {Annis}, {Bahcall}, \& {SDSS Collaboration}}]{york2000}
{York}, D.~G., {Adelman}, J., {Anderson}, John~E., J., {et~al.} 2000, \aj, 120,
  1579

\bibitem[{{Yu} \& {Jeffery}(2010)}]{yus2010}
{Yu}, S. \& {Jeffery}, C.~S. 2010, \aap, 521, A85

\bibitem[{{Yungelson} \& {Kuranov}(2017)}]{yungelson2017}
{Yungelson}, L.~R. \& {Kuranov}, A.~G. 2017, \mnras, 464, 1607

\bibitem[{{Zorotovic} {et~al.}(2010){Zorotovic}, {Schreiber}, {G{\"a}nsicke},
  \& {Nebot G{\'o}mez-Mor{\'a}n}}]{zorotovic2010}
{Zorotovic}, M., {Schreiber}, M.~R., {G{\"a}nsicke}, B.~T., \& {Nebot
  G{\'o}mez-Mor{\'a}n}, A. 2010, \aap, 520, A86

\end{thebibliography}

\clearpage
\onecolumn

\begin{appendix}

\section{Orbital changes due to wind accretion}
\label{app:A}

\begin{figure}
    \centering
    \includegraphics[width=0.6\columnwidth]{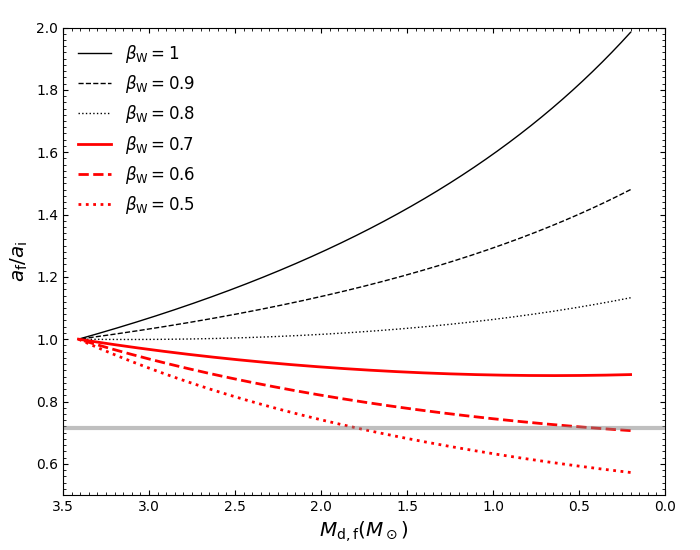}
    \caption{{Orbital changes as a function of the donor mass due to the wind accretion processes {with $M_{\rm 1,i}=3.41M_\odot$, $M_{\rm 2,i}=3.06M_\odot$. The grey line is for $a_{\rm f}/a_{\rm i}\simeq 0.71$, corresponding to the peaks of $P_{\rm 1,mt}/P_{\rm i}$ distributions in Figure \ref{fig:4}.} The results for $\beta_{\rm W}$ from $0.5$ to $1$ are shown in black and red lines, as marked in the plot. We note that the wind accretion efficiency is equal to $1-\beta_{\rm W}$.}}
    \label{fig:B1}
\end{figure}

In \citet{hurley2002}, the interaction between the wind material and the binary orbit is taken into account, which is different from that in some previous works, such as in \citet{tauris2006,postnov2014}. Assuming that $\beta_{\rm W}$ is the wind mass loss efficiency, then the wind accretion rate of the accretor is $\dot{M}_{\rm 2}=-(1-\beta_{\rm W})\dot{M}_{\rm 1}$. Eq. \ref{eq:13} can be rewritten as:
\begin{eqnarray}
  \frac{\dot{J}_{\rm orb}}{J_{\rm orb}} = \frac{M_{2}\dot{M}_{\rm 1}}{M M_1}-\frac{\dot{M}_{\rm 2}}{M}= \left(1-\frac{\beta_{\rm W}M_1}{M}\right)\frac{\dot{M}_{\rm 1}}{M_1},
\label{eq:b1}
\end{eqnarray}
where $J_{\rm orb}$ is the orbital angular momentum. Here, we only consider the wind accretion processes, then $\dot{J}_{\rm orb} = \dot{J}_{\rm W}$. The change of orbital separation is given by \citep{postnov2014}:
\begin{eqnarray}
  \frac{\dot{a}}{a} = -2\left(1+(\beta_{\rm W}-1)\frac{M_1}{M_2}-\frac{\beta_{\rm W}}{2}\frac{M_1}{M}\right)\frac{\dot{M}_{\rm 1}}{M_1}+2\frac{\dot{J}_{\rm W}}{J_{\rm orb}},
  \label{eq:b2}
\end{eqnarray}
If $\beta_{\rm W}$ is constant, substituting Eq. \ref{eq:b1} into Eq. \ref{eq:b2} and integrating over time, we arrive at 
\begin{eqnarray}
  \frac{a_{\rm f}}{a_{\rm i}} = \left(\frac{M_{\rm f}}{M_{\rm i}}\right)^{-1}\left(\frac{M_{\rm 2,f}}{M_{\rm 2,i} }\right)^{-2},
  \label{eq:b3}
\end{eqnarray}
where the subscripts "i" and "f" stand for the initial and final values, respectively. We note that $M_{\rm 2,f}$ is a function of $\beta_{\rm W}$, namely, $M_{\rm 2,f}=(1-\beta_{\rm W})(M_{\rm 1,i}-M_{\rm 1,f})+M_{\rm 2,i}$. For a given initial binary, the change of orbital separation is affected by the final masses of the binary components and the value of $\beta_{\rm W}$. {Figure \ref{fig:B1} presents the orbital evolution due to the wind accretion processes with different accretion efficiencies for a binary with $M_{\rm 1,i}=3.41M_\odot$, $M_{\rm 2,i}=3.06M_\odot$. The grey line is for $a_{\rm f}/a_{\rm i}\simeq 0.71$, corresponding to the peaks of $P_{\rm 1,mt}/P_{\rm i}$distributions in Figure \ref{fig:4}. For a wind accretion efficiency lower than $\sim 0.2$ ($\beta_{\rm W}\gtrsim 0.8$), the stellar winds always lead to a widened orbit ($a_{\rm f}/a_{\rm i}> 1$); whereas for a greater wind accretion efficiency, the wind-fed mass transfer would lead to the shrinkage of the binary orbit. The orbital separation change of $0.71$ suggests that about 40 percent transferred material needs to be accreted by the companion. For a star filling its Roche lobe at the tip of an AGB, the wind mass loss and RLOF processes happen simultaneously. In consideration of the possible increase of the orbit caused by the stable RLOF phase, the shrinkage of binary orbit then needs that the wind mass transfer dominates the binary evolution, that is, more mass is lost from the wind than that from the RLOF.}
 
{The wind accretion rate is calculated based on Eq. \ref{eq:12}, that is, Bondi-Hoyle accretion. The Bindi-Hoyle model is a good approximation for wind accretion if the wind is fast compared to the orbital velocity. However, AGB stars generally have slow winds, typically $5-30\;\rm km s^{-1}$ \citep{knapp1998}, which is on the same order of magnitude as the orbital velocities. The slow wind interaction is much more complex and is still under debate. Several 3D hydrodynamic simulations have suggested that the wind accretion efficiencies for AGB stars are in the broad range between $0.01-0.4$ (e.g., \citealt{liuz2017,saladino2018,chenz2020}). In this work, we mainly focus on the influences of mass transfer stability on DWD populations, and the basic assumptions relating to the binary evolution are same for Ge's and the polytropic models. Therefore, the wind accretion processes have limited effects on our main conclusions.}

\section{DWD mass distribution for different channels}
\label{app:B}

\begin{figure}
                \centering
                \includegraphics[width=0.6\columnwidth]{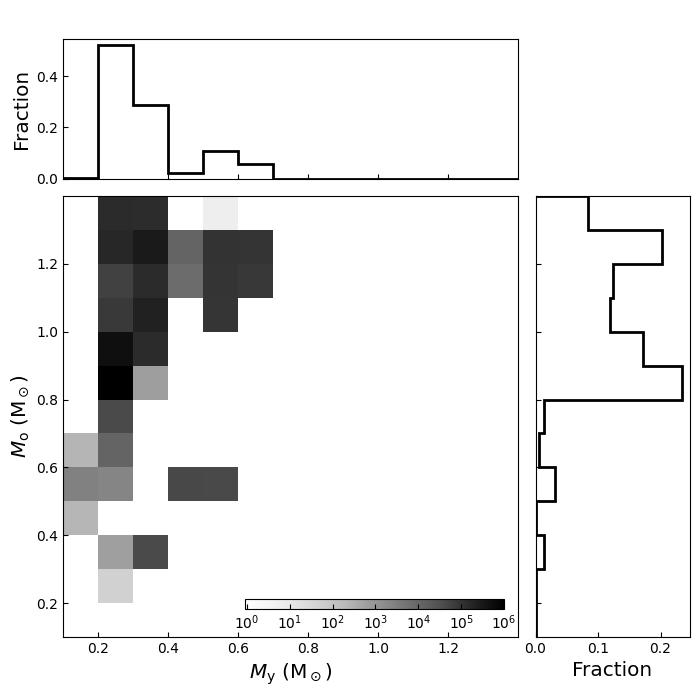}
                \centering
                \includegraphics[width=0.6\columnwidth]{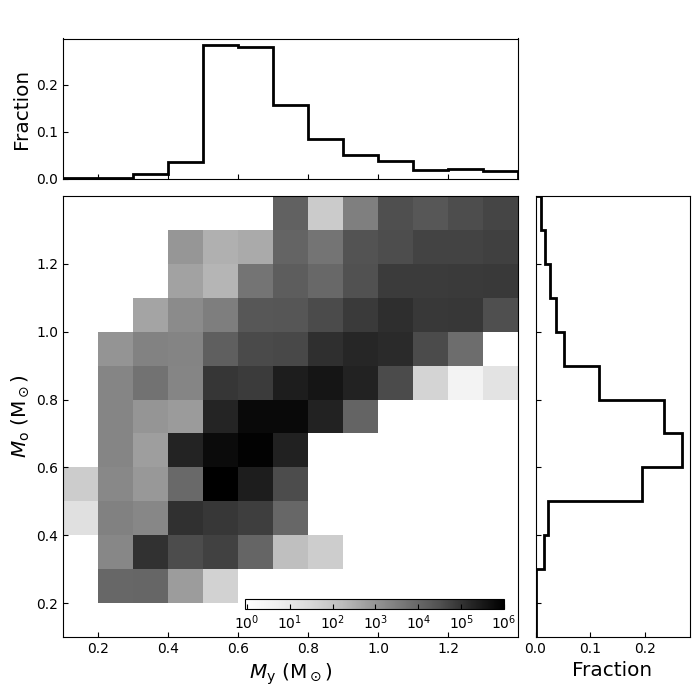}
        \caption{Number density distribution of DWDs from CE+CE channel (upper panel) and RL+CE channel (low panel) for Ge’s model. We note that the scale of the color bars is different for the two panels.}
        \label{fig:D1}
\end{figure}

\begin{figure}
                \centering
                \includegraphics[width=0.6\columnwidth]{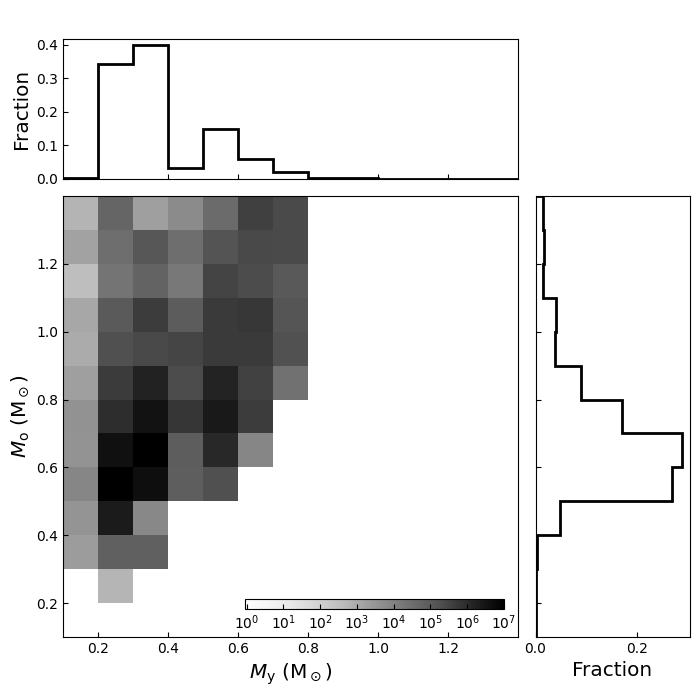}
                \centering
                \includegraphics[width=0.6\columnwidth]{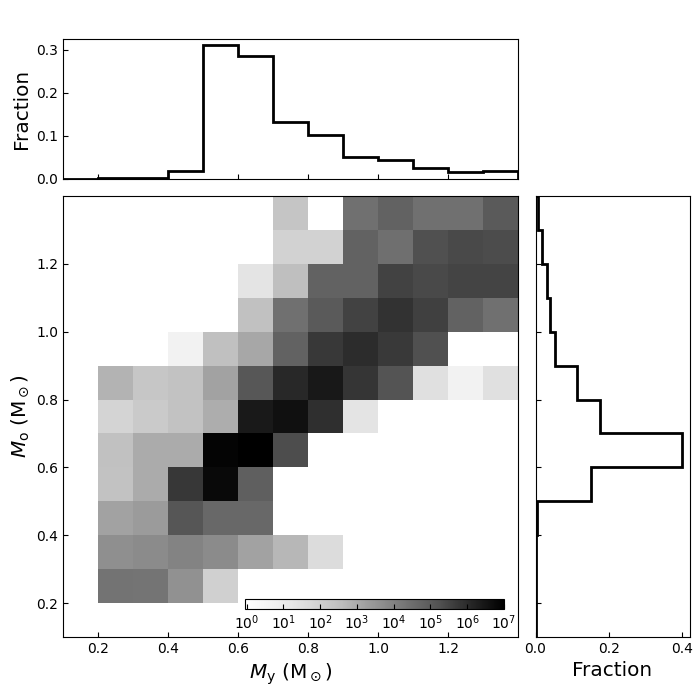}
        \caption{The number density distribution of DWDs from CE+CE channel (upper panel) and RL+CE channel (low panel) for the polytropic model.}
        \label{fig:D2}
\end{figure}

The WD mass distributions for DWDs from different channels are presented, where the results in Ge's and polytropic models are shown in Figure \ref{fig:D1} and Figure \ref{fig:D2}, respectively. In Ge's model, most DWDs are produced from the RL+CE channel, while in the polytropic model, the contribution of the RL+CE channel and CE+CE channel is comparable. 

%\newpage

\clearpage
%\FloatBarrier

\section{Observation sample of DWDs}
\label{app:C}

\renewcommand\arraystretch{1.2}
%\longtab[0]{
\begin{longtable}{lccccc}
\caption{Observation sample of DWDs.}
\label{tab:C1}\\
\hline\hline
Name & Period & $M_{\rm brighter}$ & $M_{\rm dim}$ & SDSS g$-$band & Refs\\
 & d & $M_\odot$ & $M_\odot$ & mag & \\
\hline
\endfirsthead
\caption{continued}\\
\hline
\hline
Name & Period & $M_{\rm brighter}$ & $M_{\rm dim}$ & SDSS g$-$band & Refs\\
 & d & $M_\odot$ & $M_\odot$ & mag & \\
\hline
\endhead
\hline
\endfoot
\hline 
\endlastfoot
ZTF J1539+5027 & 0.00480 & $0.21_{-0.015}^{+0.014}$ & $0.61_{-0.022}^{+0.017}$ & $20.37\pm 0.019$ & 1\\
ZTF J2243+5242 & 0.00611 & $0.349_{-0.074}^{+0.093}$ & $0.384_{-0.074}^{+0.0114}$ & $20.359\pm 0.029$ & 2\\
J0651+2844 & 0.00886 & $0.247$ & $0.49$ & $19.109\pm 0.010$ & 3,4,5 \\
ZTF J0538+1953 & 0.01003 & $0.45_{-0.05}^{+0.05}$ & $0.32_{-0.03}^{+0.03}$ & $18.734\pm 0.008$ & 6 \\
PTF J0533+0209 & 0.01428 & $0.167$ & $0.652$ & $18.980\pm 0.010$ & 7 \\
ZTF J2029+1534 & 0.01449 & $0.32$ & $0.30$ & $20.380\pm 0.024$ & 6 \\
ZTF J0722-1839 & 0.01646 & $0.38_{-0.04}^{+0.04}$ & $0.33_{-0.03}^{+0.03}$ & $18.976\pm 0.012$ & 6 \\
ZTF J1749+0924 & 0.01836 & $0.40_{-0.05}^{+0.07}$ & $0.28_{-0.04}^{+0.05}$ & $20.357\pm 0.022$ & 6 \\
J0106+1000 & 0.02715 & $0.189_{-0.012}^{+0.011}$ & $0.56_{-0.08}^{+0.22}$ & $19.728\pm 0.016$ & 8,9,5 \\
J1630+4233 & 0.02766 & $0.298_{-0.009}^{+0.019}$ & $0.76_{-0.25}^{+0.26}$ & $19.037\pm 0.009$ & 10,9,5 \\
SDSS J0822+3048 & 0.02813 & $0.304_{-0.014}^{+0.140}$ & $0.524_{-0.05}^{+0.05}$ & $20.337\pm 0.021$ & 11,12 \\
ZTF J1901+5309 & 0.02820 & $0.36_{-0.04}^{+0.04}$ & $0.36_{-0.05}^{+0.05}$ & $17.891\pm 0.006$ &  13,6\\
SDSS J1043+0551 & 0.03170 & $0.183_{-0.01}^{+0.01}$ & $\geq 0.07$ & $19.126\pm 0.011$ & 11 \\
J1235+1543 & 0.03438 & $0.363_{-0.011}^{+0.011}$ & $\geq 0.179$ & $17.328\pm 0.005$ & 14 \\
J1053+5200 & 0.04256 & $0.204_{-0.012}^{+0.012}$ & $0.77_{-0.25}^{+0.24}$ & $18.936\pm 0.009$ & 15,9,5 \\
J0056-0611 & 0.04338 & $0.18_{-0.01}^{+0.01}$ & $0.82_{-0.11}^{+0.14}$ & $17.419\pm 0.005$ & 16,17,9,5 \\
J1056+6536 & 0.04351 & $0.334_{-0.016}^{+0.016}$ & $0.77_{-0.24}^{+0.24}$ & $19.761\pm 0.015$ & 16,18,9,5 \\
J0923+3028 & 0.04495 & $0.275_{-0.015}^{+0.015}$ & $0.77_{-0.24}^{+0.25}$ & $15.677\pm 0.003$ & 19,9,5 \\
J1436+5010 & 0.04580 & $0.234_{-0.013}^{+0.013}$ & $0.79_{-0.22}^{+0.24}$ & $18.220\pm 0.006$ & 20,15,9,5 \\
J1711+2724 & 0.05200 & $0.672_{-0.044}^{+0.044}$ & $\geq 0.192$ & $17.156\pm 0.005$ & 14 \\
J0825+1152 & 0.05819 & $0.279_{-0.021}^{+0.021}$ & $0.79_{-0.21}^{+0.25}$ & $18.755\pm 0.008$ & 18,9,5 \\
WD0597-666 & 0.06099 & $0.37_{-0.02}^{+0.02}$ & $0.32_{-0.03}^{+0.03}$ & $-$ & 21,22 \\
J1741+6526 & 0.06111 & $0.17_{-0.01}^{+0.01}$ & $1.17_{-0.05}^{+0.08}$ & $18.411\pm 0.007$ & 20,23,9,5 \\
J0755+4906 & 0.06302 & $0.184_{-0.01}^{0.01}$ & $0.97_{-0.11}^{+0.18}$ & $20.246\pm 0.020$ & 19,9,5 \\
J0736+1622 & 0.06900 & $0.36_{-0.007}^{+0.007}$ & $\geq 0.033$ & $16.305\pm 0.005$ & 14 \\
J2338-2052 & 0.07644 & $0.258_{-0.016}^{+0.016}$ & $0.75_{-0.25}^{+0.27}$ & $19.655\pm 0.017$ & 16,9,5 \\
J2309+2603 & 0.07653 & $0.176_{-0.01}^{+0.01}$ & $0.93_{-0.11}^{+0.18}$ & $19.26\pm 0.010$ & 5 \\
J0849+0445 & 0.07870 & $0.179_{-0.01}^{+0.01}$ & $0.86_{-0.15}^{+0.20}$ & $19.292\pm 0.010$ & 20,15,9,5 \\
J0022-1014 & 0.07989 & $0.375_{-0.02}^{+0.02}$ & $\geq 0.23$ & $19.761\pm 0.018$ & 24,9,5 \\
J0751-0141 & 0.08001 & $0.194_{-0.01}^{+0.01}$ & $0.98_{-0.02}^{+0.02}$ & $17.453\pm 0.005$ & 16,23,9,5 \\
J2119-0018 & 0.08677 & $0.159_{-0.01}^{+0.01}$ & $0.83_{-0.05}^{+0.15}$ & $20.168\pm 0.020$ & 19,17,9,5 \\
J1234+0228 & 0.09143 & $0.227_{-0.014}^{+0.014}$ & $0.74_{-0.25}^{+0.26}$ & $17.843\pm 0.005$ & 24,20,9,5 \\
J1152+0248 & 0.09987 & $0.47_{-0.11}^{+0.11}$ & $0.442_{-0.088}^{+0.088}$ & $18.35\pm 0.020$ & 25 \\
J1005-2249 & 0.11602 & $0.378_{-0.023}^{+0.023}$ & $0.316_{-0.011}^{+0.011}$ & $17.266\pm 0.005$ & 26 \\
WD1242-205 & 0.11877 & $0.39_{-0.02}^{+0.02}$ & $0.56_{-0.03}^{+0.03}$ & $19.109\pm 0.010$ & 27 \\
WD1704+481.2 & 0.14479 & $0.39_{-0.05}^{+0.05}$ & $0.54_{-0.07}^{+0.07}$ & $14.434\pm 0.004$ & 28,22 \\
WD1101+364 & 0.14500 & $0.32$ & $0.37$ & $14.496\pm 0.003$ & 29 \\
J0112+1835 & 0.14698 & $0.16_{-0.01}^{+0.01}$ & $0.74_{-0.06}^{+0.16}$ & $17.110\pm 0.016$ & 20,17,9,5 \\
J1233+1602 & 0.15090 & $0.169_{-0.01}^{+0.01}$ & $0.98_{-0.1}^{+0.17}$ & $19.911\pm 0.017$ & 19,9,5 \\
LAMOST J0140355+392651 & 0.15863 & $0.19_{-0.08}^{+0.13}$ & $0.95_{-0.13}^{+0.14}$ & $17.64\pm 0.020$ & 30 \\
WD 2331+290 & 0.16649 & $0.44$ & $\geq 0.34$ & $15.53\pm0.04$ & 31,29 \\
J1112+117 & 0.17248 & $0.176_{-0.01}^{+0.01}$ & $0.75_{-0.25}^{+0.25}$ & $16.307\pm 0.016$ & 32,20,9,5 \\
Hen2-428 & 0.17580 & $0.42_{-0.07}^{+0.07}$ & $0.66_{-0.11}^{+0.11}$ & $-$ & 33,34 \\
J1005+3550 & 0.17652 & $0.168_{-0.01}^{+0.01}$ & $0.75_{-0.24}^{+0.25}$ & $19.004\pm 0.610$ & 18,16,9,5 \\
J2349+3553 & 0.18130 & $0.386_{-0.014}^{+0.014}$ & $\geq 0.742$ & $18.406\pm0.007$ & 14 \\
J0818+3536 & 0.18315 & $0.165_{-0.01}^{+0.01}$ & $0.76_{-0.24}^{+0.25}$ & $20.756\pm 0.026$ & 19,9,5 \\
SDSS J2357+5428 & 0.18979 & $0.15_{-}^{+0.09}$ & $1.06_{-0.05}^{+0.05}$ & $16.829\pm 0.004$ & 35,36 \\
J1443+1509 & 0.19053 & $0.201_{-0.013}^{+0.013}$ & $1.0_{-0.10}^{+0.17}$ & $18.650\pm 0.016$ & 20,9,5 \\
J1840+6423 & 0.19130 & $0.182_{-0.011}^{+0.011}$ & $0.86_{-0.15}^{+0.21}$ & $18.963\pm 0.014$ & 32,20,9,5 \\
J2103-0027 & 0.20308 & $0.161_{-0.01}^{+0.01}$ & $0.89_{-0.14}^{+0.20}$ & $18.488\pm 0.014$ & 18,16,9,5 \\
HE0225-1912 & 0.22000 & $0.545$ & $0.23$ & $-$ & 37 \\
J1238+1946 & 0.22275 & $0.21_{-0.011}^{+0.011}$ & $0.87_{-0.14}^{+0.20}$ & $17.291\pm 0.023$ & 16,9,5 \\
J0345+1748 & 0.23503 & $0.218_{-0.012}^{+0.012}$ & $0.80_{-0.03}^{+0.03}$ & $-$ & 38 \\
J0822+2753 & 0.24400 & $0.192_{-0.012}^{+0.012}$ & $0.93_{-0.12}^{+0.18}$ & $18.314\pm 0.013$ & 15,20,9,5 \\
J1717+6757 & 0.24614 & $0.185_{-0.01}^{+0.01}$ & $0.90_{-0.04}^{+}$ & $-$ & 39 \\
J1631+0605 & 0.24776 & $0.162_{-0.01}^{+0.01}$ & $0.79_{-0.20}^{+0.23}$ & $19.262\pm 0.011$ & 5 \\
J2132+0754 & 0.25056 & $0.187_{-0.01}^{+0.01}$ & $1.07_{-0.08}^{+0.14}$ & $18.110\pm 0.006$ & 16,9,5 \\
J1141+3850 & 0.25958 & $0.177_{-0.01}^{+0.01}$ & $0.92_{-0.12}^{+0.19}$ & $19.058\pm 0.017$ & 16,9,5 \\
J1630+2712 & 0.27646 & $0.17_{-0.01}^{+0.01}$ & $0.80_{-0.18}^{+0.23}$ & $20.146\pm 0.017$ & 19,9,5 \\
HE2209-1414 & 0.27693 & $0.58_{-0.03}^{+0.03}$ & $0.58_{-0.08}^{+0.08}$ & $-$ & 40 \\
WD2020-425 & 0.30000 & $0.813$ & $0.54$ & $-$ & 41 \\
J1005+0542 & 0.30560 & $0.388_{-0.02}^{+0.02}$ & $0.91_{-0.21}^{+}$ & $19.763\pm 0.023$ & 18,9 \\
J0917+4638 & 0.31642 & $0.173_{-0.01}^{+0.01}$ & $0.77_{-0.24}^{+0.24}$ & $18.764\pm 0.019$ & 41,9,5 \\
PG1114+224 & 0.31980 & $0.41$ & $\geq 0.07$ & $16.094\pm0.004$ & 29 \\
J0152+0749 & 0.32288 & $0.169_{-0.01}^{+0.01}$ & $0.83_{-0.17}^{+0.22}$ & $18.033\pm 0.009$ & 20,9,5 \\
WD0453-295 & 0.36000 & $0.399$ & $0.44$ & $-$ & 37 \\
J1422+4352 & 0.37930 & $0.181_{-0.01}^{+0.01}$ & $0.78_{-0.22}^{+0.23}$ & $19.822\pm 0.023$ & 19,9,5 \\
WD0028-474 & 0.38960 & $0.60_{-0.06}^{+0.06}$ & $0.45_{-0.04}^{+0.04}$ & $-$ & 43,37 \\
J1046-0153 & 0.39539 & $0.375_{-0.02}^{+0.02}$ & $0.23_{-0.04}^{+}$ & $18.098\pm 0.018$ & 20,16,9 \\
J1557+2823 & 0.40741 & $0.461_{-0.02}^{+0.02}$ & $0.52_{-0.1}^{+}$ & $17.712\pm 0.029$ & 20,16,9 \\
J1538+0252 & 0.41915 & $0.168_{-0.01}^{+0.01}$ & $0.92_{-0.12}^{+0.19}$ & $18.721\pm 0.014$ & 16,9,5 \\
J1439+1002 & 0.43741 & $0.181_{-0.01}^{+0.01}$ & $0.79_{-0.2}^{+0.23}$ & $17.938\pm 0.012$ & 19,9,5 \\
J0940+6304 & 0.48438 & $0.18_{-0.01}^{+0.01}$ & $0.90_{-0.13}^{+0.19}$ & $19.765\pm 0.014$ & 5 \\
J0022+0031 & 0.49135 & $0.457_{-0.02}^{+0.02}$ & $0.28_{-0.05}^{+}$ & $19.284\pm 0.033$ & 24,16,9 \\
HE0410-1137 & 0.50867 & $0.51_{-0.04}^{+0.04}$ & $0.39_{-0.03}^{+0.03}$ & $-$ & 43,37 \\
HE1414-0848 & 0.51781 & $0.55_{-0.03}^{+0.03}$ & $0.71_{-0.03}^{+0.03}$ & $-$ & 44 \\
J0840+1527 & 0.52155 & $0.192_{-0.01}^{+0.01}$ & $0.76_{-0.25}^{+0.25}$ & $19.319\pm 0.027$ & 16,9,5 \\
J0755+4800 & 0.54627 & $0.409_{-0.02}^{+0.02}$ & $1.17_{-0.28}^{+}$ & $16.039\pm 0.017$ & 16,9 \\
J0802-0955 & 0.54687 & $0.198_{-0.013}^{+0.012}$ & $0.82_{-0.18}^{+0.23}$ & $18.885\pm 0.012$ & 16,9,5 \\
J1104+0918 & 0.55319 & $0.454_{-0.02}^{+0.02}$ & $0.69_{-0.14}^{+}$ & $16.659\pm 0.014$ & 16,9,20 \\
J1157+0546 & 0.56500 & $0.186_{-0.02}^{+0.02}$ & $0.61_{-0.15}^{+}$ & $19.818\pm 0.024$ & 16,9,20 \\
J1518+1354 & 0.57655 & $0.147_{-0.18}^{+0.18}$ & $0.75_{-0.25}^{+0.28}$ & $19.121\pm 0.010$ & 5 \\
J1512+2615 & 0.59999 & $0.250_{-0.014}^{+0.014}$ & $0.76_{-0.23}^{+0.25}$ & $19.486\pm 0.012$ & 19,9,5 \\
J1518+0658 & 0.60935 & $0.224_{-0.013}^{+0.013}$ & $0.84_{-0.17}^{+0.22}$ & $17.581\pm 0.017$ & 20,32,9,5 \\
WD1210+140 & 0.64194 & $0.334$ & $\geq 0.46$ & $14.45\pm0.03$ & 29,37,45 \\
J1151+5858 & 0.66902 & $0.186_{-0.011}^{+0.011}$ & $0.84_{-0.15}^{+0.21}$ & $20.150\pm 0.025$ & 16,9,20 \\
J0730+1703 & 0.69770 & $0.182_{-0.01}^{+0.01}$ & $0.76_{-0.23}^{+0.24}$ & $20.076\pm 0.028$ & 18,16,9,5 \\
WD1534+503 & 0.71129 & $0.392_{-0.059}^{+0.069}$ & $0.617_{-0.096}^{+0.110}$ & $-$ & 46 \\
J0845+1624 & 0.75599 & $0.434_{-0.02}^{+0.02}$ & $0.23_{-0.04}^{+}$ & $19.817\pm 0.020$ & 18,9 \\
J0811+0225 & 0.82194 & $0.179_{-0.01}^{+0.01}$ & $1.27_{-0.06}^{+0.11}$ & $18.669\pm 0.024$ & 16,9,5 \\
J1039+1645 & 0.82470 & $0.458_{-0.018}^{+0.018}$ & $\geq 0.31$ & $19.065\pm0.01$ & 5 \\
PG1519+500 & 0.86030 & $0.42$ & $\geq 0.14$ & $16.210\pm0.03$ & 29 \\
HE0320-1917 & 0.86492 & $0.311$ & $\geq 0.45$ &$-$ & 37,45 \\
J0125+2017 & 0.88758 & $0.184_{-0.01}^{+0.01}$ & $\geq 0.14$ & $17.335\pm0.005$ & 5 \\
J1241+0633 & 0.95912 & $0.199_{-0.012}^{+0.012}$ & $0.80_{-0.19}^{+0.23}$ & $17.795\pm 0.006$ & 5 \\
WD2234+222 & 1.01016 & $0.186_{-0.01}^{+0.01}$ & $0.77_{-0.22}^{+0.24}$ & $17.163\pm 0.019$ & 47,20,16,9,5 \\
J0815+2309 & 1.07357 & $0.20_{-0.021}^{+0.021}$ & $0.80_{-0.21}^{+0.26}$ & $17.805\pm 0.015$ & 16,9,5 \\
PG0934+338 & 1.11420 & $0.38$ & $\geq 0.50$ & $16.121\pm0.004$ & 29 \\
PG1713+333 & 1.12740 & $0.41$ & $\geq 0.19$ & $14.258\pm0.003$ & 31,29 \\
WD1428+373 & 1.15674 & $0.348$ & $\geq 0.233$ & $15.299\pm0.003$ & 48,49 \\
WD1022+050 & 1.15700 & $0.44$ & $\geq 0.30$ & $14.132\pm0.005$ & 29 \\
PG0834+501 & 1.28490 & $0.40$ & $\geq 0.22$ & $14.994\pm0.003$ & 29 \\
PG1036+086 & 1.32830 & $0.42$ & $\geq 0.37$ & $16.177\pm0.004$ & 29 \\
WD0136+768 & 1.40722 & $0.47$ & $0.37$ & $-$ & 22,50 \\
PG1202+608 & 1.49300 & $0.40$ & $\geq 0.34$ & $13.34\pm0.08$ & 29 \\
WD0135-052 & 1.55578 & $0.47_{-0.05}^{+0.05}$ & $0.52_{-0.05}^{+0.05}$ & $13.098\pm 0.001$ & 51,52,22 \\
J1130+0933 & 1.55935 & $0.179_{-0.01}^{+0.01}$ & $\geq 0.19$ & $16.939\pm0.004$ & 5 \\
WD1204+450 & 1.60266 & $0.46$ & $0.52$ & $15.049\pm 0.003$ & 50,22 \\
WD0326-273 & 1.87540 & $0.364$ & $\geq 0.96$ & $-$ & 37,45 \\
J0318-0107 & 1.91280 & $0.40_{-0.05}^{+0.05}$ & $0.49_{-0.05}^{+0.05}$ & $14.585\pm 0.003$ & 37,43 \\
PG1632+177 & 2.04987 & $0.392_{-0.059}^{+0.059}$ & $0.526_{-0.082}^{+0.095}$ & $13.412\pm 0.002$ & 46 \\
J1128+1743 & 2.16489 & $0.183_{-0.01}^{+0.01}$ & $\geq 0.11$ & $19.494\pm 0.013$ & 5 \\
WD1349+144 & 2.21000 & $0.553$ & $0.33$ & $15.058\pm 0.010$ & 37,45 \\
HE1511-0448 & 3.22200 & $0.497$ & $\geq 0.67$ & $-$ & 37,45 \\
WD 1241-010 & 3.34741 & $0.4$ & $\geq 0.42$ & $16.327\pm0.014$ & 31,29 \\
WD 1317+453 & 4.87214 & $0.36$ & $\geq 0.44$ & $16.218\pm0.013$ & 31,29 \\
WD 2032+188 & 5.08460 & $0.406$ & $\geq 0.469$ & $-$ & 49,50 \\
WD 1824+040 & 6.26602 & $0.428$ & $\geq 0.515$ & $-$ & 49,50 \\
PG 1115+166 & 30.08730 & $0.69$ & $\geq 0.52$ & $14.92\pm0.02$ & 37,53 \\

%\hline
%\hline
\end{longtable} 
\tablefoot{{$M_{\rm brighter}$ and $M_{\rm dim}$ are for the WDs with higher and lower effective temperature. References}-- (1) \citet{burdge2019}; (2) \citet{burdge2020a}; (3) \citet{brown2011b}; (4) \citet{hermes2012}; (5) \citet{brown2016a}; (6) \citet{burdge2020b}; (7) \citet{burdge2019b}; (8) \citet{kilic2011c}; (9) \citet{gianninas2014}; (10) \citet{kilic2011b}; (11) \citet{brown2017}; (12) \citet{kosakowski2021}; (13) \citet{coughlin2020}; (14) \citet{breedt2017}; (15) \citet{kilic2010b}; (16) \citet{brown2013}; (17) \citet{hermes2014}; (18) \citet{kilic2012}; (19) \citet{brown2010}; (20) \citet{brown2012}; (21) \citet{moran1997}; (22) \citet{maxted2002c}; (23); \citet{kilic2014}; (24) \citet{kilic2011a}; (25) \citet{hallakoun2016}; (26) \citet{bours2014}; (27) \citet{debes2015}; (28) \citet{maxted2000a}; (29) \citet{brown2011a}; (30) \citet{elbadry2021}; (31) \citet{marsh1995}; (32) \citet{hermes2013}; (33) \citet{santander2015}; (34) \citet{reindl2020}; (35) \citet{marsh2011}; (36) \citet{bours2015}; (37) \citet{napiwotzki2020}; (38) \citet{parsons2013}; (39) \citet{vennes2011}; (40) \citet{karl2003}; (41) \citet{napiwotzki2007}; (42) \citet{kilic2007}; (43) \citet{rebassa2017}; (44) \citet{napiwotzki2002}; (45) \citet{nelemans2005b}; (46) \citet{kilic2021}; (47) \citet{kilic2009}; (48) \citet{marsh2000b}; (49) \citet{morales2005}; (50) \citet{maxted1999}; (51) \citet{bergeron1989}; (52) \citet{saffer1988}; (53) \citet{maxted2002a}}
%}

\end{appendix}

%----------------------------------------------------------------
\end{document}